\documentclass{LMCS}

\def\doi{7 (4:04) 2011}
\lmcsheading%
{\doi}
{1--48}
{}
{}
{Jan.~31, 2010}
{Nov.~23, 2011}
{}

\usepackage{enumerate}
\usepackage{hyperref}

\usepackage{amsmath,amssymb,amsfonts,latexsym,stmaryrd}
\usepackage{epsfig}

\usepackage{gastex,delarray,wrapfig}
\usepackage{extarrows}

\usepackage{verbatim}
\usepackage[mathscr]{euscript}
\usepackage{subfigure,multimedia,mathtools}
\usepackage{delarray}
\usepackage{graphicx}
\usepackage{tikz}
\usepackage{listings}
\usepackage{float}

\usepackage{pgf}
\usepackage{tikz}
\usetikzlibrary{automata}
\usetikzlibrary{trees}
\usetikzlibrary{shapes}
\usetikzlibrary{petri}
\usetikzlibrary{arrows}
\usetikzlibrary{snakes}
\usetikzlibrary{backgrounds}

\newcommand{\dcps}{{DCPS}}

\newcommand{\open}{\mbox{$[\![$}}
\newcommand{\close}{\mbox{$]\!]$}}

\def\by#1{\mathop{{\hbox{\setbox0=\hbox{$\scriptstyle{#1\quad}$}{$%
\mathrel{\mathop{\setbox1=\hbox to \wd0{\rightarrowfill}\ht1=3pt\dp1=-2pt\box1}\limits^{#1}}%
$}}}}}

\newcommand{\lby}[1]{{\xLongrightarrow  
{#1}}}
\newcommand{\bby}[2]{\xLongrightarrow  
[#1]{#2}}

\newcommand{\m}[1]{\mathcal{#1}}
\newcommand{\co}{\langle}
\newcommand{\cf}{\rangle}

\newcommand{\ra}{\rightarrow}

\newcommand{\vtr}{\vartriangleright}
\newcommand{\vtl}{\vartriangleleft}

\newcommand{\id}[2]{{\sf Id}_{#1}^{#2}}

\newcommand{\sub}{\hookleftarrow}
\newcommand{\swi}{\mapsto}

\begin{document}

\title[]{Context-bounded analysis for concurrent programs with dynamic
             creation of threads}

\author[M.F.~Atig]{Mohamed Faouzi Atig\rsuper a}	
\address{{\lsuper a}Uppsala University, Sweden}	
\email{mohamed\_faouzi.atig@it.uu.se}  

\author[A.~Bouajjani]{Ahmed Bouajjani\rsuper b}	
\address{{\lsuper b}LIAFA,  University Paris Diderot, France}	
\email{abou@liafa.jussieu.fr}  

\author[S.~Qadeer]{Shaz Qadeer\rsuper c}	
\address{{\lsuper c}Microsoft Research, Redmond, WA, USA}	
\email{qadeer@microsoft.com}  



\keywords{Pushdown Systems, Program Verification, Reachability Analysis}
\subjclass{D.2.4, D.3.1, F.4.3,  I.2.2}
\titlecomment{A shorter version of this paper has been published in the Proceedings of TACAS 2009, LNCS 5505}


\begin{abstract}
  \noindent Context-bounded analysis has been shown to be both efficient and effective at finding bugs in concurrent programs. According to its original definition, context-bounded analysis explores all behaviors of a concurrent program up to some fixed number of context switches between threads. This definition is inadequate for programs that create threads dynamically because bounding the number of context switches in a computation also bounds the number of threads involved in the computation. In this paper, we propose a more general definition of context-bounded analysis useful for programs with dynamic thread creation. The idea is to bound the number of context switches for each thread instead of bounding the number of switches of all threads. We consider several variants based on this new definition, and we establish decidability and complexity results for the analysis induced by them.
\end{abstract}

\maketitle

\section*{Introduction}

The verification of multithreaded programs is a challenging problem both from the theoretical and the practical point of view. (We consider here programs with parallel threads which may use local variables as well as shared (global) variables.) Assuming that the variables of the program range over a finite domain (which can be obtained using some abstraction on the manipulated data), there are several aspects in multithreaded programs which make their analysis complex or even undecidable in general \cite{Ram00}. 

Indeed, it is well known that for instance in the case where each thread can be modeled as a finite-state system, the state space of the program grows exponentially w.r.t. the number of threads, and the reachability problem is PSPACE-hard. Moreover, if threads are modeled as pushdown systems, which corresponds to allowing unbounded depth (recursive) procedure calls in the program, then the reachability problem becomes undecidable as soon as two threads are considered.

Context-bounding has been proposed in \cite{QR05}  as a suitable technique for the analysis of multithreaded programs. The idea is to consider only the computations of the program that perform at most some fixed number of context switches between threads. (At each point only one  thread is active and can modify the global variables, and a context-switch happens when the active thread terminates or is interrupted, and a pending one is  activated.)
The state space which must be explored may still be unbounded in presence of recursive procedure calls, but the context-bounded reachability problem is decidable even in this case.
In fact, context-bounding provides a very useful tradeoff between computational complexity and verification coverage. This tradeoff is based on three important properties.  First, context-bounded verification can be performed more efficiently than unbounded verification.  
From the complexity-theoretic point of view, it can be seen that context-bounded reachability is an NP-complete problem (even in the case of pushdown threads). 
Second, many concurrency errors, such as data races and atomicity violations, are manifested in executions with few context switches \cite{MQ07}.  Finally, verifying all executions of a concurrent program up to a context bound provides an intuitive and meaningful notion of coverage to the programmer.  
%

While the concept of context-bounding is adequate for multithreaded programs with a (fixed) finite number of threads, the question we consider in this paper is whether this concept is still adequate when dynamic creation of threads is considered.

Dynamic thread creation is useful for modeling several important aspects, e.g., (1) unbounded number of concurrent executions  of software modules such as file systems, device drivers, non-blocking data structures etc., or (2) creation of asynchronous activity such as forking a thread, queuing a closure to a threadpool with or without timers, callbacks, etc.
Both these sources are very important for modeling operating system components; they are likely to become important even for application software as it becomes increasingly parallel in order to harness the power of multi-core architectures.

We argue that the ``classical'' notion of context-bounding  which has been used so far in the existing work is actually too restrictive in this case. Indeed,
bounding the number of context switches in a computation also bounds the number of threads involved. 
In this paper, we propose a more general definition of context-bounded analysis useful for programs with dynamic thread creation. The idea is to bound the number of context switches for each thread instead of bounding the number of switches of all threads. We consider several variants based on this new definition, and we establish decidability and complexity results for the analysis induced by them.

We introduce a notion of $K$-bounded computations where each of the involved threads can be interrupted and resumed at most $K$ times. In fact, 
we  consider that when a thread is created, the number of context switches it can perform is the one of its ancestor (at the moment of the creation) minus 1.
Notice that the number of context switches by all threads in a computation is not bounded since the number of threads involved  is not bounded.

In the case of finite-state threads, we prove that this problem is as hard as the coverability problem for vector addition systems with states (or, Petri nets) (which is EXPSPACE-complete). The reduction from our problem to the coverability problem of  vector addition systems with states  is based on the simple idea of counting the number of pending threads for different values of the global and local states, as well as of the number of switches that these threads are allowed to perform. 
Conversely, we prove that the coverability problem of vector addition systems with states  can be reduced to the 2-bounded reachability problem. These results show that in the case of dynamic thread creation,
considering the  notion of context-bounding for each individual thread makes the complexity jumps from NP-completeness to EXPSPACE-completeness, even in  the case of finite-state threads. 
Then, an interesting question is whether it is possible to have a notion of context-bounding with a lower complexity. We propose for that the notion of stratified context-bounding. The idea is to consider computations where the scheduling of the threads is ordered according to their  number of allowed switches: First, threads of level $K$ (the level means here the number of allowed switches) are scheduled generating threads of level $K-1$, then threads of level $K-1$ are scheduled, and so on.
 Again, notice that  $K$-stratified computations may have an unbounded number of context switches since it is possible to schedule an unbounded number of threads at each level. This concept generalizes obviously the ``classical'' notion of context-bounding. We prove that, for finite-state threads, the $K$-stratified context-bounded reachability problem is NP-complete (i.e., it matches the complexity of the ``classical'' context-bounded reachability problem). The proof is by a reduction to the satisfiability problem of existential Presburger formulas.

Then, we consider the case of dynamic creation of pushdown threads. We prove that, surprisingly, the $K$-bounded reachability problem is in fact decidable, and that the same holds also for the $K$-stratified context-bounded reachability problem. To establish these results, we prove that these problems (for pushdown threads) can be reduced to their corresponding problems for finite-state threads. This reduction is not trivial. The main ideas behind the reduction are as follows:
First, the $K$-bounded behaviors of each single thread can be represented by a labeled pushdown system which (1) makes visible (as labels) on its transitions the created threads, and (2) guesses points of interruption-resumption and the corresponding values of the global states. (These guesses are also made visible on the transitions.) Then, the main problem is to ``synchronize'' these labeled pushdown systems so that their guesses can be validated. The key observation is that it is possible to abstract these systems without loss of preciseness by finite-state systems. This is due to the fact that we can consider that some of the generated threads can be lost (since they can be seen as threads that are never activated), and therefore we can reason about the downward closure of the languages of the labeled pushdown systems mentioned above (w.r.t.  suitable sub-word relation). This downward closure is in fact always regular and effectively constructible.

\bigskip

\paragraph{\bf Related work}

In the last few years, several implementations and algorithmic improvements have been proposed for context-bounded verification \cite{BESS05,MQ07,SES08,LTKR08,LR08,TMP09}. For instance, context-bounded verification has been implemented in explicit-state model checkers such as CHESS \cite{MQ07} and SPIN \cite{ZJ08}; it has also been implemented in symbolic model checkers such as SLAM \cite{QW04}, jMoped \cite{SES08}, and in \cite{LR08}. In this paper, we propose  more general definitions of context-bounded analysis useful for programs with dynamic thread creation.

\medskip

Several models based on rewriting systems or networks of pushdown systems have been considered to model multithreaded programs 
\cite{LS98,EP00,SS00,Markus,BT03,BT05}. While these models allow to model dynamic thread creation, they  only allow communication between processes in a very restrictive way.

\medskip

In \cite{BMOT05},  a model based on networks of pushdown systems  called CDPN was proposed.
While this model allows dynamic creation of processes, it allows only a restricted form of synchronization where a process has the right 
to read only the control states  of its immediate children (i.e., the processes it has created).

\medskip
 
  A symbolic algorithm for over-approximating reachability in Boolean programs with   unboundedly many threads was given in \cite{CKS06,CKS07}. Our approach complements these techniques since they are able to prove that a safety property of interest holds. While our work is useful for effectively detecting  bad behaviors of the analyzed programs.

A recent paper proposes an algorithm for the verification problem for parametrized concurrent  programs with procedural calls under  a k-round-robin schedule \cite{TMP10}. Our  work  is more  powerful than  this framework as long as  the data domain is bounded.

\section{Preliminary definitions and notations}

In this  section, we  introduce  some basic definitions and notations that will be  used in the rest of the paper.

\subsection{Integers,  functions, and vectors}

\subsubsection*{Integers} Let $\mathbb{Z}$ be    the set of integers and   $\mathbb{N}$ be  the set of positive integers (or natural numbers). For every  $i,j \in \mathbb{Z}$  such that $i \leq j$,   we use  $[i,j]$ and $[i,j[$ to denote   respectively the sets $\{k \in \mathbb{Z}\,|\, i\leq k \leq j\}$ and  $\{k \in \mathbb{Z}\,|\, i\leq k < j\}$.

\subsubsection*{Functions}\label{}
 Let $A$ and $B$ be two sets.  We denote by   $[A \rightarrow B]$     the set of all functions from $A$ to $B$. If   $f,g $ are  two functions from $A$ to $\mathbb{N}$, then  we write $g \leq f$ if and only if  $g(a) \leq f(a)$ for all $a \in A$. We use $f+g$ (resp. $f-g$ if $g \leq f$) to denote   the function from $A$ to $\mathbb{N}$  defined as follows: $(f+g)(a)=f(a)+g(a)$ (resp. $(f-g)(a)=f(a)-g(a)$) for all $a \in A$.  
 For every subset $C \subseteq A$, we  use  $\id{A}{C}$ to denote the function from $A$ to $\mathbb{N}$  defined as follows: 
 
 \begin{equation}
 \label{funct.000}
\id{A}{C}(a) = \left\{ \begin{array}{ll}
1 & \textrm{if $a \in C$}\\
0 & \textrm{if $a  \in (A \setminus C)$}\\
 \end{array} \right.
\end{equation}
 
 In particular, $\id{A}{\emptyset}$  denotes the function that maps any element of $ A$ to $0$.

\subsubsection*{Vectors}

Let $n$ be a natural number and $A$ be  a set. An $n$-{\em dim} vector {\bf v} over $A$ is an element of $A^n$. For every $i \in [1,n]$, we denote by ${\bf v}[i] \in A$ the $i^{th}$ component   of ${\bf v}$. 
Given $j \in [1,n]$ and $a \in A$, we denote  by ${\bf v}[j \sub a]$    the $n$-{\em dim} vector  ${\bf v'}$ over $A$  such that ${\bf v'}[j]=a$ and ${\bf v'}[k]={\bf v}[k]$ for all $k \in [1,n]$ and $k \neq j$.

\subsubsection*{Vectors of integers} The order relation $\leq$ between integers is generalized in a pointwise manner to vectors of integers. We   write  ${\bf 0}^n$ to denote  the $n$-{\em dim} vector ${\bf v}$ over $\mathbb{Z}$ such that ${\bf v}[i]=0$ for all $i \in [1,n]$. We trivially extend  the addition and subtraction operations over integers to vectors of integers.

\subsection{Words and languages}

Given a finite set $\Sigma$ called an alphabet and whose elements are called letters or symbols, a word $u$ over $\Sigma$ is either a finite sequence of letters in $\Sigma$ or the empty word $\epsilon$. The length of $u$ is denoted by $|u|$. (We assume that $|\epsilon|=0$.) For every      $a \in   \Sigma$, we use $|u|_a$  to denote the number of occurrences of $a$ in $u$. For every    $j \in  [1,|u|]$, we use $u(j)$ to denote the $j^{th} $ letter  of $u$. 

A language $L$ over $\Sigma$ is a (possibly infinite) set of words over $\Sigma$. We adopt the  widespread notations  $\Sigma^*$ and $\Sigma^+$ to represent respectively the languages containing    all  words and all non-empty  words over $\Sigma$. We use also $\Sigma_{\epsilon}$ to denote  the set $\Sigma \cup \{\epsilon\}$.

We denote by $\preceq \subseteq \Sigma^* \times \Sigma^*$ the {\em subword relation} defined as follows: For every  $u, v \in \Sigma^*$, $u \preceq v$ if and only if: $(1)$ $u=\epsilon$, or $(2)$  there are   $i_1,i_2\ldots,i_{|u|} \in [1,|v|]$ such that $i_1<i_2<\cdots<i_{|u|} $ and $u(j)=v(i_j)$ for all $j \in [1,|u|]$.  Given a language $L \subseteq \Sigma^*$, the {\em downward closure} of $L$  is the language $L \downarrow=\{u \in \Sigma^*\,|\, \exists v \in L,\, u \preceq v\}$.

Let $\Theta $ be a subset of $\Sigma$.  Given  a word $u \in \Sigma^*$, we denote by $u|_{\Theta}$ the projection of $u$ over $\Theta$, i.e., the word obtained from $u$ by erasing all the symbols that are not in $\Theta$.  
This definition is extended to languages as follows: If $L$  is a language over $\Sigma$, then $L|_{\Theta}=\{u|_{\Theta}\,\mid \, u \in L\}$.

The {\em Parikh image} of a word $u \in \Sigma^*$ is  a function from $\Sigma$ to $\mathbb{N}$ such that: For every $a\in \Sigma$,  $\mathit{Parikh}(u)(a)=|u|_a$. Accordingly, the Parikh image of a language $L \subseteq \Sigma^*$, written $\mathit{Parikh}(L)$, is the set of Parikh images of $u \in L$.

Let $\Sigma_1$ and $\Sigma_2$ be two alphabets. A homomorphism ${\sf h}$ is a function from $\Sigma_1^*$ to $\Sigma_2^*$ such that ${\sf h}(\epsilon)=\epsilon$ and ${\sf h}(uv)={\sf h}(u) {\sf h}(v)$ for all $u, v \in \Sigma_1^*$. By definition,  the homomorphism ${\sf h}$ is  completely characterized by the  function $f_{h}: \Sigma_1 \rightarrow \Sigma_2^*$ s.t. for any $a \in \Sigma_1$, $f_h(a)=h(a)$.

%

\subsection{Transition systems}

A transition system is a triplet  $\m{T}=(C,\Sigma,\rightarrow)$ where: $(1)$ $C$ is a (possibly infinite)  set of configurations (also called states), $(2)$
 $\Sigma$ is a finite set of labels (or actions), and $(3)$ $\rightarrow \subseteq C \times \Sigma_{\epsilon} \times C$ is a transition relation.

Given two configurations $c,c' \in C$ and an action $a \in \Sigma$, we write  $c \by{a}_{\m{T}} c'$ if  $(c,a,c') \in \rightarrow$. A finite run  $\rho$ of $\m{T}$ from $c$ to $c'$ is a finite sequence $c_0 a_1 c_1a_2 \cdots a_n c_n$, for some $n \geq 1$, such that: $(1)$  $c_0=c$ and $c_n=c'$, and $(2)$  $c_i \by{a_{i+1}}_{\m{T}} c_{i+1}$ for all $i \in [0,n[$. 
In this case, we say that  $\rho$ has length $n$  and is labelled by the word $a_1 a_2 \cdots a_n$.

Let   $u \in \Sigma^*$ be an input word. We  write $c \,{\bby{n}{u}} {}_{\m{T}} \,c'$ if one of the following two cases holds: (1)  $n=0$, $c=c'$,  and $u=\epsilon$, and (2)   there is a run $\rho$ of length $n$ from $c$ to $c'$ labelled  by $u$.  We also write $c \, \lby{u}{}_{\m{T}}^*\, {c'}$ to denote   that $c \,{\bby{n}{u}} {}_{\m{T}} \,c'$ for some $n\geq 0$.  Finally, for every   $C_1, C_2 \subseteq C$,   we have $\mathit{Traces}_{\m{T}}(C_1,C_2)=\{u\in \Sigma^* \,|\, \exists (c_1,c_2) \in C_1 \times C_2\,,\, c_1 \, \lby{u}{}_{\m{T}}^* \, {c_2}\}$.

\subsection{Finite state automata}

\label{sec.fsa}
A finite state automaton (FSA for short) is a quintuple $\m{A}=(Q,\Sigma,\Delta,I,F)$ where: $(1)$ $Q$ is the  finite non-empty set of states, $(2)$ $\Sigma$ is the  finite set of input symbols (called also the input alphabet), $(3)$
 $\Delta \subseteq (Q \times \Sigma_{\epsilon} \times Q)$ is the transition relation,  $(4)$ $I \subseteq Q$ is the  set of initial states, and $(5)$
$F \subseteq Q$ is the  set of final states.
We use  $q\by{a}_{\m{A}} q'$ to denote that $(q,a,q')$ is in $\Delta$ .

The size of $\m{A}$,  denoted  by  $|\m{A}|$, is defined by $(|Q|+|\Sigma|)$. We  denote by  $\m{T}({\m{A}})=(Q,\Sigma,\Delta)$  the  transition system associated to $\m{A}$. The 
language  accepted (or recognized) by  $\m{A}$ is   defined as follows  $L(\m{A})=\mathit{Traces}_{\m{T}(\m{A})}(I,F)$.

It is well known that 
the class of  languages accepted by finite state automata  (the class of rational  (or regular) languages) is  effectively closed under union, intersection, homomorphism, and  projection operations \cite{HU79}.

\subsection{Pushdown automata}
\label{pda}

A pushdown automaton (PDA for short) is a 7-tuple  $\m{P}=(P,\Sigma,\Gamma,\Delta,p_0,\gamma_0,F)$ where:

\begin{iteMize}{$\bullet$}
\item $P$ is the  finite  non-empty set of  states,
\item $\Sigma$ is the finite set of  input symbols (called also the input alphabet),
\item  $\Gamma$ is the finite set of stack symbols (called also the stack alphabet),
\item $\Delta \subseteq \big((P \times \Gamma) \times \Sigma_{\epsilon} \times (P \times \Gamma^{\leq 2})\big)  $ is the transition relation (where $\Gamma^{\leq 2}= \Gamma_{\epsilon} \cup \Gamma^2$).

\item $p_0 \in P$ is the initial state,
\item $\gamma_0 \in \Gamma$ is the initial stack symbol, and 
\item $F \subseteq P$ is the set of final states.

\end{iteMize}

The size of $\m{P}$, denoted by    $|\m{P}|$, is defined as $(|P|+|\Sigma|+|\Gamma|)$. We  use   $\co p,\gamma\cf \by{a}_{\m{P}} \co p', u \cf$ to denote that $((p,\gamma),a,(p',u))$ is in  $\Delta$.

A configuration of $\m{P}$ is a pair  $(p,w)$ where $p \in P$ and $w \in \Gamma^*$.
The set of all configurations of $\m{P}$ is denoted by $\mathit{Conf}({\m{P}})$.
The transition system associated to $\m{P}$, denoted by  $\m{T}(\m{P})$, is given  by the tuple $(\mathit{Conf}(\m{P}),\Sigma,\ra)$  where $\ra$ is the smallest transition relation such that:  if $\co p,\gamma \cf \by{a}_{\m{P}} \co p',u \cf$, then $(p, \gamma w) \by{a}_{\m{T}(\m{P})}$ $ (p' ,u w)$ for all $w \in \Gamma^*$.  The language of  $\m{P}$ is defined as follows $L(\m{P})=\mathit{Traces}_{\m{T}(\m{P})}(\{(p_0,\gamma_0)\},F \times \Gamma^*)$.

It is well known that the class of context-free languages (i.e., accepted by pushdown automata) are closed under  concatenation, union, Kleene star, homomorphism,  projection, and intersection with a rational language. However,  context-free languages are not closed under complement and intersection \cite{HU79}. 

Let us recall now  that the downward closure  of a context-free language, with respect to the subword relation, is effectively a rational language.

\begin{thm}[\cite{Cour91}]
\label{prop-down-pda}
If  $\m{P}$ is a PDA, then,  it is possible to  construct, in time and space exponential in $|\m{P}|$,  a  finite state automaton $\m{A}$ such that $L(\m{A})=L(\m{P})\downarrow$ and  the size of $|\m{A}|$ is exponential in $|\m{P}|$ in the worst case.
\end{thm}

We can prove  that the exponential blow-up in Theorem \ref{prop-down-pda}  can not be avoided. 
This is due to the fact  that  pushdown automata are more succinct than finite state automata. 
To show  that,   let us  consider   the following pushdown automaton  $\m{P}=(\{p_0,p_1,p_2\},$ $\{a\},$ $\{\bot,\gamma_0,\ldots,\gamma_{n}\},\Delta,p_0,\bot,\{p_2\})$ where $n \in \mathbb{N}$   and   $\Delta$ is the transition relation composed  from  the following transitions: 

\begin{enumerate}[(1)]
\item $\co p_0, \bot \cf \by{\epsilon}_{\m{P}}  \co p_1, \gamma_0 \bot \cf $,

\item  for every $i \in [0,n[$,  $\co p_1, \gamma_i \cf \by{\epsilon} \co p_1, \gamma_{i+1} \gamma_{i+1}\cf$,

\item  $\co p_1, \gamma_n \cf \by{a}_{\m{P}} \co p_1,\epsilon \cf$, and 

\item $\co p_1, \bot \cf \by{\epsilon}_{\m{P}}  \co p_2, \epsilon  \cf $.
\end{enumerate}

\noindent
It is easy to observe that $L(\m{P})=\{a^{2^n}\}$ and therefore the minimal finite state automaton $\m{A}$ recognizing $L({\m{P}})\downarrow$ has at least  ${2^n}$ states whereas the size of $\m{P}$ is $(n+5)$.

\section{Dynamic  network of concurrent pushdown systems}

In this section, we introduce  dynamic network of  concurrent pushdown systems. 
Intuitively,  a  dynamic network of  concurrent pushdown systems  $\m{M}$ models dynamic multithreaded programs with (potentially) recursive procedure calls. Threads are modeled as pushdown processes which may spawn new threads (or processes).  Each thread may  have its local variables and has also access to global variables. The values of local variables are modeled using the stack alphabet $\Gamma$, whereas the values of the global variables are modeled using a finite non-empty set of states  $Q$. Transitions of the form  $\co q, \gamma \cf \by{}_{\m{M}}  \co q', u \cf \vtr\epsilon$ correspond to standard transitions of pushdown systems (popping $\gamma$ and then pushing $u$ while changing the state from $q$ to $q'$).  Transitions of the form  $\co q, \gamma \cf \by{}_{\m{M}}  \co q', u \cf \vtr \gamma'$ correspond to standard transitions of pushdown systems  with a creation  of a  thread  whose   initial stack content is  $\gamma'  \in \Gamma$.  Transitions of the form $\co {q}, \gamma \cf \swi_{\m{M}} \co q', u \cf$ correspond  to  interrupt the execution of the active thread after   the performing   the standard pushdown operations, and 
transitions of the form $ q \swi_{\m{M}} q' \vtl\gamma$ correspond to  start/resume the execution of  a pending   thread with topmost stack symbol $\gamma' \in \Gamma$ after  changing the state from $q$ to $q'$.

\subsection{Syntax}

\begin{defi}[\dcps]
A dynamic network of   concurrent pushdown system  (DCPS for short) is a tuple $\m{M}=(Q,\Gamma,\Delta,q_0,\gamma_0)$ where:

\begin{iteMize}{$\bullet$}
\item $Q$ is the finite non-empty set of states,
\item $\Gamma$ is a finite set of stack symbols (called also stack alphabet),
\item $\Delta= \Delta_{\sf cr} \cup \Delta_{\sf in} \cup \Delta_{\sf rs}$  where:

\begin{iteMize}{$-$}
\item  $\Delta_{\sf cr}  \subseteq  \big((Q \times \Gamma) \times (Q \times \Gamma^{\leq 2})  \times \Gamma_{\epsilon}\big)$ is a finite set of  ({\em creation }) transitions.
\item  $\Delta_{\sf in} \subseteq \big((Q \times \Gamma) \times (Q \times \Gamma^{\leq 2})\big)$ is a finite set of  ({\em interruption}) transitions.

\item $\Delta_{\sf rs} \subseteq \big(Q \times \Gamma \times Q \big)$  is a finite set of  ({\em resumption}) transitions. 

\end{iteMize}

\item $q_0$ is the initial state, and
\item $\gamma_0$ is the initial stack symbol.

\end{iteMize}

\end{defi}

 In the rest of the paper, we adopt the following notations:  $(1)$     $\co q,\gamma \cf \by{}_{\m{M}} \co q',u \cf \vtr \alpha$   to denote that  $\big((q,\gamma),(q',u),\alpha\big) \in \Delta_{\sf cr}$, $(2)$        $ \co q,\gamma \cf \swi_{\m{M}} \co q',u\cf$   to denote that  $\big((q,\gamma),(q',u)\big) \in \Delta_{\sf in}$, and $(3)$       $ q \swi_{\m{M}} q' \vtl \gamma$   to denote that  $\big(q,\gamma,q'\big) \in \Delta_{\sf rs}$. The size of   $\m{M}$ is given by   $|\m{M}|=|Q|+|\Gamma|$.


When unbounded recursion is not considered, threads can be modeled as finite state processes instead of pushdown systems. This corresponds to the special case where, for all $((q,\gamma), (q',u), \alpha) \in \Delta_{cr}$ and  $((q,\gamma),(q',u)) \in \Delta_{in}$, the pushed word $u$ is of length at most $1$.

\begin{defi}[DCFS]
A dynamic concurrent finite-state systems (DCFS for short) is a DCPS $\m{M}=(Q,\Gamma,\Delta,q_0,\gamma_0)$ where, for all  $((q,\gamma), (q',u),  \alpha) \in \Delta$ and  $((q,\gamma),(q',u)) \in \Delta$, we have $|u|\leq 1$.
\end{defi}

\subsection{Semantics}

\begin{defi}[Local configurations of a DCPS]
Let $\m{M}=(Q,\Gamma,\Delta,q_0,\gamma_0)$ be a DCPS. A local configuration of  a thread of $\m{M}$ is a pair $(w,i)$ where $w \in \Gamma^*$  is its call stack and  $i \in \mathbb{N}$ is its switch number. Let $\mathit{Loc}(\m{M})$ denote the set of local configurations of $\m{M}$.
\end{defi}

Intuitively, the switch number of a thread is  the  number of interruptions/resumptions together with   the switch number of its creator  (at the moment of the creation) plus one.

\begin{defi}[Configurations of a DCPS]
Let $\m{M}=(Q,\Gamma,\Delta,q_0,\gamma_0)$ be a DCPS.
A configuration $c$ of  a $\m{M}$ is  an element of $Q \times (\mathit{Loc}(\m{M}) \cup \{\bot\}) \times  [\mathit{Loc}(\m{M}) \ra \mathbb{N}]$.  We use $\mathit{Conf}(\m{M})$ to denote the set of all configurations of $\m{M}$.
\end{defi}

A configuration of the form   $(q,(w,i),\mathit{Val})$ (resp. $(q,\bot,\mathit{Val})$) of $\m{M}$ means that: $(1)$    $q \in Q$ is the value of the global store, $(2)$
$(w,i)$   is the  local configuration of  the active thread (resp. there is no active thread), and   $(3)$ $ \mathit{Val}\,:\, \mathit{Loc}(\m{M}) \ra \mathbb{N}$ is a  function that associates for each $(w',i')\in \mathit{Loc}(\m{M})$, the number of  pending   threads with local configuration $(w',i')$.

Given a configuration  $c=(q,\eta,\mathit{Val}) \in \mathit{Conf}(\m{M})$,  let   $\mathit{State}(c)=q$,  $\mathit{Active}(c)=\eta$, and $\mathit{Idle}(c)=\mathit{Val}$.  We use  $c^{\sf init}_{\m{M}}=(q_0,\bot,\id{{\mathit{Loc}(\m{M})}}{\{(\gamma_0,0)\}} )$ to denote the initial configuration of $\m{M}$.

%

\begin{defi}[Transition system of a DCPS]
Let $\m{M}=(Q,\Gamma,\Delta,q_0,\gamma_0)$ be a DCPS. The transition system associated with $\m{M}$ is given by   $\m{T}(\m{M})=(\mathit{Conf}(\m{M}),\Sigma,\ra)$ where $\Sigma=\Delta$ and $\ra$ is the smallest relation such that:

\begin{iteMize}{$\bullet$}
\item if $t=\co q, \gamma \cf \by{}_{\m{M}} \co q',u \cf \vtr \alpha$, then  $(q,(\gamma w,i),\mathit{Val}) \by{t}_{\m{T}(\m{M})} (q',(u w,i),\mathit{Val}')$  for all $w \in \Gamma^*$,  $i \in \mathbb{N}$, and $\mathit{Val}, \mathit{Val}' \in [{\mathit{Loc}(\m{M})} \ra \mathbb{N}]$ such that:

\begin{iteMize}{$-$}
\item  If $\alpha \in \Gamma$, then $\mathit{Val'}=\mathit{Val}+\id{{\mathit{Loc}(\m{M})}}{\{(\alpha,i+1)\}}$ .
\item If $\alpha=\epsilon$, then $\mathit{Val'}=\mathit{Val}$.
\end{iteMize}

\item if $t= \co q,\gamma \cf \swi_{\m{M}}  \co q',u\cf$, then  $(q,(\gamma w,i),\mathit{Val}) \by{t}_{\m{T}(\m{M})} (q',\bot,\mathit{Val}+\id{\mathit{Loc}(\m{M})}{\{(u w,i+1)\}}) $  for all $w \in \Gamma^*$,  $i \in \mathbb{N}$, and $\mathit{Val} \in [{\mathit{Loc}(\m{M})} \ra \mathbb{N}]$.

\item if $t= q \swi_{\m{M}}  q' \vtl \gamma$, then  $(q,\bot,\mathit{Val}+\id{\mathit{Loc}(\m{M})}{\{(\gamma w,i)\}}) \by{t}_{\m{T}(\m{M})} (q',(\gamma w,i),\mathit{Val}) $  for all $w \in  \Gamma^*$,  $i \in \mathbb{N}$, and $\mathit{Val} \in [{\mathit{Loc}(\m{M})} \ra \mathbb{N}]$.
\end{iteMize}

\noindent
where for every sets $A$ and $C$ such that $C \subseteq A$,   $\id{A}{C}$  denotes the function from $A$ to $\mathbb{N}$  such that  $\id{A}{C}=1$ if $a \in C$ and $\id{A}{C}(a)=0$  if  $a \in (A \setminus C)$ (see Equation. \ref{funct.000}). 

\end{defi}

The transition $(q,(\gamma w,i),\mathit{Val}) \by{t}_{\m{T}(\m{M})} (q',(u w,i),\mathit{Val}')$, with $t=\co q, \gamma \cf \by{}_{\m{M}} \co q', u \cf \vtr \alpha$, corresponds to the execution of pushdown  operation (pop or  push) with the possibility of a creation of  a new thread (if $\alpha \in \Gamma$) which is added to the set  of pending  threads. The created thread gets the switch number $i+1$. The transition $(q,(\gamma w,i),\mathit{Val}) \by{t}_{\m{T}(\m{M})} (q',\bot,\mathit{Val}')$, with $t=\co q,\gamma \cf \swi_{\m{M}}  \co q',u\cf$, corresponds to interrupt the execution of the current active thread after performing the pushdown operation: The local configuration    $(uw,i)$ of the  active thread is added to the set of the idle threads after incrementing its switch number. The transition $(q,\bot,\mathit{Val}) \by{t}_{\m{T}(\m{M})} (q',(\gamma w,i),\mathit{Val}')$, 
 with $t=q \swi_{\m{M}}  q' \vtl \gamma$, corresponds to start/resume (from the state $q'$) the execution of a pending thread with local configuration $(\gamma w,i)$.

\subsection{Bounded semantics} Let $\m{M}=(Q,\Gamma,\Delta,q_0,\gamma_0)$ be a DCPS.  For every $I \subseteq  \mathbb{N}$, let  $\mathit{Conf}_I(\m{M})$  denote  the set of  configurations of $\m{M}$ such that $c \in \mathit{Conf}_I(\m{M})$ if and only if $\mathit{Active}(c) \in \Gamma^* \times I$. 
In the following, we restrict the behavior of $\m{T}(\m{M})$  to the set of runs where  the switch numbers of the    active  threads are  always  in  $I$.

\begin{defi}[Bounded  transition system of a DCPS]
For every $I \subseteq    \mathbb{N}$,  $\m{T}_{I}(\m{M})$ denotes the transition system $(\mathit{Conf}(\m{M}),\Delta,\ra_{I})$ where: For every $c,c' \in \mathit{Conf}(\m{M})$, $c  \by{t}_{\m{T}_I(\m{M})} c'$ if and only if: $(1)$ $c \by{t}_{\m{T}(\m{M})} c'$,  and $(2)$ $c \in \mathit{Conf}_{I}(\m{M})$ or $c' \in \mathit{Conf}_{I}(\m{M})$.
\end{defi}

\subsection{Reachability problems}

Let $\m{M}=(Q,\Gamma,\Delta,q_0,\gamma_0)$ be a DCPS.  We consider  the  following three notions of reachability:

\begin{defi}[The state reachability problem]
A state $q \in Q$ is  reachable by $\m{M}$ if and only if there are  $c \in \mathit{Conf}(\m{M})$  and $\tau \in \Delta^*$ such that $c^{\sf init}_{\m{M}}\, \lby{\tau}{}{}_{\m{T}(\m{M})}^* \, c$, $\mathit{Active}(c)=\bot$, and $\mathit{State}(c) =q$. The  state reachability  (SR for short) problem  for $\m{M}$ consists in deciding, for a given  set  $F \subseteq Q$,  whether  there is a state $q \in F$ such that $q$ is reachable by $\m{M}$.

\end{defi}

Notice that we consider, in the definition of the state  reachability problem, that the set of reachable configurations that we are interested in are those with    no active thread. This is only for the sake of simplicity and does not constitute at all a restriction. Indeed, we can show that the problem of checking whether there are $c \in \mathit{Conf}(\m{M})$  and $\tau \in \Delta^*$ such that $c^{\sf init}_{\m{M}}\, \lby{\tau}{}{}_{\m{T}(\m{M})}^* \, c$ and $\mathit{State}(c)  \in F$ can be reduced to the state reachability problem for a DCPS $\m{M}'=(Q,\Gamma,\Delta',q_0,\gamma_0)$ built up from $\m{M}$ by adding to $\Delta$ some transition rules  that   interrupt the execution of  the  active thread when   the current  state  is in $F$.

\begin{defi}[The $k$-bounded state reachability problem]
Let  $k \in \mathbb{N}$. A state $q \in Q$ is $k$-bounded reachable by $\m{M}$ if and only if there are  $c \in \mathit{Conf}(\m{M})$  and $\tau \in \Delta^*$ such that $c^{\sf init}_{\m{M}}  \;\lby{\tau}{}{}_{\m{T}_{[0,k]}(\m{M})}^* \, c$, $\mathit{Active}(c)=\bot$, and $\mathit{State}(c)=q$.  The  $k$-bounded state reachability (BSR[$k$] for short) problem  for $\m{M}$ consists   in deciding, for a given set $F \subseteq Q$,  whether   there is a state $q \in F$ such that $q$ is $k$-bounded reachable by $\m{M}$.
\end{defi}

Observe that, in BSR$[k]$ problem, a bound $k+1$ is imposed on the number of switches (interruptions/resumptions) performed by each thread (together with the switch number of its ancestor (at the moment of its creation) plus one). However, due to dynamic creation of threads, bounding the number of switches of each thread does not bound the number of switches in the whole computation of the system (since an arbitrary large number of threads can be involved in these computations). 

\begin{defi}[The $k$-stratified  state reachability problem]
\label{straified}
Let  $k \in \mathbb{N}$. A state $q \in Q$  is $k$-stratified reachable by $\m{M}$ if and only if there are    $\tau_0,\tau_1,\ldots, \tau_k \in \Delta^*$, and  $c_1, \ldots, c_{k+1} \in \mathit{Conf}(\m{M})$ such that $\mathit{State}(c_{k+1})=q$, $\mathit{Active}(c_{k+1})=\bot$, and we have:

$$c^{\sf init}_{\m{M}}  \;\lby{\tau_0}{}_{\m{T}_{\{0\}}(\m{M})}^* \, c_1 \;\lby{\tau_1}{}_{\m{T}_{\{1\}}(\m{M})}^*\,  \cdots \;\lby{\tau_{k-1}}{}_{\m{T}_{\{k-1\}}(\m{M})}^* \, c_k \;\lby{\tau_k}{}_{\m{T}_{\{k\}}(\m{M})}^* \, c_{k+1}$$  The  $k$-stratified  state reachability   (SSR[$k$]  for short) problem for $\m{M}$    consists   in deciding, for a given $F \subseteq Q$, whether  there is a state $q \in F$ s.t.  $q$ is $k$-stratified reachable by $\m{M}$.
\end{defi}

In the SSR[$k$] problem, a special kind of $k$-bounded computations (called stratified computations) are considered: In such a computation, threads are scheduled according to their increasing switch number (from $0$ to $k$): First, threads with switch number $0$ are scheduled generating threads with switch number  $1$, then threads with switch number $1$ are scheduled  generating threads with switch number $2$, and so on. 

Observe that even in the case of stratified computations, an arbitrarily large
number of context switches may occur along a computation due to dynamic creation of
threads. Very particular stratified computations
are those where the whole number of context switches is bounded \cite{QR05}.

\section{The SR problem and the \texorpdfstring{BSR$[k]$}{BSR[k]}
  problem for DCFSs}

In the following, we show that the SR problem and the BSR$[k]$ problem
for dynamic networks of concurrent finite-state systems are as hard as
the coverability problem for vector addition systems with states
(which is EXPSPACE-complete).

\begin{thm}
\label{dcfs-exp-complete}
The SR problem and the BSR[$k$] problem, with $k \geq 2$, for DCFSs are EXPSPACE-complete.
\end{thm}

Next, we recall some basic definitions and notations about vector addition systems with states (or equivalently, Petri nets). Then, this proof of Theorem \ref{dcfs-exp-complete}  is  structured as follows:  First, we show that the BSR$[k]$ problem for  DCFSs is polynomially reducible to the SR problem for DCFSs (Proposition \ref{chap3.prop.dcfs01}). Then, we show that the SR problem for DCFSs is polynomially reducible to the coverability problem for VASSs (Proposition \ref{chap3.prop.dcfs-cover01}).
Finally, we prove that the coverability problem for VASSs is polynomially reducible to the BSR$[2]$ problem for DCFSs (Proposition \ref{chap3.prop.dcfs03}). As an immediate consequence of these results and Theorem \ref{petri_1}, we obtain that the SR problem and the BSR$[k]$ problem for  DCFSs are EXPSPACE-complete.

\subsection{Vector addition systems with states}

A vector addition system with states (VASS for short) is a tuple $\m{V}=(n,Q, \Sigma,\delta,q_0,{\bf u}_0)$ where:

\begin{iteMize}{$\bullet$}
\item $n \in \mathbb{N}$ is the dimension,
\item $Q$ is the  finite non-empty set of states,
\item $\Sigma$ is the  finite set of actions (or labels), 
\item $\delta: Q \times \Sigma \rightarrow Q \times {([-1,1])}^n$ is the displacement function, 
\item $q_0 \in Q$ is the initial state,  and
\item ${\bf u}_0 $ is the initial $n$-{\em dim} vector over $\mathbb{N}$ such that $0 \leq {\bf u}_0(i)\leq 1$ for all $i \in [1,n]$.
\end{iteMize}

The size of $\m{V}$, denoted  by $|V|$, is  defined as   $(n+|Q|+|\Sigma|)$.  A configuration of $\m{V}$ is a pair $(q,{\bf u})$ where $q \in Q$ and ${\bf u} \in \mathbb{N}^n$. Given a configuration $c=(q,{\bf u})$, we let $\mathit{State}(c)=q$ and $\mathit{Val}(c)={\bf u}$. The set of all configurations of $\m{V}$ is denoted by $\mathit{Conf}(\m{V})$.

 The transition system associated to $\m{V}$, denoted  by $\m{T}(\m{V})$,  is given  by $(\mathit{Conf}(\m{V}), \Sigma, \ra)$, where  $\ra$ is  the smallest transition relation satisfying the following condition:  For every $q_1,q_2 \in Q$ and ${\bf u_1}, {\bf u_2} \in \mathbb{N}^n$, $(q_1,{\bf u_1}) \by{a}_{\m{T}(\m{V})} (q_2,{\bf u_2})$ if and only if $\delta((q_1,a))=(q_2,{\bf u_2}-{\bf u_1})$.


 A state $q \in Q$ is reachable by $\m{V}$ if and only if  there are  $w \in \Sigma ^*$ and $c \in \mathit{Conf}(\m{V})$ such that $(q_0,{\bf u}_0) \,\lby{w}{}_{\m{T}(\m{V})}^*\, c$ and   $\mathit{State}(c) =q$.  The coverability problem for  $\m{V}$ consists  in  deciding, for a given set $F \subseteq Q$, whether there is  $q \in F$  such that $q$ is reachable by $\m{V}$.
 

\begin{thm}[{\cite{lipton,Rackoff78}}]
\label{petri_1}
The coverability problem for vector addition systems with states is  EXPSPACE-complete. 
\end{thm}

\subsection{From the BSR$[k]$ problem   for DCFSs to the SR problem for DCFSs}
In the following, we show that, for every $k \in \mathbb{N}$, the BSR$[k]$ for DCFSs  is polynomially reducible to the SR problem for DCFSs.  Intuitively, given a DCFS $\m{M}=(Q,\Gamma,\Delta,q_0,\gamma_0)$ and a natural number $k$, we construct a DCFS $\m{M}'$ that records for each thread its switch number and can execute  only threads with recorded switch number less than $k$. Formally, the DCFS $\m{M}'=(Q',\Gamma',\Delta',q'_0,\gamma'_0)$ is defined as follows:

\begin{iteMize}{$\bullet$}

\item $Q'=Q$ is a finite set of states,
\item $\Gamma'=\Gamma_{\epsilon} \times [0,k+1]$ is a finite set of stack symbols. A stack symbol $(\alpha,i)$ corresponds to a thread with stack content $\alpha$ and switch number $i$.

\item $\Delta'$ is the smallest transition relation satisfying the following conditions:

\begin{iteMize}{$-$}
\item For every   $i \in [0,k]$ and $\co q,\gamma \cf \by{}_{\m{M}} \co q',u \cf \vtr \epsilon$, then  $\co q,(\gamma,i) \cf \by{}_{\m{M}'} \co q',(u,i) \cf \vtr \epsilon$. 
\item For every $i \in [0,k]$ and $ \co q,\gamma \cf \by{}_{\m{M}} \co q',u \cf \vtr \alpha$  for some  stack symbol $\alpha \in \Gamma$,  then  $\co q,(\gamma,i) \cf \by{}_{\m{M}'}  \co q',(u,i)  \cf \vtr (\alpha,i+1)$. 

\item For every  $i \in [0,k]$ and $ \co q,\gamma \cf \swi_{\m{M}}\,  \co q',u \cf $,  then  $\co q,(\gamma,i) \cf \swi_{\m{M}'} \, \co q',(u,i+1)\cf$.

\item For every  $i \in [0,k]$ and $ q \swi_{\m{M}}\, q' \vtl \gamma $,  then  $ q \swi_{\m{M}'}\,   q' \vtl (\gamma,i)$.

\end{iteMize}

\item $q'_0=q_0$ is the initial state, and 
\item $\gamma'_0=(\gamma_0,0)$ is the initial stack symbol.

\end{iteMize}


 \medskip

Observe that the size of the DCFS $\m{M}'$ is polynomial in the size of $\m{M}$. Moreover,  the relation between $\m{M}$  and $\m{M}'$ is given by the following lemma:

\begin{lem}
\label{chap3.lemma3.dcfs017}
Let $q \in Q$.  $q$ is $k$-bounded reachable by $\m{M}$ iff $q$ is reachable by $\m{M}'$.
\end{lem}

The proof of Lemma \ref{chap3.lemma3.dcfs017} is done  by induction on the length of the runs and is given in Appendix \ref{lemm.appendix.-1}.

As an immediate consequence of Lemma \ref{chap3.lemma3.dcfs017}, we obtain  the following result:

\begin{prop}
\label{chap3.prop.dcfs01}
Let  $k \geq 1$. The BSR$[k]$ problem for DCFSs is polynomially reducible to the SR problem for DCFSs.
\end{prop}

\subsection{From  the SR problem for DCFSs to the coverability problem for VASSs}
In the following, we show  that the SR problem for DCFSs is polynomially reducible to the  coverability problem for VASSs.  For a given  DCFS $\m{M}=(Q,\Gamma,\Delta,q_0,\gamma_0)$, with $\Gamma=\{\gamma_0,\ldots,\gamma_n\}$, we can construct a VASS $\m{V}=(m,P,\Sigma,\delta,p_0,{\bf u}_0)$  which has the following structure:

\begin{iteMize}{$\bullet$}
\item $m=n+2$ is the dimension of $\m{V}$. It is easy to observe that the dimension   of $\m{V}$  is  equal to $|\Gamma_{\epsilon}|$ which is  the number of all possible stack contents of threads of $\m{M}$.

\item $P=(Q \times (\Gamma_{\epsilon}\cup \{\bot\})) \cup \{p_{\sf halt}\}$ is the set of  states of $\m{V}$ (with $p_{\sf halt} \notin Q$). A state of the form $(q,w) \in Q \times \Gamma_{\epsilon}$ (resp. $(q,\bot)$) of $\m{V}$  means that the state of $\m{M}$ is $q$ and that the stack content of the active thread is $w$ (resp. there is no active thread).  The state $p_{\sf halt}$ is used  in order to  interrupt  the simulation of $\m{M}$ by $\m{V}$. 
 
 \item $\Sigma=\Delta$ is the input alphabet of $\m{V}$.
 
 \item $\delta\,: \, P \times \Sigma \ra P \times ([-1,1])^m$ is the transition function of $\m{V}$ defined as follows: For every $p \in P $ and $t \in \Sigma$, we have:

 \begin{iteMize}{$-$}
 \item $\delta(p,t)=(p',{\bf 0}^m)$ if   there are $q,q' \in Q$, $\gamma \in \Gamma$, and $u \in \Gamma_{\epsilon}$ such that $t=\co q, \gamma \cf \by{}_{\m{M}} $ $\co q', u \cf \vtr \epsilon$,  $p=(q,\gamma)$, and $p'=(q',u)$. This corresponds to  the simulation of  a transition rule  of $\m{M}$ without thread creation.
 
 \item $\delta(p,t)=(p',{\bf 0}^m[i \sub 1])$ if   $i \in [1,m[$ and there are  $q,q' \in Q$, $\gamma \in \Gamma$, and $u \in \Gamma_{\epsilon}$ such that $t=\co q, \gamma \cf \by{}_{\m{M}} $ $\co q', u \cf \vtr \gamma_{i-1}$, $p=(q,\gamma)$, and $p'=(q',u)$. This corresponds to  the simulation of  a transition rule  of $\m{M}$ with thread creation.

 \item     $\delta(p,t)=(p',{\bf 0}^m[ j\sub 1])$ if  $j \in [1,m]$, and there are $q,q' \in Q$, $\gamma \in \Gamma$, and $u \in \Gamma_{\epsilon}$ such that $t=\co q, \gamma \cf \swi_{\m{M}} $ $\co q', u \cf $, $p=(q,\gamma)$, $p'=(q',\bot)$, $u=\epsilon$ if $j=m$,  and $u=\gamma_{j-1}$ if $j<m$. This corresponds to  the  interruption of   the execution of the current active thread.

 \item     $\delta(p,t)=(p',{\bf 0}^m[i \sub -1])$ if $i \in [1,m[$, and there are $q,q' \in Q$,  such that $t= q\,\swi_{\m{M}}\, q' \vtl \gamma_{i-1}$, $p=(q,\bot)$, and $p'=(q',\gamma_{i-1})$. This corresponds  to   the execution of a pending   thread with topmost stack symbol $\gamma_{i-1}$.

 \item $\delta(p,t)=(p_{\sf halt},{\bf 0}^m)$ otherwise. This indicates the end of the simulation of $\m{M}$ by $\m{V}$ whenever the transition $t$ can not be applied from  the state $p$.

 \end{iteMize}
 \item ${\bf u}_0=(1,0,\ldots,0)$. This corresponds  to the initial  pending thread of $\m{M}$ (i.e., initially $\m{M}$ has one pending thread with local configuration $(\gamma_0,0)$).

 \item $p_0=(q_0,\bot)$ is the initial state of $\m{V}$. This corresponds to the initial state $q_0$ of $\m{M}$.

\end{iteMize}

 \medskip

Observe that the size of $\m{V}$ is polynomial in the size of $\m{M}$. Moreover, the relation between $\m{M}$ and $\m{V}$ is given by  the following lemma:

\begin{lem}
\label{chap3.lem.dcfs-vass-rel1}
Let $q \in Q$. $q $ is reachable by $\m{M}$ if and only if $(q,\bot)$ is reachable by $\m{V}$.
\end{lem}

The proof of Lemma \ref{chap3.lem.dcfs-vass-rel1} is  done by induction on the length of the runs  and is given in Appendix \ref{appendix.ref.chap3.lem.dcfs-vass-rel1}.

As an immediate consequence of Lemma \ref{chap3.lem.dcfs-vass-rel1}, we obtain the following result:

\begin{prop}
\label{chap3.prop.dcfs-cover01}
The SR problem for DCFSs is polynomially reducible to the  coverability problem for VASSs.
\end{prop}

\subsection{From the coverability problem for VASSs to the BSR$[2]$ problem for DCFSs}
\label{vass-bsr}
In the following, we prove that the coverability problem for VASSs is polynomially reducible to the BSR$[2]$ for DCFSs. Given a VASS $\m{V}=(n,Q,\Sigma,\delta,q_0,{\bf u}_0)$, we construct a DCFS $\m{M}$ such that  the coverability problem for $\m{V}$ is reducible to the BSR$[2]$ problem for $\m{M}$. We assume w.l.o.g that for every $q \in Q$ and $a \in \Sigma$, $\delta(q,a) \in Q \times \{{\bf u} \in \mathbb{N}^n\mid \sum_{i=1}^n {\mathit{abs}}({\bf u}[i]) \leq 1\}$ and ${\bf u}_0={\bf 0}^n$. Intuitively,   $\m{M}$ has, for each $i \in [1,n]$,  a stack symbol $\gamma_i$ such that the number of pending threads with local configuration $(\gamma_i,2)$ denotes the current value of the $i$-th counter of $\m{V}$. The system $\m{M}$ has also 
  a special stack symbol $\gamma'_0$ such that the  pending threads with local configuration $(\gamma'_0,1)$  are used to create threads with local configuration $(\gamma_i,2)$ where $i \in [1,n]$ (which corresponds to  the increment of  the value of a counter of $\m{V}$). We now sketch the behavior of $\m{M}$. First, $\m{M}$ creates  an arbitrary  number of threads with local configuration $(\gamma'_0,1)$ from the initial configuration. Then, the simulation of a rule $\delta(q,a)=(q',{\bf u})$ depends on the value of the vector ${\bf u}$: $(1)$ If ${\bf u}={\bf 0}^n$, then $\m{M}$ moves its  state from  $q$ to $q'$, $(2)$ If ${\bf u}={\bf 0}^n[i \sub 1]$ for some $i \in [1,n]$, then $\m{M}$ uses a thread with local configuration $(\gamma'_0,1)$ to create a thread with local configuration $(\gamma_i,2)$ while moving its state from $q$ to $q'$, and $(3)$  If ${\bf u}={\bf 0}^n[i \sub -1]$ for some $i \in [1,n]$, then $\m{M}$ transforms the local configuration of a pending thread from $(\gamma_i,2)$ to $(\epsilon,3)$.
 Formally $\m{M}=(P,\Gamma,\Delta,p_0,\gamma_0)$ is built from $\m{V}$ as follows:

 \begin{iteMize}{$\bullet$}
 \item $P= \{p_0\} \cup Q $ is the set of states such that $p_0 \notin Q$. $p_0$ is the initial state. A  state  $q \in Q$ represents  the current state of $\m{V}$.

 \item $\Gamma=\{\gamma_0,\gamma_1, \cdots, \gamma_n\} \cup \{\gamma'_0\}$ is the finite set of stack symbols. The symbol $\gamma_0$ represents the initial stack symbol. The  symbol $\gamma'_0$ represents the stack content of  auxiliary threads that are ``consumed"  in order  to simulate  an operation  of   $\m{V}$. For every $i \in [1,n]$, the number of pending threads with stack content $\gamma_i$ denotes the current value of the $i$-th counter of $\m{V}$.
 
 \item $\Delta$ is the smallest transition relation satisfying the following conditions:
 
 \begin{iteMize}{$-$}
 
 \item 
 $\co p_0, \gamma_0 \cf \by{}_{\m{M}} \co p_0,\gamma_0 \cf \vtr \gamma'_0$ and $\co p_0,\gamma_0 \cf \swi_{\m{M}} \co q_0,\epsilon \cf $. These transitions create an arbitrary number of threads with local configuration $(\gamma'_0,1)$ before moving the state from $p_0$ to $q_0$.

 \item For every $q \in Q$, we have that  $q \swi_{\m{M}} q \vtl \gamma'_0$. This transition corresponds to start  the execution of a pending thread  with stack content $\gamma'_0$ to simulate an operation   of   $\m{V}$ that  increments the value of  a counter.
  
   \item For every $q \in Q$, we have that  $\co q, \gamma'_0 \cf \swi_{\m{M}} \co q,\epsilon \cf $. This transition corresponds to the interruption of    the execution of the current active  thread with stack content $\gamma'_0$ in order to permit   the  simulation by $\m{M}$  of an operation of $\m{V}$ that decrements a counter.

 \item 
 For every $q,q' \in Q$ and $a \in \Sigma$, if $\delta(q,a)=(q',{\bf 0}^n)$, then $\co q, \gamma'_0 \cf \by{}_{\m{M}} \co q',\gamma'_0 \cf \vtr \epsilon$. This transition simulates an operation of $\m{V}$ that moves the state from $q$ to $q'$.

 \item 
 For every $q,q' \in Q$, $a \in \Sigma$, and each $i \in [1,n]$, if $\delta(q,a)=(q',{\bf 0}^n[i \sub 1])$, then $\co q, \gamma'_0 \cf \by{}_{\m{M}} \co q',\gamma'_0 \cf \vtr \gamma_i$. This transition simulates an operation that  increments the $i$-th counter of $\m{V}$. Notice that the switch number of the created thread with stack content $\gamma_i$ is $2$ since the switch number of the active thread (with stack content $\gamma'_0$) is always equal to $1$.

 \item 
 For every $q,q' \in Q$, $a \in \Sigma$, and $i \in [1,n]$, if $\delta(q,a)=(q',{\bf 0}^n[i \sub -1])$, then $q \swi_{\m{M}} q' \vtl \gamma_i$,  and  $\co q', \gamma_i \cf \swi_{\m{M}} \co q',\epsilon \cf$. These transitions simulate an operation that  decrements  the value of the $i$-th counter of $\m{V}$. 
\end{iteMize}
 \end{iteMize}
 
 \medskip

\noindent Observe that the size of $\m{M}$ is polynomial in the size of $\m{V}$. Moreover,  the relation between $\m{V}$  and $M$ is given by the following lemma:
 
 \begin{lem}
 \label{chap3.lem.vass.dcfs.dir1}
 Let $q \in Q$. $q$ is reachable by $\m{V}$ if and only if $q$ is $2$-bounded reachable by $\m{M}$.
 \end{lem}
 
 The proof of Lemma \ref{chap3.lem.vass.dcfs.dir1} is done by induction on the length of the runs  and is given in Appendix \ref{appendix.ref.chap3.lem.vass.dcfs.dir1}.

As an immediate consequence of Lemma \ref{chap3.lem.vass.dcfs.dir1}, we obtain the following result:

\begin{prop}
\label{chap3.prop.dcfs03}
The coverability problem for VASSs is polynomially reducible to the BSR$[2]$ for DCFSs.
\end{prop}

\section{The SSR$[k]$ problem for DCFSs}
In this section, we   consider  the problem SSR$[k]$ for  $k \in \mathbb{N}$. We show that  the problem SSR$[k]$  for DCFSs is NP-complete. But before going into the details, let us recall the definition of the existential Presburger arithmetic and some related results.
 
 \subsection{Existential Presburger arithmetic}

Let $\m{V}$ be a set of variables. We use $x,y,\ldots$ to range over variables in $\m{V}$.  The set of terms of the Presburger arithmetic is  defined by:
 $$t\,::= \,\,0\,|\,1\,|\,x\,|\,t+t $$

Then, the  class  of  existential formulae is defined  as follows:

  $$ \varphi\,\,::=\,\, t \leq t  \,|\, \varphi\vee \varphi\,|\, \varphi\wedge \varphi \,|\, \exists x. \,\varphi$$

 The length of a Presburger formula $\varphi$, denoted  by $|\varphi|$,  is the number of letters used in writing $\varphi$.  
The notion of free variables for an existential  Presburger formula is defined as usual.
  We write  $\mathit{FV}(\varphi) \subseteq \m{V}$  to denote that the formula $\varphi$ has $\mathit{FV}(\varphi)$ as a set of free variables. The semantics of existential Presburger formulae is defined in the standard way. Given   a function $f$ from $\mathit{var}(\varphi)$ to $\mathbb{N}$, we write $f \models \varphi$ if $\varphi$ holds for $f$ (in the obvious sense) and, in this case, we say    that   $f$ satisfies $\varphi$.
We use    $\open\varphi\close$  to denote the set $\{f \in [\mathit{FV}(\varphi)\ra \mathbb{N}]\mid f \models \varphi\} $.

An existential  Presburger formula $\varphi$ is  satisfiable  if and only if $\open \varphi \close \neq \emptyset$. The satisfiability problem  for $\varphi$ consists in checking whether $\varphi$ is satisfiable. It is  well-known that the satisfiability problem for existential Presburger formulae is NP-complete \cite{VSS05}.

 \begin{thm}
 \label{pres.np-complete}
The satisfiability problem for existential Presburger formulae  is   NP-complete.
 \end{thm}

 We recall  that the Parikh image of a context-free language is definable by an existential Presburger formula.

\begin{thm}[\cite{SSMH04}]
\label{pres-pda}
If  $\m{P}$ is a PDA with input alphabet $\Sigma$, then, it is possible  to  construct, in time and space polynomial in  $|\m{P}|$, an existential Presburger formula $\varphi$ with free variables $\Sigma$ such that  $\open \varphi \close =\mathit{Parikh}(L(\m{P}))$.
\end{thm}

 
 \subsection{The SSR$[k]$ problem  for DCFSs  is NP-complete}
 In this section, we mainly prove the following result:

\begin{thm}
\label{srpkl_dcfs}
For every $k \in \mathbb{N}$, the problem SSR$[k]$  for DCFSs is NP-complete.
\end{thm}

The NP-hardness is proved by  a reduction from  the coverability  problem of  acyclic Petri nets \cite{Ste95} to SSR$[k]$. This is done by  a simple  adaptation  of the construction given in  Section \ref{vass-bsr}. The upper-bound is obtained by a reduction to the satisfiability problem  for \emph{existential} Presburger formulae.

Let $\m{M}=(Q,\Gamma,\Delta,q_0,\gamma_0)$  be a DCFS, $k$ be a natural number, and $F \subseteq Q$ be a set of target states.  To reduce  the $k$-stratified state  reachability problem  for $\m{M}$ to the satisfiability problem of an existential formula $\varphi$, we proceed in two steps: First,  we construct a bounded stack pushdown automaton $\m{P}$  that simulates the $k$-stratified computations of $\m{M}$ without taking into account the causality constraints. (The use of a pushdown automaton here is for  technical convenience. In principle, $\m{P}$ can be encoded as a finite state automaton, but this will make the construction cumbersome.) In fact,  $\m{P}$ assumes that there is an unbounded number of pending threads  for any local configurations in $\Gamma_{\epsilon} \times [0,k]$. Intuitively $\m{P}$ performs the same pushdown operations  as the ones specified by $\Delta$ while making visible as transition labels: $(1)$ $(\gamma,i,\vtr) $ if the local configuration of the  created (or the interrupted)  thread is $(\gamma,i)$, $(2)$ $(\gamma,i,\vtl)$ if the  local configuration of  the   pending thread that has been   activated is $(\gamma,i)$, and $(3)$ $(\epsilon, i, -)$  if there no  thread creation and the switch number of the current active thread is $i$.

Then, we show that there is a $k$-stratified computation of $\m{M}$ if and only if  there is a computation $\pi$ of $\m{P}$  that satisfies the following two conditions: 

\begin{iteMize}{$\bullet$}
\item {\em The stratified condition}: Threads in $\pi$ are  scheduled according to their increasing switch number (from $0$ to $k$).

\item   {\em The flow condition}:  For every stack content $\gamma \in \Gamma$  and switch number $i \in [0,k]$, the number of occurrences of  $(\gamma,i,\vtr)$ in $\pi$ is greater than the number of occurrences of $(\gamma,i,\vtl)$ in $\pi$ (i.e., the number of created (or interrupted) threads  with local configuration $(\gamma,i)$  is greater than the number of  threads with local configuration $(\gamma,i)$ that has been activated).
 \end{iteMize}

Since  the set of traces  that satisfies the stratified condition  is   a regular one, we can construct a pushdown automaton $\m{P}'$  (of bounded stack depth) that recognizes the set of traces of $\m{P}$ that satisfies the first condition. Therefore, we can  use  Theorem \ref{pres-pda} to  construct an existential Presburger formula $\varphi'$ that characterizes the Parikh image of  the set of  traces of $\m{P}'$.
On the other hand,  the  flow condition can be expressed as an existential Presburger formula $\varphi''$ over the set of variables  $\{(\gamma,i,\vtr)\,|\, \gamma \in \Gamma, i \in [0,k] \}$ and  $\{(\gamma,i,\vtl)\,|\, \gamma \in \Gamma, i \in [0,k] \}$.   Armed with these results, we can show that      the $k$-stratified state  reachability problem  for $\m{M}$ is reducible to the satisfiability problem of  the existential formula $\varphi=\varphi' \wedge \varphi''$.

\medskip

Let us give more details about the constructions described above.

\medskip

\noindent
{\bf From the DCFS $\m{M}$ to the pushdown automaton $\m{P}$:} The pushdown automaton  $\m{P}=(P,\Sigma,\Gamma_{\m{P}},\Delta_{\m{P}},$ $p_0,\gamma_{\m{P}},F_{\m{P}})$  is built up from $\m{M}$ as follows:

\begin{iteMize}{$\bullet$}
\item $P=Q  $ is the finite set of states. A state $q$ represents  the global state of  $\m{M}$.

\item $\Sigma= \bigcup_{i=0}^{k} \Sigma_i $  is the finite set  of input symbols where $\Sigma_i=\Sigma^{\sf cr}_i \cup \Sigma_i^{\sf r} \cup \Sigma_i^{\sf l}$ with $\Sigma^{\sf cr}_i=\Gamma_{\epsilon} \times \{i+1\} \times \{\vtr\} $,  $\Sigma_i^{\sf r}= \Gamma \times \{i\} \times \{\vtl\}$, and  $\Sigma_i^{\sf l} = \{(\epsilon,i,-)\}$ for all $i \in [0,k]$. A  transition labeled with $(\alpha,i,\vtr)$ corresponds to  a rule  of $\m{M}$ that: (1) creates  a thread with  local configuration $(\alpha,i)$, or $(2)$ interrupts the execution of the active thread with stack content  is  $\alpha$. A  transition labeled with $(\alpha,i,\vtl)$ corresponds to  a rule  of $\m{M}$ that  activates   a pending thread with  local configuration $(\alpha,i)$. A transition labeled with $(\epsilon,i,-)$ corresponds to a  rule of $\m{M}$ without thread creation and where the switch number of the current active thread is $i$.

\item $\Gamma_{\m{P}}=(\Gamma_{\epsilon} \times [0,k]) \cup \{\bot\} $ is the finite set of stack symbols. Each symbol in $\Gamma_{\m{P}}$  corresponds to the local configuration of the  active thread of $\m{M}$.

\item $\Delta_{\m{P}}$ is the smallest transition relation satisfying the following conditions:

\begin{iteMize}{$-$}

\item  For every $i \in [0,k]$ and   $\co q, \gamma  \cf \by{}_{\m{M}} \co q', u \cf \vtr \epsilon $,   $\co q,(\gamma,i) \cf \by{(\epsilon,i,-)}_{\m{P}}$ $ \co q',(u,i)\cf$. This transition corresponds to the simulation of  a transition of $\m{M}$ without thread creation.

\item  For every $i \in [0,k]$ and $\co q, \gamma  \cf \by{}_{\m{M}} \co q', u \cf \vtr \alpha$ with $\alpha \in \Gamma$,  $\co q,(\gamma,i)  \cf\,  \by{(\alpha,i+1,\vtr)}_{\m{P}} \, $ $\co q',(u,i)\cf$. This corresponds to  the simulation of    a transition of $\m{M}$ with thread creation.

\item  For every $i \in [0,k]$ and  $\co q, \gamma  \cf \swi_{\m{M}} \co q', u \cf $, $\co q,(\gamma,i)  \cf\,  \by{(u,i+1,\vtr)}_{\m{P}} \,$ $\co q',\bot\cf$. This corresponds to  the interruption of   the execution of the active thread of $\m{M}$.
\item  For every $i \in [0,k]$ and  $q   \swi_{\m{M}}  q' \vtl \gamma $, $\co q,\bot\cf\,  \by{(\gamma,i,\vtl)}_{\m{P}} \,$ $\co q', (\gamma,i) \cf$. This corresponds to  the activation of   a   pending  thread of $\m{M}$ with local configuration  $(\gamma,i)$.

\end{iteMize}
\item $p_0=q_0$ is the initial state.
\item $\gamma_{\m{P}}=\bot$ is the initial stack symbol.
\item $F_{\m{P}}=F$ is the set of final states.
\end{iteMize}\smallskip

\noindent Observe that the size of the pushdown automaton $\m{P}$ is polynomial in the size of the DCFS $\m{M}$. Moreover, the   depth  of the stack of  $\m{P}$ is always bounded by one.

{The relation between the DCFS $\m{M}$ and the pushdown automaton  $\m{P}$} is established by Lemma \ref{ssrproblem-pda-relation} which states that there is a state $q \in F$  such that $q$ is  $k$-stratified  reachable by $\m{M}$  if and only if  there is a computation $\pi$ of $\m{P}$  that satisfies the stratified condition and the flow condition. 

\begin{lem}
\label{ssrproblem-pda-relation}
A state $q \in F$  is $k$-stratified reachable by $\m{M}$ if and only if there is  $\sigma_i \in \Sigma_{i}^*$ for all $i \in [0,k]$ such that:

\begin{iteMize}{$\bullet$}
\item $\sigma_0 \sigma_1 \cdots \sigma_k \in \mathit{Traces}_{\m{T}(\m{P})}(\{(q_0,\bot)\},F \times \{\bot\})$, and
\item $|\sigma_i|_{(\gamma,i,\vtl)} \leq |\sigma_{i-1}|_{(\gamma,i,\vtr)}$ for all $\gamma \in \Gamma$ and $i \in [0,k]$ where $\sigma_{-1}=(\gamma_0,0,\vtr)$.
\end{iteMize}
\end{lem}

The proof of Lemma \ref{ssrproblem-pda-relation} is done by induction and is given in the Appendix \ref{appendix.ssrproblem-pda-relation}.
\bigskip

\noindent
{\bf From the PDA $\m{P}$ to the {existential} Presburger formula $\varphi$:} In the following, we show that
the problem of checking  whether  there is  $\sigma_i \in \Sigma_{i}^*$ for all $i \in [0,k]$ such that $\sigma_0 \sigma_1 \cdots \sigma_k \in \mathit{Traces}_{\m{T}(\m{P})}(\{(q_0,\bot)\},F \times \{\bot\})$ and $|\sigma_i|_{(\gamma,i,\vtl)} \leq |\sigma_{i-1}|_{(\gamma,i,\vtr)}$ for all $\gamma \in \Gamma$ and $i \in [0,k]$ with  $\sigma_{-1}=(\gamma_0,0,\vtr)$ is polynomially reducible to the satisfiability problem  of an existential Presburger formula $\varphi$.  This implies that the SSR$[k]$ problem for $\m{M}$ is polynomially reducible to the satisfiability problem for $\varphi$ (see  Lemma \ref{ssrproblem-pda-relation}).

\begin{lem}
\label{pres.pda.chap3.lem}
It is possible to construct an existential Presburger formula $\varphi$ with  $\open \varphi \close \neq \emptyset$ if and only if there is  $\sigma_i \in \Sigma_{i}^*$ for all $i \in [0,k]$ such that $\sigma_0 \sigma_1 \cdots \sigma_k \in \mathit{Traces}_{\m{T}(\m{P})}(\{(q_0,\bot)\},F \times \{\bot\})$ and $|\sigma_i|_{(\gamma,i,\vtl)} \leq |\sigma_{i-1}|_{(\gamma,i,\vtr)}$ for all $\gamma \in \Gamma$ and $i \in [0,k]$ with  $\sigma_{-1}=(\gamma_0,0,\vtr)$.
\end{lem}
\proof
Let $\m{P}'$ be the pushdown automaton such that  $L(\m{P}')= \mathit{Traces}_{\m{T}(\m{P})}(\{(q_0,\bot)\},F \times \{\bot\}) \cap ({\Sigma^*_0} \cdot \Sigma_1^* \cdots \Sigma^*_k)$. Such pushdown automaton $\m{P}'$ is effectively constructible from $\m{P}$ since the class of pushdown automata is closed under intersection with a regular language.

Now, we can use Theorem \ref{pres-pda} to construct a Presburger formula $\varphi'$  with free variables  $\Sigma$ such that $\open \varphi \close=\mathit{Parikh}(L(\m{P}'))$. In addition, for every $i \in [1,k]$, we construct an existential Presburger formula $\varphi_i$ with free variables $\Sigma$ such that $\varphi_i= \,\bigwedge_{\gamma \in \Gamma}\, \big( (\gamma,i,\vtl) \leq (\gamma,i,\vtr)\big)$. Let $\varphi_0= \big( \bigwedge_{\gamma \in \Gamma \setminus \{\gamma_0\}} \big( (\gamma,0,\vtl) \leq 0 \big) \big) \wedge  \big((\gamma_0,0,\vtl) \leq 1\big) $ and $\varphi''= \bigwedge_{i=0}^k  \varphi_i$.

Then, it is not hard  to see  that the existential Presburger formula $\varphi= \varphi' \wedge \varphi'' $ is satisfiable if and only if for every $i \in [0,k]$,  there are there are $\sigma_i \in \Sigma_{i}^*$ for all $i \in [0,k]$ such that $\sigma_0 \sigma_1 \cdots \sigma_k \in \mathit{Traces}_{\m{T}(\m{P})}(\{(q_0,\bot)\},F \times \{\bot\})$ and $|\sigma_i|_{(\gamma,i,\vtl)} \leq |\sigma_{i-1}|_{(\gamma,i,\vtr)}$ for all $\gamma \in \Gamma$ and $i \in [0,k]$ with  $\sigma_{-1}=(\gamma_0,0,\vtr)$. \qed

As an immediate consequence of  Theorem \ref{pres.np-complete} and  Lemma \ref{pres.pda.chap3.lem}, we obtain the following result: 
 
\begin{lem}
For every $k \in \mathbb{N}$, the problem SSR$[k]$  for DCFSs is in NP.
\end{lem}

\section{Reachability analysis for  dynamic networks of concurrent pushdown systems}
In this section, we consider the case of DCPSs. It is well-known that the SR problem is undecidable already for networks with two concurrent pushdown processes. We show however that both problems BSR$[k]$ and SSR$[k]$ are decidable, for any given bound $k \in \mathbb{N}$. For that, we prove the following fact.

\begin{thm}
\label{srpkl_dcps}
For every $k \in \mathbb{N}$, the  problems BSR$[k]$ and the SSR$[k]$ for DCPS are exponentially reducible to the corresponding problems for DCFS.
\end{thm}

A corollary of  Theorem \ref{dcfs-exp-complete}, Theorem \ref{srpkl_dcfs}, and Theorem \ref{srpkl_dcps}, we obtain the following results:

\begin{cor}
\label{srpkl_dcps_coro}
For every $k \in \mathbb{N}$, the  BSR$[k]$ problem  for DCPSs is in 2-EXPSPACE, and the SSR$[k]$ problem  for DCPSs  is in NEXPTIME. 
\end{cor}

The rest of this section is devoted to the proof of Theorem \ref{srpkl_dcps}. Let us fix a DCPS $\m{M}=(Q,\Gamma,\Delta,q_0,\gamma_0)$.  We show that it is possible to construct a DCFS $\m{M}_{\sf fs}$ such that the problems BSR$[k]$ and SSR$[k]$ for $\m{M}$ can be reduced to their corresponding problems for $\m{M}_{\sf fs}$. Let us present the main steps of this construction. 
For that, let us consider the problem BSR$[k]$, for some fixed $k \in \mathbb{N}$. Then, let us concentrate on the computations of one  thread, and assume that this thread will be interrupted  $i$ times (with $i \leq k+1$) during its execution starting from some initial global state $q$ and initial local state $\gamma$. The computations of such a thread correspond to runs of a pushdown automaton,  built out of $\m{M}$, which (1) performs the same operations on the stack and the global state as the ones specified by $\Delta$, (2) makes visible as transition labels the local state (element of $\Gamma$) of spawned threads,    and (3) nondeterministically guesses jumps from a global state to another one corresponding to the effect of context switches. These jumps are also made visible as transition labels under the form of $(q,\alpha,q') \in (Q \times \Gamma_{\epsilon} \times  Q)$ (meaning that the computation of the thread is interrupted at the  state $q$ with  stack content $\alpha w$ for some $w \in \Gamma^*$,   and is resumed at the state $q'$). In fact, if a thread  fires  a transition labeled by a  symbol  of the form $(q,\epsilon,q')$ then its execution  will be  definitely interrupted  (i.e., the execution of this thread will never be resumed again). The number of such jumps in each run is precisely  $i$. 

Then, the problem is to handle the composition of all the computations   of the generated threads and to make sure that the guesses made by each one of them (on their control state jumps due to context switches) are correct. In fact, handling this composition is very a hard task in general when threads are  modeled as  pushdown automata.  To overcome this difficulty, the key observation is that it is possible to assume without loss of preciseness that some of  the  generated threads can be ignored (or lost).  Indeed, these threads can always be considered as threads which will never be scheduled. Therefore, the behaviors of each thread can be modeled using a finite-state automaton which recognizes the downward closure of the language of the pushdown automaton of a thread with respect to  the subword relation. We know by Theorem \ref{prop-down-pda} that this automaton is effectively constructible. So, 
let $\m{A}_{(q,\gamma)}$ be the automaton modeling the computations of threads starting from the state $q$ and initial stack content $\gamma$, and performing at most  $k+1$ interruptions.  We assume w.l.o.g that $\m{A}_{(q,\gamma)}$ has no $\epsilon$-transitions.

The next step is to synchronize the so-defined finite-state automata in order to represent valid computations of the whole system. For that, we define a DCFS $\m{M}_{\sf fs}$ which simulates the composition of these automata as follows:

\begin{iteMize}{$\bullet$}

\item A pending thread with  stack content $\gamma$ which has never been activated can be dispatched by $\m{M}_{\sf fs}$ at the moment of a context switch. For that, $\m{M}_{\sf fs}$ has a rule 
$\co q ,\gamma \cf \by{}{}_{\m{M}_{\sf fs}} \co \sharp, s_0 \cf \vtr \epsilon$ where  $s_0$ is the initial state of $\m{A}_{(q,\gamma)}$, for every possible  starting  $q$ and  every stack symbol $\gamma \in \Gamma$.   This rule allows to check that the control state is $q$, and to move the  system  to a special state $\sharp$ corresponding to  the simulation of  a phase without  context switches.

\item
During the simulation, when a transition $s \by{\gamma}_{\m{A}_{(q,\gamma)}} s'$, with $\gamma \in \Gamma$, is encountered, a new thread  is   spawned by $\m{M}_{\sf fs}$ with initial stack content $\gamma$. This is done using a rule of the form
$\co \sharp, s \cf \by{}_{\m{M}_{\sf fs}} \co \sharp, s' \cf  \vtr \gamma$. The new thread will stay pending until $\m{M}_{\sf fs}$ can dispatch it.

\item
Encountering a transition $s \by{(q_1,\alpha,q_2)}_{\m{A}_{(q,\gamma)}} s'$ means that the computation of the simulated thread  is interrupted at  the global store $q_1$ with   stack content $\alpha w$ for some $w \in \Gamma^*$, and will be resumed later when the global state will become $q_2$ (due to the execution of some other threads).  
 Then,  $\m{M}_{\sf fs}$ moves from its global  state $\sharp$ to the  global  state $q_1$   so that the control can be taken by   another  pending   thread), and transforms the stack  configuration of the current thread (which may be interrupted) to $(q_2,(s',\alpha))$. This is done by a rule of the form  $\co \sharp, s \cf \swi_{\m{M}_{\sf fs}} \co q_1,(q_2,(s',\alpha))\cf$.

\item  To simulate a transition $q \swi_{\m{M}} q' \vtl \gamma$  that starts/resumes the execution of a pending thread with topmost stack symbol $\gamma \in \Gamma$, $\m{M}_{\sf fs}$ has the  rules $q \swi_{\m{M}_{\sf fs}} q' \vtl \gamma$ and $q \swi_{\m{M}_{\sf fs}} q' \vtl (q',(s,\gamma))$. 
In this  case, we observe  that the only action that can be done by $\m{M}_{\sf fs}$ after executing these rules is to activate   some pending thread with topmost stack symbol $\gamma'$  (either dispatched for the first time, or resumed after some interruption). 

We have seen above how $\m{M}_{\sf fs}$ dispatches pending threads for the first time. The resumption of threads at state $q'$ is done by having rules of the form $\co q',  (q',(s,\gamma)) \cf  \by{}_{\m{M}_{\sf fs}} \co \sharp, s \cf \vtr \epsilon$. Such a rule means that if a pending thread $(q',(s,\gamma))$ exists, then it can be activated and the simulation of its behaviors is  resumed  from the state $s$ (at which it was stopped at the last interruption). 
\end{iteMize}\smallskip

\noindent Let us give in more details the construction described above.

\subsection{Simulation of  threads of $\m{M}$ with finite-state automata}
Next, we  give  the construction of the finite state automaton $\m{A}_{(q,\gamma)}$ for some given $q \in Q$ and $\gamma \in \Gamma$.
For that, we start by considering a  pushdown automaton $\m{P}_{(q,\gamma)}$ simulating the behaviors of a  thread that starts its execution from  the global  state $q$ and the initial  stack configuration $\gamma$
after some number of jumps in the global  state (representing guesses on the effect of context switches). The spawned thread as well as the guesses on the  global state jumps made during the computation are made visible as  transition labels.

Then, let $\m{P}_{(q,\gamma)}=(P,\Sigma,\Gamma, \Delta_{\m{P}},q,\gamma,Q)$ be the pushdown automaton  where:

\begin{iteMize}{$\bullet$}
\item $P=Q \cup (Q \times \Gamma)$ is the finite set of states,
\item $\Sigma= \Gamma \cup \Sigma_{\sf sw} \cup \Sigma_{\sf inr}$ is the finite set of input symbols with $\Sigma_{\sf sw}=Q \times \Gamma \times Q $ and $\Sigma_{\sf inr}=Q \times \{\epsilon\} \times Q$,
\item $\Delta_{\m{P}}$ is the smallest  transition relation  such that: 

\begin{iteMize}{$-$}
\item For every  $\co q_1, \gamma_1\cf  \by{}_{\m{M}} \co q_2, u\cf  \vtr \alpha$,   $\co q_1, \gamma_1\cf  \by{\alpha}_{\m{P}_{(q,\gamma)}} \co q_2, u\cf $. This rule simulates a pushdown operation on  the active thread with the possibility of a thread creation.

\item For every  $\co q_1, \gamma_1\cf  \swi_{\m{M}} \co q_2, u\cf $ and $q'_2 \in Q$,   $\co q_1, \gamma_1\cf  \by{(q_2,\epsilon,q'_2)}_{\m{P}_{(q,\gamma)}} \co q'_2, u\cf $. This rule  corresponds to   interrupt  the execution of the active  thread at the state $q_2$. In addition,  the execution of this thread will never be resumed again.

\item For every  $\co q_1, \gamma_1\cf  \swi_{\m{M}} \co q_2, u\cf $, $q'_2 \in Q$, and $\gamma' \in \Gamma$,   $\co q_1, \gamma_1\cf  \by{(q_2,\gamma',q'_2)}_{\m{P}_{(q,\gamma)}} $ $\co (q'_2,\gamma'), u\cf $ and $\co (q'_2,\gamma') , \gamma' \cf$ $  \by{\epsilon}_{\m{P}_{(q,\gamma)}} \co q'_2, \gamma' \cf $. This rule simulates  the interruption of the execution of the active  thread at the state $q_2$. In addition,  the execution of this   thread will be resumed at the state $q'_2$ with topmost stack symbol $\gamma'$.

\end{iteMize}

\end{iteMize}\smallskip

\noindent Then, the set of behaviors represented by this pushdown automaton which correspond to precisely  $i\geq 1$ context switches (or interruptions)  is given by the following  language:

$$L'_{((q,\gamma),i)}=L({\m{P}_{(q,\gamma)}}) \cap \big(\big(  \Gamma^* \cdot \Sigma_{\sf sw} \big)^{i-1} \big(  \Gamma^* \cdot \Sigma_{\sf inr} \big)\big)$$

%

The set $L'_{((q,\gamma),i)}$ is a context-free language in general (since it is the intersection of a context-free language with a  regular one). Due to the fact that some of the generated threads can be ignored (or lost), we can consider without loss of preciseness 
the downward closure of $L'_{(q,\gamma)}$ w.r.t. the sub-word relation corresponding to the deletion of symbols in $\Gamma$ while preserving all symbols in $\Sigma_{\sf sw} \cup \Sigma_{\sf inr}$, i.e., the set

$$L'_{(q,\gamma)}=\bigcup_{i=1}^{k+1} \bigg(L'_{((q,\gamma),i)} \downarrow  \cap  \big(\big(  \Gamma^* \cdot \Sigma_{\sf sw} \big)^{i-1} \big(  \Gamma^* \cdot \Sigma_{\sf inr} \big)\big) \bigg)$$

By Theorem \ref{prop-down-pda},   the language $L'_{(q,\gamma)}$ is  regular  and can be effectively represented by a finite-state automaton
$\m{A}_{(q,\gamma)}=(S_{(q,\gamma)}, \Sigma,\Delta_{(q,\gamma)},I_{(q,\gamma)},F_{(q,\gamma)})$. We assume w.l.o.g that all the  states in the automaton $\m{A}_{(q,\gamma)}$ are co-reachable from the final states. We assume also that $\Delta_{(q,\gamma)} \subseteq S_{(q,\gamma)} \times \Sigma \times S_{(q,\gamma)}$ (i.e., there is no transition of $\m{A}_{(q,\gamma)}$ labeled by the empty word).

\subsection{From the DCPS $\m{M}$ to the  DCFS $\m{M}_{\sf fs}$} In the following,  
we give the formal definition of  the DCFS $\m{M}_{\sf fs} $. The system $\m{M}_{\sf fs}$ is defined by the tuple  $ (Q_{\sf fs},\Gamma_{\sf fs},\Delta_{\sf fs},q_0,\gamma_0)$ where:

\begin{iteMize}{$\bullet$}
\item $Q_{\sf fs}=Q \cup \{\sharp\} $ is the finite set of states.
\item $\Gamma_{\sf fs}=\Gamma \cup S_{\sf fs}^{\sf sm} \cup S_{\sf fs}^{\sf sw}$ is the finite set of stack alphabet where   $S_{\sf fs}^{\sf sm}=\bigcup_{(q,\gamma) \in Q \times \Gamma} S_{(q,\gamma)}$  and  $S_{\sf fs}^{\sf sw}=Q \times S_{\sf fs}^{\sf sm} \times \Gamma_{\epsilon}$.

\item $\Delta_{\sf fs}$ is the smallest set of transitions such that
\begin{iteMize}{$-$}

\item {\em Initialize:} For every $\gamma \in \Gamma$ and $q \in Q$,   we have $\co q, \gamma\cf  \by{}_{\m{M}_{\sf fs}} \co \sharp, s_0 \cf \vtr \epsilon$
where  $s_0$ is the initial state of $\m{A}_{(q,\gamma)}$.

\item {\em Spawn:}  For every  $q \in Q$, $\gamma \in \Gamma$, and  $s \by{\alpha}_{\m{A}_{(q,\gamma)}} s'$,   we have $\co \sharp, s \cf  \by{}_{\m{M}_{\sf fs}} \co \sharp, s' \cf \vtr \alpha$. (Notice that, from the definition of $\m{A}_{(q,\gamma)}$, $\alpha$ is necessarily in $\Gamma$.)

\item {\em Interrupt:} For every $q \in Q $, $\gamma \in \Gamma$, and $s \by{(q_1,\alpha,q_2)}_{\m{A}_{(q,\gamma)}} s'$, we have $\co \sharp, s \cf  \swi_{\m{M}_{\sf fs}} \co q_1, (q_2,(s',\alpha)) \cf $.

\item {\em Dispatch:} For every $s \in S_{\sf fs}^{\sf sm}$ and $q_1 \swi_{\m{M}} q_2 \vtl \gamma'$, we have $q_1\swi_{\m{M}_{\sf fs}} q_2 \vtl \gamma'$ and $q_1\swi_{\m{M}_{\sf fs}} q_2 \vtl (q_2,(s,\gamma'))$.

\item {\em Resume:}  For every $q \in Q $, $\gamma\in \Gamma$, and   $ s \in S_{\sf fs}^{\sf sm}$,  we have $\co q, (q,(s,\gamma)) \cf \by{}_{\m{M}_{\sf fs}}$ $ \co \sharp, s \cf \vtr \epsilon$.

\end{iteMize}

\end{iteMize}

 Theorem \ref{srpkl_dcps} is an immediate consequence of Lemma \ref{dcfs->dcps-1}.

\begin{lem}
\label{dcfs->dcps-1}
For every $k \in \mathbb{N}$, a control state $q \in Q$ is $k$-bounded reachable (resp. $k$-stratified) reachable by  $\m{M}$ iff $q$ is  $k$-bounded (resp. $k$-stratified) reachable  by  $\m{M}_{\sf fs}$.
\end{lem}

 The proof of Lemma \ref{dcfs->dcps-1} is given in Appendix \ref{sec.proof.lemma}.

 \section{Conclusion}

 We have proposed new concepts for context-bounded verification we believe that are natural and suitable for programs with dynamic thread creation. These concepts are based on the idea of bounding the number of switches for each thread and not for all the threads in a computation. 
 
 First, we have proved that even for  finite-state threads, adopting such a notion of context-bounding leads in general to a problem which is as hard as the coverability problem of Petri nets. This means that, in theory, the complexity of this problem is high, but in practice, there are quite efficient techniques (based on iterative computation of under/upper approximations) developed recently for solving this problem which have been implemented and used successfully in \cite{GRB06,GRVB06}. Moreover, we have proposed a notion of stratified context-bounding for which the verification is in NP, i.e., as hard as in the case without dynamic thread creation. An interesting question is how to implement efficiently the analysis in this case using clever encodings in SMT solvers.
 
Moreover, we have proved that the considered problems are still decidable for the case of pushdown threads. This is done by a nontrivial reduction to the corresponding problems for finite-state threads. This reduction is based on computing the regular downward closure of context-free languages w.r.t. the sub-word relation. The downward closure computation may lead in general to an unavoidable exponential blow-up. This is due to the succinctness of context-free grammars w.r.t.  finite state automata: For instance,  the finite language $\{a^{2^N}\}$, for a fixed $N\geq 1$, can be defined with  a context-free grammar of size $N$ whereas a finite-state automaton   representing it (or its downward closure) is necessarily of size at least $2^N$. An interesting open problem is whether there is an alternative proof technique which allows to avoid the downward closure construction. In practice, we believe that it would be possible to overcome this problem by for instance designing algorithms allowing to generate efficiently and incrementally (parts of the) downward closure.

Finally, in our models, we consider that each created thread inherits a switch number from its father (the one of its father plus $1$). An alternative definition can be obtained by considering that each created thread is given the switch number $0$. (Therefore, each thread can perform up to $k$ switches.)  However, the problem  SSR$[k]$   for finite state threads (resp. pushdown threads) becomes EXPSPACE-complete  (in 2-EXPSPACE) instead of NP-complete (NEXPTIME) for this definition.


\bibliographystyle{alpha}

\bibliography{main.bib}

\begin{thebibliography}{BMOT05}

\bibitem[BESS05]{BESS05}
A.~Bouajjani, J.~Esparza, S.~Schwoon, and J.~Strejcek.
\newblock Reachability analysis of multithreaded software with asynchronous
  communication.
\newblock In {\em FSTTCS'05}, LNCS 3821, pages 348--359. Springer, 2005.

\bibitem[BMOT05]{BMOT05}
Ahmed Bouajjani, Markus M{\"u}ller-Olm, and Tayssir Touili.
\newblock Regular symbolic analysis of dynamic networks of pushdown systems.
\newblock In {\em CONCUR'05}, LNCS, 2005.

\bibitem[BT03]{BT03}
Ahmed Bouajjani and Tayssir Touili.
\newblock {Reachability Analysis of Process Rewrite Systems}.
\newblock In {\em FSTTCS'03}. LNCS 2914, 2003.

\bibitem[BT05]{BT05}
Ahmed Bouajjani and Tayssir Touili.
\newblock {On Computing Reachability Sets of Process Rewrite Systems}.
\newblock In {\em RTA'05}. LNCS, 2005.

\bibitem[CKS06]{CKS06}
Byron Cook, Daniel Kroening, and Natasha Sharygina.
\newblock Over-approximating boolean programs with unbounded thread creation.
\newblock {\em Formal Methods in Computer Aided Design}, 0:53--59, 2006.

\bibitem[CKS07]{CKS07}
Byron Cook, Daniel Kroening, and Natasha Sharygina.
\newblock Verification of boolean programs with unbounded thread creation.
\newblock {\em Theoretical Computer Science}, 388(1-3):227 -- 242, 2007.

\bibitem[Cou91]{Cour91}
Bruno Courcelle.
\newblock On construction obstruction sets of words.
\newblock {\em EATCS'91}, 44:178--185, June 1991.

\bibitem[EP00]{EP00}
J.~Esparza and A.~Podelski.
\newblock Efficient algorithms for pre* and post* on interprocedural parallel
  flow graphs.
\newblock In {\em POPL'00}. ACM, 2000.

\bibitem[GRB06a]{GRVB06}
P.~Ganty, J.~F. Raskin, and L.~Van Begin.
\newblock A complete abstract interpretation framework for coverability
  properties of {WSTS}.
\newblock In {\em VMCAI'06}, LNCS 3855, pages 49--64. Springer, 2006.

\bibitem[GRB06b]{GRB06}
G.~Geeraerts, J.~F. Raskin, and L.~Van Begin.
\newblock Expand, enlarge and check: New algorithms for the coverability
  problem of {WSTS}.
\newblock {\em J. Comput. Syst. Sci.}, 72(1):180--203, 2006.

\bibitem[HU79]{HU79}
John~E. Hopcroft and Jeffrey~D. Ullman.
\newblock {\em Introduction to Automata Theory, Languages and Computation}.
\newblock Addison-Wesley, 1979.

\bibitem[Lip76]{lipton}
R.~Lipton.
\newblock The reachability problem requires exponential time.
\newblock Technical Report TR 66, 1976.

\bibitem[LMP09]{TMP09}
Salvatore {La Torre}, P.~Madhusudan, and Gennaro Parlato.
\newblock Reducing context-bounded concurrent reachability to sequential
  reachability.
\newblock In {\em CAV}, volume 5643 of {\em Lecture Notes in Computer Science},
  pages 477--492. Springer, 2009.

\bibitem[LMP10]{TMP10}
Salvatore {La Torre}, P.~Madhusudan, and Gennaro Parlato.
\newblock Model-checking parameterized concurrent programs using linear
  interfaces.
\newblock In {\em CAV}, volume 6174 of {\em Lecture Notes in Computer Science},
  pages 629--644. Springer, 2010.

\bibitem[LR08]{LR08}
A.~Lal and T.~W. Reps.
\newblock Reducing concurrent analysis under a context bound to sequential
  analysis.
\newblock In {\em CAV'08}, LNCS 5123, pages 37--51. Springer, 2008.

\bibitem[LS98]{LS98}
D.~Lugiez and Ph. Schnoebelen.
\newblock The regular viewpoint on {PA}-processes.
\newblock In {\em Proc.\ 9th Int.\ Conf.\ Concurrency Theory ({CONCUR}'98),
  Nice, France, Sep.\ 1998}, volume 1466, pages 50--66. Springer, 1998.

\bibitem[LTKR08]{LTKR08}
A.~Lal, T.~Touili, N.~Kidd, and T.~W. Reps.
\newblock Interprocedural analysis of concurrent programs under a context
  bound.
\newblock In {\em TACAS'08}, LNCS 4963, pages 282--298. Springer, 2008.

\bibitem[Mo02]{Markus}
M.~Muller-olm.
\newblock Variations on constants.
\newblock Habilitation thesis, Dortmund University, 2002.

\bibitem[MQ07]{MQ07}
M.~Musuvathi and S.~Qadeer.
\newblock Iterative context bounding for systematic testing of multithreaded
  programs.
\newblock In {\em PLDI'07}, pages 446--455. ACM, 2007.

\bibitem[QR05]{QR05}
S.~Qadeer and J.~Rehof.
\newblock Context-bounded model checking of concurrent software.
\newblock In {\em TACAS'05}, LNCS 3440, pages 93--107. Springer, 2005.

\bibitem[QW04]{QW04}
S.~Qadeer and D.~Wu.
\newblock {KISS}: keep it simple and sequential.
\newblock In {\em PLDI'04}, pages 14--24. ACM, 2004.

\bibitem[Rac78]{Rackoff78}
Charles Rackoff.
\newblock The covering and boundedness problems for vector addition systems.
\newblock {\em Theor. Comput. Sci.}, 6:223--231, 1978.

\bibitem[Ram00]{Ram00}
G.~Ramalingam.
\newblock Context-sensitive synchronization-sensitive analysis is undecidable.
\newblock {\em ACM Trans. Program. Lang. Syst.}, 22(2):416--430, 2000.

\bibitem[SES08]{SES08}
D.~Suwimonteerabuth, J.~Esparza, and S.~Schwoon.
\newblock Symbolic context-bounded analysis of multithreaded java programs.
\newblock In {\em SPIN'08}, LNCS 5156, pages 270--287. Springer, 2008.

\bibitem[SS00]{SS00}
Helmut Seidl and Bernhard Steffen.
\newblock Constraint-based inter-procedural analysis of parallel programs.
\newblock In {\em 9th European Symposium on Programming (ESOP)}, 2000.

\bibitem[SSMH04]{SSMH04}
H.~Seidl, T.~Schwentick, A.~Muscholl, and P.~Habermehl.
\newblock Counting in trees for free.
\newblock In {\em ICALP'04}, LNCS 3142, pages 1136--1149. Springer, 2004.

\bibitem[Ste95]{Ste95}
Iain~A. Stewart.
\newblock Reachability in some classes of acyclic petri nets.
\newblock {\em Fundam. Inform.}, 23(1):91--100, 1995.

\bibitem[VSS05]{VSS05}
Kumar~Neeraj Verma, Helmut Seidl, and Thomas Schwentick.
\newblock On the complexity of equational {H}orn clauses.
\newblock In {\em CADE'05}, LNCS 3632, pages 337--352. Springer, 2005.

\bibitem[ZJ08]{ZJ08}
A.~Zaks and R.~Joshi.
\newblock Verifying multi-threaded {C} programs with {SPIN}.
\newblock In {\em SPIN'08}, LNCS 5156, pages 325--342. Springer, 2008.

\end{thebibliography}

\appendix

\section{The proof of Lemma  \ref{chap3.lemma3.dcfs017}}
\label{lemm.appendix.-1}

\noindent
{\bf Lemma  \ref{chap3.lemma3.dcfs017}}
Let $q \in Q$.  $q$ is $k$-bounded reachable by $\m{M}$ iff $q$ is reachable by $\m{M}'$.

\medskip

\proof
To proof  Lemma \ref{chap3.lemma3.dcfs017} we proceed as follows: First, we show that   for every reachable configuration $c$ by $\m{M}'$, the local configuration  $((w',i'),j') \in \mathit{Loc}(\m{M}')$ of any thread satisfies the condition that  the  switch number $j'$   is equal to the recored switch number  $i'$ (i.e., $i'=j'$).  This property is established by Lemma \ref{cha3.lemm.rech/dcfs}. Then, we prove  that if a state $q$ is $k$-bounded reachable by  $\m{M}$, then $q$ is  reachable  by  $\m{M}'$ (see  Lemma \ref{ref.chap3.lemma.001.dcfs}).  Finally, we show  that if a state $q$ is reachable by a  computation of $\m{M}'$, then $q$ is $k$-bounded reachable by $\m{M}$ (see Lemma \ref{chap3.lemm3.dcfs.009}).

\paragraph{\bf The switch number of any thread of $\m{M'}$ is equal to its recorded switch number:}
 In the following, we show that   for every reachable configuration $c$ by $\m{M}'$, the local configuration  $((w',i'),j') \in \mathit{Loc}(\m{M}')$ of any thread satisfies the condition that  the  switch number $j'$   is equal to the recored switch number  $i'$.

\begin{lem}
\label{cha3.lemm.rech/dcfs}
 If  $c_{\m{M}'}^{\sf init} \; \lby{\tau}{}_{\m{T}(\m{M}')}^* \, c$,  then $\mathit{Active}(c) \in \big( \{\bot\} \cup \{((w,i),i)\,|\, w \in \Gamma_{\epsilon}, i \in [0,k]\}\big)$,  $\mathit{Idle}(c)((\epsilon,l))=0$ for  all $l \in \mathbb{N}$,  and  $\mathit{Idle}(c)((w',i'),j'))=0$ for  all $w' \in \Gamma_{\epsilon}$ and  $i',j' \in \mathbb{N}$ such that $i' \neq j'$.
\end{lem}

\proof
Assume that  $c_{\m{M}'}^{\sf init} \,\bby{n}{\tau}{}_{\m{T}(\m{M}')} \, c$ for some $n \in \mathbb{N}$. We proceed by induction on $n$.

\noindent
{\bf Basis.} $n=0$. Then $c_{\m{M}'}^{\sf init}=c=(q_0,\bot,\id{\mathit{Loc}(\m{M}')}{\{((\gamma_0,0),0)\}})$. Hence, Lemma \ref{cha3.lemm.rech/dcfs} holds.


\noindent
{\bf Step.} $n>0$. Then, there is a configuration $c' \in \mathit{Conf}(\m{M}')$, $\tau' \in  (\Delta')^*$,  and $t \in \Delta'$ such that $\tau=\tau' t$, and $c_{\m{M}'}^{\sf init} \,\bby{n-1}{\tau'}{}_{\m{T}(\m{M}')}\, c' \, \by{t}{}_{\m{T}(\m{M}')}\, c$.

\noindent
Now, we apply the induction hypothesis to the run $c_{\m{M}'}^{\sf init} \,\bby{n-1}{\tau'}{}_{\m{T}(\m{M}')}\, c'$, and we obtain $\mathit{Active}(c') \in \big( \{\bot\} \cup \{((w,i),i)\,|\, w \in \Gamma_{\epsilon}, i \in [0,k]\}\big)$,  $\mathit{Idle}(c')((\epsilon,l))=0$ for  all $l \in \mathbb{N}$,  and  $\mathit{Idle}(c')((w',i'),j'))=0$ for  all $w' \in \Gamma_{\epsilon}$ and  $i',j' \in \mathbb{N}$ such that $i' \neq j'$.

\medskip
\noindent
Since $c' \, \by{t}{}_{\m{T}(\m{M}')}\, c$, then there are four  cases to study depending on the type of the  transition $t \in \Delta'$:

\begin{iteMize}{$\bullet$}
\item {\bf Case 1:} $t= \, \co q,(\gamma,r) \cf \by{}_{\m{M}'} \co q',(u,r) \cf \vtr \epsilon $ with $r \in [0,k]$. Then,  $\mathit{Active}(c')=((\gamma,r),r)$ (using the induction hypothesis). This implies that $\mathit{Active}(c)=((u,r),r)$ and $\mathit{Idle}(c)=\mathit{Idle}(c')$. Hence, all the  conditions of  Lemma \ref{cha3.lemm.rech/dcfs} are satisfied.

\item {\bf Case 2:} $t= \, \co q,(\gamma,r) \cf \by{}_{\m{M}'} \co q',(u,r) \cf \vtr (\alpha, r+1) $ with $r \in [0,k]$ and $\alpha \in \Gamma$. Then, $\mathit{Active}(c')=((\gamma,r),r)$, $\mathit{Active}(c)=((u,r),r)$, and $\mathit{Idle}(c)=\mathit{Idle}(c')+ \id{\mathit{Loc}(\m{M}')}{\{((\alpha,r+1),r+1)\}}$. This implies that all  the conditions of  Lemma \ref{cha3.lemm.rech/dcfs} are satisfied.

\item {\bf Case 3:} $t= \, \co q,(\gamma,r) \cf \swi_{\m{M}'} \co q',(u,r+1) \cf $ with $r \in [0,k]$. Then, $\mathit{Active}(c')=((\gamma,r),r)$, $\mathit{Active}(c)=\bot$, and $\mathit{Idle}(c)=\mathit{Idle}(c')+ \id{\mathit{Loc}(\m{M}')}{\{((u,r+1),r+1)\}}$. This implies that all  the conditions of  Lemma \ref{cha3.lemm.rech/dcfs} are satisfied.

\item {\bf Case 4:} $t= \, q  \swi_{\m{M}'}  q' \vtl (\gamma,r) $ with $r \in [0,k]$ and $\gamma \in \Gamma$. Then, there is $j \in \mathbb{N}$ such that $\mathit{Active}(c')=\bot$, $\mathit{Active}(c)=((\gamma,r),j)$, $\mathit{Idle}(c')((\gamma,r),j) \geq 1$, and $\mathit{Idle}(c)=\mathit{Idle}(c')- \id{\mathit{Loc}(\m{M}')}{\{((\gamma',r),j)\}}$. Since $\mathit{Idle}(c')((\gamma,r),j) \geq 1$, this implies that necessarily we have $r=j$ (from the induction hypothesis). Thus,  all  the conditions of  Lemma \ref{cha3.lemm.rech/dcfs} are satisfied.
\end{iteMize} \qed

\paragraph{\bf The Only if direction of  Lemma  \ref{chap3.lemma3.dcfs017}:}
In the following, we show that if a state $q$ is $k$-bounded reachable by  $\m{M}$, then $q$ is also reachable  by  $\m{M}'$. 

\begin{lem}
\label{ref.chap3.lemma.001.dcfs}
If $c_{\m{M}}^{\sf init}\; \lby{\tau}{}_{T_{[0,k]}(\m{M})}^* \, c$, then  there is  $\tau' \in (\Delta')^*$ such that $ c_{\m{M}'}^{\sf init}\;  \lby{\tau'}{}_{T(\m{M}')}^* \, c'$ where the configuration $c'\in \mathit{Conf}(\m{M}')$ is defined as follows:
\begin{iteMize}{$\bullet$}
\item $\mathit{State}(c')=\mathit{State}(c)$.
\item  If $\mathit{Active}(c)=\bot$, then $\mathit{Active}(c')=\bot$.

\item  If $\mathit{Active}(c)=(w,i)$ for some $w \in \Gamma_{\epsilon}$ and $i \in [0,k]$, then $\mathit{Active}(c')=((w,i),i)$.
\item     $\mathit{Idle}(c')$ is defined from $\mathit{Idle}(c)$ as
  follows:
\begin{enumerate}[\em(1)]
\item $\mathit{Idle}(c')(((w',j'),j'))=\mathit{Idle}(c)((w',j'))$  for
  all $w' \in \Gamma_{\epsilon}$ and $j' \in [0,k+1]$, and 
\item $0$ otherwise.
\end{enumerate}

\end{iteMize}
\end{lem}

\proof
First, we observe that $c_{\m{M}}^{\sf init}\; \lby{\tau}{}_{T_{[0,k]}(\m{M})}^* \, c$ implies  $\mathit{Active}(c)=\bot$ or $\mathit{Active}(c)=(w,i)$ for some $w \in \Gamma_{\epsilon}$ and $i \in [0,k]$ by definition. 
Let us assume that  $c_{\m{M}}^{\sf init}\; \bby{n}{\tau}{}_{T_{[0,k]}(\m{M})} \, c$ for some $n \in \mathbb{N}$. We proceed by induction on $n$.

\noindent
{\bf Basis.} $n=0$. This implies that  $\tau=\epsilon$ and $c_{\m{M}}^{\sf init}=c=(q_0,\bot,\id{\mathit{Loc}(\m{M})}{\{(\gamma_0,0)\}})$. Then, by taking $c'=c_{\m{M}'}^{\sf init}$ and $\tau'=\epsilon$,  all the conditions of Lemma \ref{ref.chap3.lemma.001.dcfs} are satisfied.

\medskip

\noindent
{\bf Step.} $n>0$. Then there are $c_1 \in \mathit{Conf}(\m{M})$, $\tau_1 \in \Delta^*$, and $t \in \Delta$ such that:

\begin{equation}
c_{\m{M}}^{\sf init}\; \bby{n-1}{\tau_1}{}_{T_{[0,k]}(\m{M})}\, c_1 \,\by{t}{}_{\m{T}_{[0,k]}(\m{M})} \, c
\end{equation}

\noindent
We apply the induction hypothesis to the run  $c_{\m{M}}^{\sf init}\; \bby{n-1}{\tau_1}{}_{T_{[0,k]}(\m{M})}\, c_1$, and we obtain that there are $c'_1 \in \mathit{Conf}(\m{M}')$ and  $\tau'_1 \in (\Delta')^*$ such that:

\begin{iteMize}{$\bullet$}

\item  $ c_{\m{M}'}^{\sf init}\;  \lby{\tau'_1}{}_{T(\m{M}')}^* \, c'_1$.
\item $\mathit{State}(c'_1)=\mathit{State}(c_1)$.
\item  If $\mathit{Active}(c_1)=\bot$, then $\mathit{Active}(c'_1)=\bot$.

\item  If $\mathit{Active}(c_1)=(w,i)$ for some $w \in \Gamma_{\epsilon}$ and $i \in [0,k]$, then $\mathit{Active}(c'_1)=((w,i),i)$.
\item    The function $\mathit{Idle}(c'_1)$ is defined from
  $\mathit{Idle}(c_1)$ as follows: 
\begin{enumerate}[(1)]
\item $\mathit{Idle}(c'_1)(((w',j'),j'))=\mathit{Idle}(c_1)((w',j'))$
  for all $w' \in \Gamma_{\epsilon}$ and $j' \in [0,k+1]$, and 
\item $0$ otherwise.
\end{enumerate}
\end{iteMize}

\medskip
\noindent
Since  $c_1 \,\by{t}{}_{\m{T}_{[0,k]}(\m{M})} \, c$,   one  of the following  four cases  holds:

\begin{iteMize}{$\bullet$}
\item {\bf Case 1:} $t= \co q, \gamma \cf  \by{}_{\m{M}} \co q', u \cf \vtr \epsilon$. Then, there is $i \in [0,k]$ such that $\mathit{State}(c_1)=q$, $\mathit{State}(c)=q'$, $\mathit{Active}(c_1)=(\gamma,i)$, $\mathit{Active}(c)=(u,i)$, and $\mathit{Idle}(c_1)=\mathit{Idle}(c)$. From the definition of $\m{M}'$,  $t'= \co q,(\gamma,i) \cf  \by{}_{\m{M}'} \co q', (u,i) \cf \vtr \epsilon$.  Moreover, we have $\mathit{State}(c'_1)=\mathit{State}(c_1)=q$ and  $\mathit{Active}(c_1)=((\gamma,i),i)$.  Then,  by taking  $c'=(q',((u,i),i),\mathit{Idle}(c'_1))$ and $\tau'=\tau'_1 t'$,   we can show that Lemma \ref{ref.chap3.lemma.001.dcfs} holds.

\item {\bf Case 2:} $t= \co q, \gamma\cf  \by{}_{\m{M}} \co q', u \cf \vtr \alpha$ with $\alpha \in \Gamma$. Then, there is $i \in [0,k]$ such that $\mathit{State}(c_1)=q$, $\mathit{State}(c)=q'$, $\mathit{Active}(c_1)=(\gamma,i)$, $\mathit{Active}(c)=(u,i)$, and $\mathit{Idle}(c)=\mathit{Idle}(c_1) + \id{\mathit{Loc}(\m{M})}{\{(\alpha,i+1)\}}$. From the definition of $\m{M}'$, we have $t'= \co q,(\gamma,i) \cf  \by{}_{\m{M}'} \co q' , (u,i)\cf $ $ \vtr (\alpha,i+1)$. Then, by taking  $c'=(q',((u,i),i),\mathit{Idle}(c'_1)+\id{\mathit{Loc}(\m{M}')}{\{((\alpha,i+1),i+1)\}})$ and $\tau'=\tau'_1 t'$,  we can show that Lemma \ref{ref.chap3.lemma.001.dcfs} holds.

\item {\bf Case 3:} $t=  \co q, \gamma\cf  \swi_{\m{M}} \co q', u \cf$. Then, there is $i \in [0,k]$ such that $\mathit{State}(c_1)=q$, $\mathit{State}(c)=q'$, $\mathit{Active}(c_1)=(\gamma,i)$, $\mathit{Active}(c)=\bot$, and $\mathit{Idle}(c)=\mathit{Idle}(c_1) + \id{\mathit{Loc}(\m{M})}{\{(u,i+1)\}}$. From the definition of $\m{M}'$, we have $t'= \co q,(\gamma,i) \cf  \swi_{\m{M}'} \co q', (u,i+1) \cf$. Then, by taking  $c'=(q',\bot,\mathit{Idle}(c'_1)+\id{\mathit{Loc}(\m{M}')}{\{((u,i+1),i+1)\}})$ and $\tau'=\tau'_1 t'$,  we can show that Lemma \ref{ref.chap3.lemma.001.dcfs} holds.

\item {\bf Case 4:} $t=   q  \swi_{\m{M}}  q' \vtl \gamma$ with $\gamma \in \Gamma$. Then, there is $i \in [0,k]$ such that $\mathit{State}(c_1)=q$, $\mathit{State}(c)=q'$, $\mathit{Active}(c_1)=\bot$, $\mathit{Active}(c)=(\gamma,i)$, $\mathit{Idle}(c_1)((\gamma,i)) \geq 1$, and $\mathit{Idle}(c)=\mathit{Idle}(c_1) - \id{\mathit{Loc}(\m{M})}{\{(\gamma,i)\}}$. From the definition of $\m{M}'$, we have $t'=  q \swi_{\m{M}'}  q' \vtl (\gamma,i)$. Then,  by taking  $c'=(q',((\gamma,i),i),\mathit{Idle}(c'_1)-\id{\mathit{Loc}(\m{M}')}{\{((\gamma,i),i)\}})$ and $\tau'=\tau'_1 t'$,  we can  show   that all the conditions of Lemma \ref{ref.chap3.lemma.001.dcfs} are satisfied. This is possible since $\mathit{Idle}(c'_1)(((\gamma,i),i))=\mathit{Idle}(c_1)((\gamma,i)) \geq 1$.

\end{iteMize}
\qed

\paragraph{\bf The { If direction} of Lemma \ \ref{chap3.lemma3.dcfs017} :}
In the following, we shows that if a state $q$ is reachable by a  computation of $\m{M}'$, then $q$ is $k$-bounded reachable by $\m{M}$.

\begin{lem}
\label{chap3.lemm3.dcfs.009}
If $c_{\m{M}'}^{\sf init} \;  \lby{\tau'}{}_{\m{T}(\m{M'})}^* c'$, then there is  $\tau \in \Delta^*$ such that $c_{\m{M}}^{\sf init} \; \lby{\tau}{}_{\m{T}_{[0,k]}(\m{M})}^* \; c$ where the configuration $c \in \mathit{Conf}(\m{M})$ is defined as follows:

\begin{iteMize}{$\bullet$}
\item $\mathit{State}(c)=\mathit{State}(c')$.
\item  If $\mathit{Active}(c')=\bot$, then $\mathit{Active}(c)=\bot$.
\item  If $\mathit{Active}(c')=((w,i),i)$ for some $w \in \Gamma_{\epsilon}$ and $i \in [0,k]$, then $\mathit{Active}(c)=(w,i)$.
 \item  $\mathit{Idle}(c)$ is defined from $\mathit{Idle}(c')$ as
   follows: 
\begin{enumerate}[\em(1)]
\item $\mathit{Idle}(c)((w',j'))=$ $\mathit{Idle}(c')(((w',j'),j'))$
  for all $w' \in \Gamma_{\epsilon}$ and $j' \in [0,k+1]$, and 
\item $0$ otherwise.
\end{enumerate}
\end{iteMize}
\end{lem}

\proof
First, we observe that if $c_{\m{M}'}^{\sf init} \;  \lby{\tau'}{}_{\m{T}(\m{M'})}^* c'$, then by Lemma \ref{cha3.lemm.rech/dcfs} $\mathit{Active}(c')=\bot$ or $\mathit{Active}(c')=((w,i),i)$ for some $w \in \Gamma_{\epsilon}$ and $i \in [0,k]$.
Let us assume that $c_{\m{M}'}^{\sf init} \;  \bby{n}{\tau'}{}_{\m{T}(\m{M'})} \;c'$ for some $n \in \mathbb{N}$. We proceed by induction on $n$.

\medskip
\noindent
{\bf Basis.} $n=0$. Then, $\tau'=\epsilon$ and $c_{\m{M}'}^{\sf init}=c'=(q_0,\bot,\id{\mathit{Loc}(\m{M}')}{\{((\gamma_0,0),0)\}})$. By taking $c=c_{\m{M}}^{\sf init}=(q_0,\bot,\id{\mathit{Loc}(\m{M})}{\{(\gamma_0,0)\}})$ and $\tau=\epsilon$, we can show that all the conditions of Lemma  \ref{chap3.lemm3.dcfs.009} are fulfilled.


\medskip
\noindent
{\bf Step.} $n>1$. Then, there are $\tau'_1 \in (\Delta')^*$, $t' \in \Delta'$, and $c'_1 \in \mathit{Conf}({\m{M}'})$ such that:

%

\begin{equation}
c_{\m{M}'}^{\sf init} \,\bby{n-1}{\tau'_1}{}_{\m{T}(\m{M}')} \, c'_1 \, \by{t'}{}_{\m{T}(\m{M}')} \,  c'
\end{equation}

\noindent
We apply  Lemma \ref{cha3.lemm.rech/dcfs} to  $c_{\m{M}'}^{\sf init} \,\bby{n-1}{\tau'_1}{}_{\m{T}(\m{M}')} \, c'_1$ and $c_{\m{M}'}^{\sf init} \,\bby{n}{\tau'}{}_{\m{T}(\m{M}')} \, c'$, and we obtain that:

\begin{iteMize}{$\bullet$}

\item  $\mathit{Active}(c'), \mathit{Active}(c'_1) \in \big( \{\bot\} \cup \{((w,i),i)\,|\, w \in \Gamma_{\epsilon}, i \in [0,k]\}\big)$, 

\item  $\mathit{Idle}(c'_1)((\epsilon,l))=\mathit{Idle}(c')((\epsilon,l))=0$ for  all $l \in \mathbb{N}$,  and
\item  $\mathit{Idle}(c'_1)((w',i'),j'))=\mathit{Idle}(c')((w',i'),j'))=0$ for  all $w' \in \Gamma_{\epsilon}$ and  $i' \neq j'$.
\end{iteMize}

\medskip
\noindent
We apply also the induction hypothesis to $c_{\m{M}'}^{\sf init} \,\bby{n-1}{\tau'_1}{}_{\m{T}(\m{M}')}  \; c'_1$, and we obtain that there are  $\tau_1 \in \Delta^*$ and $c_1 \in \mathit{Conf}(\m{M})$ such that:

\begin{iteMize}{$\bullet$}
\item $c_{\m{M}}^{\sf init} \; \lby{\tau_1}{}_{\m{T}_{[0,k]}(\m{M})}^* \; c_1$.
\item $\mathit{State}(c_1)=\mathit{State}(c'_1)$.
\item  If $\mathit{Active}(c'_1)=\bot$, then $\mathit{Active}(c_1)=\bot$.
\item  If $\mathit{Active}(c'_1)=((w,i),i)$ for some $w \in \Gamma_{\epsilon}$ and $i \in [0,k]$, then $\mathit{Active}(c_1)=(w,i)$.
 \item The function $\mathit{Idle}(c_1)$ is defined from
   $\mathit{Idle}(c'_1)$ as follows:
\begin{enumerate}[(1)]
\item $\mathit{Idle}(c_1)((w',j'))=$
  $\mathit{Idle}(c'_1)(((w',j'),j'))$  for all $w' \in
  \Gamma_{\epsilon}$ and $j' \in [0,k+1]$, and
\item $0$ otherwise.
\end{enumerate}
\end{iteMize}

\medskip
\noindent
On the other hand,  $ c'_1 \, \by{t'}{}_{\m{T}(\m{M}')} \,  c'
$ implies that one of the following four cases holds:

\begin{iteMize}{$\bullet$}
\item {\bf Case 1:} $t'= \, \co q, (\gamma,i) \cf \by{}_{\m{M}'} \co q', (u,i)\cf \vtr \epsilon$ with $i \in [0,k]$. Then,   $\mathit{State}(c'_1)=q$, $\mathit{State}(c')=q'$, $\mathit{Active}(c'_1)=((\gamma,i),i)$, $\mathit{Active}(c')=((u,i),i)$, and $\mathit{Idle}(c'_1)=\mathit{Idle}(c')$. We can use   the definition of $\m{M}'$ to show  that $t=\co q, \gamma\cf \by{}_{\m{M}} \co q',u\cf \vtr \epsilon$. Then, by taking $c=(q',(u,i),\mathit{Idle}(c_1))$ and $\tau=\tau_1 t$, we can  show that  Lemma \ref{chap3.lemm3.dcfs.009} holds.

\item {\bf Case 2:} $t'= \, \co q, (\gamma,i) \cf \by{}_{\m{M}'} \co
  q', (u,i)\cf \vtr (\alpha,i+1)$ with $i \in [0,k]$ and $\alpha \in
  \Gamma$. Then,   $\mathit{State}(c'_1)=q$, $\mathit{State}(c')=q'$,
  $\mathit{Active}(c'_1)=((\gamma,i),i)$,
  $\mathit{Active}(c')=((u,i),i)$, and
  $\mathit{Idle}(c')=\mathit{Idle}(c'_1)+\id{\mathit{Loc}(\m{M}')}{\{((\alpha,i+1),i+1)\}}$.
  The definition of $\m{M}'$ implies $t=\co q, \gamma\cf \by{}_{\m{M}} \co q',u\cf \vtr \alpha$. Then, by taking $c=(q',(u,i),\mathit{Idle}(c_1)+\id{\mathit{Loc}(\m{M})}{\{(\alpha,i+1)\}})$ and $\tau=\tau_1 t$, we can show that  Lemma \ref{chap3.lemm3.dcfs.009} holds.

\item {\bf Case 3:} $t'= \, \co q, (\gamma,i) \cf \swi_{\m{M}'} \co q', (u,i+1)\cf $ with $i \in [0,k]$. Then,   $\mathit{State}(c'_1)=q$, $\mathit{State}(c')=q'$, $\mathit{Active}(c'_1)=((\gamma,i),i)$, $\mathit{Active}(c')=\bot$, and $\mathit{Idle}(c')=\mathit{Idle}(c'_1)+\id{\mathit{Loc}(\m{M}')}{\{((u,i+1),i+1)\}}$. We can use   the definition of $\m{M}'$  to show that $t=\co q, \gamma\cf \swi_{\m{M}} \co q',u\cf$. Then,  by taking $c=(q',\bot,\mathit{Idle}(c_1)+\id{\mathit{Loc}(\m{M})}{\{(u,i+1)\}})$ and $\tau=\tau_1 t$, we can  show that all the conditions of Lemma \ref{chap3.lemm3.dcfs.009} are fulfilled.

\item {\bf Case 4:} $t'= \,  q \swi_{\m{M}'}  q' \vtl (\gamma,i) $ with $i \in [0,k]$. Then,   $\mathit{State}(c'_1)=q$, $\mathit{State}(c')=q'$, $\mathit{Active}(c'_1)=\bot$, $\mathit{Active}(c')=((\gamma,i),i)$, $\mathit{Idle}(c'_1)(((\gamma,i),i)) \geq 1$, and $\mathit{Idle}(c')=\mathit{Idle}(c'_1)-\id{\mathit{Loc}(\m{M}')}{\{((\gamma,i),i)\}}$. This is due to the fact that  $\mathit{Idle}(c'_1)((\gamma,i),j))=0$ for  all $j \in \mathbb{N}$ such that $i \neq j$.
We can use   the definition of $\m{M}'$ to show that $t=q \swi_{\m{M}}  q' \vtl \gamma$. Then,  by taking $c=(q',(\gamma,i),\mathit{Idle}(c_1)-\id{\mathit{Loc}(\m{M})}{\{(\gamma,i)\}})$ and $\tau=\tau_1 t$, we can easily  show that all the conditions of Lemma \ref{chap3.lemm3.dcfs.009} are fulfilled. This is possible since $\mathit{Idle}(c_1)((\gamma,i))=\mathit{Idle}(c'_1)((\gamma,i),i)$ and $\mathit{Idle}(c'_1)(((\gamma,i),i)) \geq 1$.
\end{iteMize}
\qed

As an immediate consequence of   Lemma \ref{ref.chap3.lemma.001.dcfs} and  \ref{chap3.lemm3.dcfs.009}, we obtain  that for every state $q \in Q$,  $q$ is $k$-bounded reachable by $\m{M}$ iff $q$ is reachable by $\m{M}'$.
\qed

\section{The proof of Lemma  \ref{chap3.lem.dcfs-vass-rel1}}
\label{appendix.ref.chap3.lem.dcfs-vass-rel1}

\noindent
{\bf Lemma  \ref{chap3.lem.dcfs-vass-rel1}}
Let $q \in Q$. $q $ is reachable by $\m{M}$ if and only if $(q,\bot)$ is reachable by $\m{V}$.

\medskip

\proof
To prove Lemma \ref{chap3.lem.dcfs-vass-rel1}, we proceed as follows: First, we introduce   the   function $\mu$ which defines a simulation relation between  $\m{M}$ and $\m{V}$ (see Definition \ref{chap3.def.dcfs.vass.1}). Then, we show that if  a state $q$ is reachable by $\m{M}$, then $(q,\bot)$ is also reachable by $\m{V}$ (see Lemma \ref{chap3.dcfs-vass-dir1}). Finally, we prove that if $(q,\bot)$ is reachable by $\m{V}$, then $q$ is reachable by $\m{M}$ (see Lemma \ref{chap3.dcfs-vass-dir2}).

\paragraph{\bf The simulation relation between $\m{V}$ and $\m{M}$:}
Let us define  the function $\mu$ which maps     every  configuration of $\m{M}$  to  a configuration of  $\m{V}$.

\begin{defi}
\label{chap3.def.dcfs.vass.1}
Let $\mu$ be a function from $ \mathit{Conf}(\m{M})$ to $\mathit{Conf}(\m{V})$  such that: For every $c \in \mathit{Conf}(\m{M})$, we have $\mu(c)=((q,\eta),{\bf u})$ where:

\begin{iteMize}{$\bullet$}
\item $q=\mathit{State}(c)$,
\item $\eta=\bot$ if $\mathit{Active}(c)=\bot$,
\item $\eta=w$ if $\mathit{Active}(c)=(w,i)$ for some $w \in \Gamma_{\epsilon}$ and $i \in \mathbb{N}$,
\item  ${\bf u}[i]=\sum_{j \in \mathbb{N}} \mathit{Idle}(c)((\gamma_{i-1},j))$ for all $i \in [1,m[$, and $(2)$  ${\bf u}[m]=\sum_{j \in \mathbb{N}} \mathit{Idle}(c)((\epsilon,j))$. 
 \end{iteMize}
\end{defi}

\paragraph{\bf The Only  if direction of Lemma \ref{chap3.lem.dcfs-vass-rel1}
:} In the following, we show that if  a state $q$ is reachable by $\m{M}$, then $(q,\bot)$ is also reachable by $\m{V}$.

\begin{lem}
\label{chap3.dcfs-vass-dir1}
 If   $c_{\m{M}}^{\sf init}\, \lby{\tau}{}_{\m{T}(\m{M})}^*c$, then $((p_0,\bot),{\bf u}_0) \lby{\tau}{}_{\m{T}(\m{V})}^* \mu(c)$. 
\end{lem}

\proof We use induction on the length of the run $c_{\m{M}}^{\sf
  init}\, \lby{\tau}{}_{\m{T}(\m{M})}^*c$.  For some $\ell \in
\mathbb{N}$ assume that $c_{\m{M}}^{\sf init}\,
\bby{\ell}{\tau}{}_{\m{T}(\m{M})}\, c$. We proceed by induction on
$\ell$.

\medskip
\noindent
{\bf Basis.} $\ell=0$. Then $c=c_{\m{M}}^{\sf init}$ and $\tau= \epsilon$. Moreover, we have $\mu(c_{\m{M}}^{\sf init})=((q_0,\bot),{\bf u}_0)$. This implies that  $((q_0,\bot),{\bf u}_0)\, \lby{\tau}{}_{\m{T}(\m{V})}^*\, \mu(c)$ holds.

\medskip
\noindent
{\bf Step.} $\ell>0$. Then there are  $c'\in \mathit{Conf}(\m{M})$, $\tau' \in \Sigma^*$, and $t \in \Sigma$ such that $\tau=\tau' t$ and:

\begin{equation}
c_{\m{M}}^{\sf init} \, \bby{\ell-1}{\tau'}{}_{\m{T}(\m{M})} \,c' \, \by{t}{}_{\m{T}(\m{M})}\, c
\end{equation}

\noindent
We apply the induction  hypothesis to the run $c_{\m{M}}^{\sf init} \, \bby{\ell-1}{\tau'}{}_{\m{T}(\m{M})} \,c'$, and we obtain:

\begin{equation}
\label{chap3.eq.dcfs-cov005}
((q_0,\bot),{\bf u}_0)\, \lby{\tau'}{}_{\m{T}(\m{V})}^*\, \mu(c')
\end{equation}

\noindent
Let us assume that $\mu(c')=\varsigma'$ and $\mu(c)=\varsigma$. Since $c' \, \by{t}{}_{\m{T}(\m{M})}\, c$, then one of  the following cases holds:

\begin{iteMize}{$\bullet$}
\item {\bf Case 1:} If $t= \co q, \gamma \cf \by{}_{\m{M}}  \co q', u \cf \vtr \epsilon$.  Then, there is $i \in \mathbb{N}$ such that  $\mathit{State}(c')=q$, $\mathit{State}(c)=q'$, $\mathit{Active}(c')=(\gamma,i)$, $\mathit{Active}(c)=(u,i)$, and $\mathit{Idle}(c)=\mathit{Idle}(c')$. We can use  the definition of $\mu$ to  show that  $\mathit{State}(\varsigma')=(q,\gamma)$, $\mathit{State}(\varsigma)=(q',u)$, and $\mathit{Val}(\varsigma)=\mathit{Val}(\varsigma')$. Moreover, from the definition of $\m{V}$, we have $\delta((q,\gamma),t)=((q',u),{\bf 0}^m)$.  This implies that $\varsigma' \, \by{t}{}_{\m{T}(\m{V})} \, \varsigma$, and so we obtain $((p_0,\bot),{\bf u}_0) \lby{\tau}{}_{\m{T}(\m{V})}^* \mu(c)$.

\item {\bf Case 2:} If $t=\co q,\gamma \cf \by{}_{\m{M}} \co q',u \cf \vtr \gamma_{j-1}$ for some $j \in [1,m[$. Then,  there is $i \in \mathbb{N}$ such that  $\mathit{State}(c')=q$, $\mathit{State}(c)=q'$, $\mathit{Active}(c')=(\gamma,i)$, $\mathit{Active}(c)=(u,i)$, and $\mathit{Idle}(c)=\mathit{Idle}(c')+\id{\mathit{Loc}(\m{M})}{\{(\gamma_{j-1},i+1)\}}$. We can use  the definition of $\mu$ to show  that  $\mathit{State}(\varsigma')=(q,\gamma)$, $\mathit{State}(\varsigma)=(q',u)$, and $\mathit{Val}(\varsigma)=\mathit{Val}(\varsigma')[j \sub (\mathit{Val}(\varsigma')[j]+1)]$. Moreover, from the definition of $\m{V}$, we have $\delta((q,\gamma),t)=((q',u),{\bf 0}^m[j \sub 1])$.  This implies that $\varsigma' \, \by{t}{}_{\m{T}(\m{V})} \, \varsigma$, and so we obtain $((p_0,\bot),{\bf u}_0) \lby{\tau}{}_{\m{T}(\m{V})}^* \mu(c)$.

\item {\bf Case 3:} If $t=\co q,\gamma \cf \swi_{\m{M}} \co q',\epsilon \cf$. Then,  there is $i \in \mathbb{N}$ such that  $\mathit{State}(c')=q$, $\mathit{State}(c)=q'$, $\mathit{Active}(c')=(\gamma,i)$, $\mathit{Active}(c)=\bot$, and $\mathit{Idle}(c)=\mathit{Idle}(c')+\id{\mathit{Loc}(\m{M})}{\{(\epsilon,i+1)\}}$. We can use  the definition of $\mu$ to show that  $\mathit{State}(\varsigma')=(q,\gamma)$, $\mathit{State}(\varsigma)=(q',\bot)$, and $\mathit{Val}(\varsigma)=\mathit{Val}(\varsigma')[m \sub (\mathit{Val}(\varsigma')[m]+1)]$. Moreover, from the definition of $\m{V}$, we have $\delta((q,\gamma),t)=((q',\bot),{\bf 0}^m[m \sub 1])$.  This implies that $\varsigma' \, \by{t}{}_{\m{T}(\m{V})} \, \varsigma$, and so we obtain   $((p_0,\bot),{\bf u}_0) \lby{\tau}{}_{\m{T}(\m{V})}^* \mu(c)$.

\item {\bf Case 4:} If $t=\co q,\gamma \cf \swi_{\m{M}} \co q',\gamma_{j-1} \cf$ for some $j \in [1,m[$. Then,  there is $i \in \mathbb{N}$ such that  $\mathit{State}(c')=q$, $\mathit{State}(c)=q'$, $\mathit{Active}(c')=(\gamma,i)$, $\mathit{Active}(c)=\bot$, and $\mathit{Idle}(c)=\mathit{Idle}(c')+\id{\mathit{Loc}(\m{M})}{\{(\gamma_{j-1},i+1)\}}$. We can use  the definition of $\mu$ to show  that  $\mathit{State}(\varsigma')=(q,\gamma)$, $\mathit{State}(\varsigma)=(q',\bot)$, and  $\mathit{Val}(\varsigma)=\mathit{Val}(\varsigma')[j \sub (\mathit{Val}(\varsigma')[j]+1)]$. Moreover, from the definition of $\m{V}$, we have $\delta((q,\gamma),t)=((q',\bot),{\bf 0}^m[j \sub 1])$.  This implies that $\varsigma' \, \by{t}{}_{\m{T}(\m{V})} \, \varsigma$, and so we obtain  $((p_0,\bot),{\bf u}_0) \lby{\tau}{}_{\m{T}(\m{V})}^* \mu(c)$.

\item {\bf Case 5:} If $t= q \swi_{\m{M}}  q' \vtl \gamma_{j-1}$ for some $j \in [1,m[$. Then,  there is $i \in \mathbb{N}$ such that  $\mathit{State}(c')=q$, $\mathit{State}(c)=q'$, $\mathit{Active}(c')=\bot$, $\mathit{Active}(c)=(\gamma_{j-1},i)$, ${\mathit{Idle}(c')}((\gamma_{j-1},i)) \geq 1$, and $\mathit{Idle}(c)=\mathit{Idle}(c')-\id{\mathit{Loc}(\m{M})}{\{(\gamma_{j-1},i)\}}$. We can use  the definition of $\mu$ to show   $\mathit{State}(\varsigma')=(q,\bot)$, $\mathit{State}(\varsigma)=(q',\gamma_{j-1})$, $\mathit{Val}(\varsigma')[j] \geq 1$, and $\mathit{Val}(\varsigma)=\mathit{Val}(\varsigma')[j \sub (\mathit{Val}(\varsigma')[j]-1)]$. Moreover, from the definition of $\m{V}$, we have $\delta((q,\bot),t)=((q',\gamma_{j-1}),{\bf 0}^m[j \sub -1])$.  This implies that $\varsigma' \, \by{t}{}_{\m{T}(\m{V})} \, \varsigma$, and so we obtain  $((p_0,\bot),{\bf u}_0) \lby{\tau}{}_{\m{T}(\m{V})}^* \mu(c)$.
\end{iteMize}
\qed

\paragraph{\bf The If direction of  Lemma \ref{chap3.lem.dcfs-vass-rel1}:} In the following, we prove that if $(q,\bot)$ is reachable by $\m{V}$, then $q$ is reachable by $\m{M}$.

\begin{lem}
\label{chap3.dcfs-vass-dir2}
Let $\varsigma \in (Q \times (\Gamma_{\epsilon}\cup \{\bot\})) \times \mathbb{N}^m$ and $\tau \in \Sigma^*$. If    $((q_0,\bot),{\bf u}_0)\, \lby{\tau}{}_{\m{T}(\m{V})}^*\, \varsigma$, then there is  $c \in \mathit{Conf}(\m{M})$ such that $\varsigma=\mu(c)$ and $c_{\m{M}}^{\sf init}\, \lby{\tau}{}_{\m{T}(\m{M})}^*\,c$. 
\end{lem}

\proof We use induction on the length of the run $p_0\,
\lby{\tau}_{\m{T}(\m{V})}^*\, \varsigma$.  For some $\ell \in
\mathbb{N}$ assume $((q_0,\bot),{\bf u}_0)\,
\bby{\ell}{\tau}{}_{\m{T}(\m{V})} \, \varsigma$. We proceed by
induction on $\ell$.

\noindent
{\bf Basis.} $\ell=0$. Then, $((q_0,\bot),{\bf u}_0)=\varsigma$ and $\tau=\epsilon$. By taking $c=c_{\m{M}}^{\sf init}$, we have $c_{\m{M}}^{\sf init}\, \lby{\tau}{}_{\m{T}(\m{M})}^*\,c$. Moreover, using the definition of $\mu$, we have $\mu(c_{\m{M}}^{\sf init})=\mu(c)=((q_0,\bot),{\bf u}_0)=\varsigma$.

\noindent
{\bf Step.} $\ell >0$. Then, there are $\varsigma' \in \mathit{Conf}(\m{V})$, $\tau' \in \Sigma^*$, and $t \in \Sigma$ such that $\tau=\tau' t$ and
$((q_0,\bot),{\bf u}_0)\, \bby{\ell-1}{\tau'}{}_{\m{T}(\m{V})} \, \varsigma' \,\by{t}{}_{\m{T}(\m{V})} \; \varsigma$.
Moreover, we can assume that  $\mathit{State}(\varsigma') \in Q \times (\Gamma_{\epsilon}\cup \{\bot\})$  since   $\mathit{State}(\varsigma) \in Q \times (\Gamma_{\epsilon}\cup \{\bot\})$  (see the definition of the transition function of $\m{V}$).

We apply now  the induction hypothesis to the run $((q_0,\bot),{\bf u}_0)\, \bby{\ell-1}{\tau'}{}_{\m{T}(\m{V})} \, \varsigma' $, and we obtain that there is a configuration $c'  \in \mathit{Conf}(\m{M})$ such that  $\mu(c')=\varsigma' $ and $c_{\m{M}}^{\sf init} \, \lby{\tau'}{}^*_{\m{T}(\m{M})}\, c'$.
 On the other hand,  the run $\varsigma' \, \by{t}{}_{\m{T}(\m{M})}\, \varsigma$ implies that one of the following cases holds:

\begin{iteMize}{$\bullet$}
\item {\bf Case 1:}  If $t= \co q, \gamma \cf \by{}_{\m{M}}  \co q',u \cf \vtr \epsilon$. Then, from the definition of $\m{V}$, we have $\mathit{State}(\varsigma')=(q,\gamma)$, $\mathit{State}(\varsigma)=(q',u)$, and $\mathit{Val}(\varsigma)=\mathit{Val}(\varsigma')$. Moreover, from the definition of the function $\mu$, we know that there is $i \in \mathbb{N}$ such that $\mathit{State}(c')=q$ and $\mathit{Active}(c')=(\gamma,i)$. Let $c=(q,(u,i),\mathit{Idle}(c'))$. Then,  $c' \by{t}_{\m{T}(\m{M})} c$ and $\mu(c)=\varsigma$. So, we obtain $\varsigma=\mu(c)$ and $c_{\m{M}}^{\sf init}\, \lby{\tau}{}_{\m{T}(\m{M})}^*\,c$.

\item {\bf Case 2:}  If   $t= \co q, \gamma \cf \by{}_{\m{M}}  \co q', u \cf \vtr \gamma_{j-1}$ for some $j \in [1,m[$. Then, from the definition of $\m{V}$, we have $\mathit{State}(\varsigma')=(q,\gamma)$, $\mathit{State}(\varsigma)=(q',u)$, and  $\mathit{Val}(\varsigma)=\mathit{Val}(\varsigma')[j \sub (\mathit{Val}(\varsigma')[j]+1)]$. Moreover, from the definition of the function $\mu$, we know that there is $i \in \mathbb{N}$ such that $\mathit{State}(c')=q$ and $\mathit{Active}(c')=(\gamma,i)$. Let $c=(q,(u,i),\mathit{Idle}(c')+\id{\mathit{Loc}(\m{M})}{\{(\gamma_{j-1},i+1)\}})$. Then,  $c' \by{t}_{\m{T}(\m{M})} c$ and $\mu(c)=\varsigma$. So, we obtain $\varsigma=\mu(c)$ and $c_{\m{M}}^{\sf init}\, \lby{\tau}{}_{\m{T}(\m{M})}^*\,c$.

\item {\bf Case 3:}  If   $t= \co q, \gamma \cf \swi_{\m{M}}  \co q', \epsilon \cf $. Then, from the definition of $\m{V}$, we have $\mathit{State}(\varsigma')=(q,\gamma)$, $\mathit{State}(\varsigma)=(q',\bot)$, and  $\mathit{Val}(\varsigma)=\mathit{Val}(\varsigma')[m \sub (\mathit{Val}(\varsigma')[m]+1)]$. Moreover, from the definition of the function $\mu$, we know that there is $i \in \mathbb{N}$ such that $\mathit{State}(c')=q$ and $\mathit{Active}(c')=(\gamma,i)$. Let $c=(q,\bot,\mathit{Idle}(c')+\id{\mathit{Loc}(\m{M})}{\{(\epsilon,i+1)\}})$. Then,   $c' \by{t}_{\m{T}(\m{M})} c$ and $\mu(c)=\varsigma$. So, we obtain $\varsigma=\mu(c)$ and $c_{\m{M}}^{\sf init}\, \lby{\tau}{}_{\m{T}(\m{M})}^*\,c$.

\item {\bf Case 4:}  If   $t= \co q, \gamma \cf \swi_{\m{M}}  \co q', \gamma_{j-1} \cf $ for some $j \in [1,m[$. Then, from the definition of $\m{V}$, we have $\mathit{State}(\varsigma')=(q,\gamma)$, $\mathit{State}(\varsigma)=(q',\bot)$, and  $\mathit{Val}(\varsigma)=\mathit{Val}(\varsigma')[j \sub (\mathit{Val}(\varsigma')[j]+1)]$. Moreover, from the definition of the function $\mu$, we know that there is $i \in \mathbb{N}$ such that $\mathit{State}(c')=q$ and $\mathit{Active}(c')=(\gamma,i)$. Let $c=(q,\bot,\mathit{Idle}(c')+\id{\mathit{Loc}(\m{M})}{\{(\gamma_{j-1},i+1)\}})$. Then,   $c' \by{t}_{\m{T}(\m{M})} c$ and $\mu(c)=\varsigma$. So, we obtain $\varsigma=\mu(c)$ and $c_{\m{M}}^{\sf init}\, \lby{\tau}{}_{\m{T}(\m{M})}^*\,c$.

\item {\bf Case 5:}  If   $t=  q \swi_{\m{M}}   q'  \vtl \gamma_{j-1} $ for some $j \in [1,m[$. Then, from the definition of $\m{V}$, we have $\mathit{State}(\varsigma')=(q,\bot)$, $\mathit{State}(\varsigma)=(q',\gamma_{j-1})$, $\mathit{Val}(\varsigma')[j]\geq 1$, and  $\mathit{Val}(\varsigma)=\mathit{Val}(\varsigma')[j \sub (\mathit{Val}(\varsigma')[j]-1)]$. Moreover, from the definition of the function $\mu$, we know that there is $i \in \mathbb{N}$ such that $\mathit{State}(c')=q$, $\mathit{Active}(c')=\bot$, and $\mathit{Idle}(c')((\gamma_{j-1},i)) \geq 1$. Let $c=(q,(\gamma_{j-1},i),\mathit{Idle}(c')-\id{\mathit{Loc}(\m{M})}{\{(\gamma_{j-1},i)\}})$. Then,   $c$ is well defined,  $c' \by{t}_{\m{T}(\m{M})} c$, and $\mu(c)=\varsigma$. So, we obtain $\varsigma=\mu(c)$ and $c_{\m{M}}^{\sf init}\, \lby{\tau}{}_{\m{T}(\m{M})}^*\,c$.\qed
\end{iteMize}

Hence Lemma   \ref{chap3.lem.dcfs-vass-rel1} is an immediate consequence of Lemma \ref{chap3.dcfs-vass-dir1} and Lemma \ref{chap3.dcfs-vass-dir2}.\qed

\section{The proof of Lemma \ref{chap3.lem.vass.dcfs.dir1}}
\label{appendix.ref.chap3.lem.vass.dcfs.dir1}

\noindent
{\bf Lemma \ref{chap3.lem.vass.dcfs.dir1}}
 Let $q \in Q$. $q$ is reachable by $\m{V}$ if and only if $q$ is $2$-bounded reachable by $\m{M}$.

\medskip

\proof
 To prove Lemma \ref{chap3.lem.vass.dcfs.dir1}, we proceed as follows: First, we prove that  if   $q\in Q$ is reachable by  $\m{V}$, then $q$ is $2$-bounded reachable by   $\m{M}$ (see Lemma  \ref{chap3.lemm.cov.dcfs.vssa-dir1} ). Then,  we show that  if $q \in Q$ is $2$-bounded reachable by   $\m{M}$, then  $q$  is reachable by  $\m{V}$ (see Lemma \ref{chap3.lemm.cov.dcfs.vssa-dir2}).  
 
 \subsubsection*{The If direction of Lemma \ref{chap3.lem.vass.dcfs.dir1}:}
 In the following, we show  if   $q\in Q$ is reachable by  $\m{V}$, then $q$ is $2$-bounded reachable by   $\m{M}$.

\begin{lem}
\label{chap3.lemm.cov.dcfs.vssa-dir1}
If $(q_0,{\bf 0}^n) \,\lby{\sigma}{}_{\m{T}(\m{V})}^* \, (q,{\bf u})$, then for every $m \in \mathbb{N}$, there are $\tau \in \Delta^*$ and  $\mathit{Val} \in [\mathit{Loc}(\m{M}) \ra \mathbb{N}]$ such that: $(1)$ $\mathit{Val}((\gamma_i,2))={\bf u}[i]$ for all $i \in [1,n]$, $(2)$ $\mathit{Val}((\gamma'_0,1))=m$, and $(3)$ $c_{\m{M}}^{\sf init}\, \lby{\tau}{}_{\m{T}_{[0,2]}(\m{M})}^* \, (q,\bot,\mathit{Val})$.
\end{lem}

\proof We use induction on the length of the run $(q_0,{\bf 0}^n)
\,\lby{\sigma}{}_{\m{T}(\m{V})}^* \, (q,{\bf u})$.  For some $\ell \in
\mathbb{N}$ assume that $(q_0,{\bf 0}^n)
\,\bby{\ell}{\sigma}{}_{\m{T}(\m{V})} \, (q,{\bf u})$. We proceed by
induction on $\ell$.

\noindent
{\bf Basis.} $\ell=0$. Then $\sigma=\epsilon$, $q=q_0$ and ${\bf u}={\bf 0}^n$. It is easy to observe that for every $m \in \mathbb{N}$, $\m{T}_{[0,2]}(\m{M})$, from the initial configuration $c_{\m{M}}^{\sf init}$, can apply $m$-times  the transition $t_0=\co p_0,\gamma_0 \cf \by{}_{\m{M}} \co p_0,\gamma_0 \cf \vtr \gamma'_0$ followed by  the transition $t'_0=\co p_0,\gamma_0 \cf \swi_{\m{M}} \co q_0,\epsilon \cf $ to reach the configuration $(q_0,\bot,\mathit{Val})$ (i.e., $c_{\m{M}}^{\sf init} \, \lby{t_0^{m} \cdot t'_0}{}_{\m{T}_{[0,2]}(\m{M})}^* \, (q_0, \bot,\mathit{Val})$) with  $\mathit{Val}((\gamma'_0,1))=m$ and $\mathit{Val}((\gamma_i,2))={\bf u}[i]$ for all $i \in [1,n]$.

\noindent
{\bf Step.} $\ell>0$. Then,  there are $q' \in Q$, ${\bf u'}\in \mathbb{N}^n$,  $\sigma' \in \Sigma^* $, and $a \in \Sigma$ such that $\sigma=\sigma'a$ and:

\begin{equation}
(q_0,{\bf 0}^n) \,\bby{\ell-1}{\sigma'}{}_{\m{T}(\m{V})} \, (q',{\bf u'}) \, \by{a}{}_{\m{T}(\m{V})} \, (q,{\bf u}) 
\end{equation}

\noindent
We apply the induction hypothesis to $(q_0,{\bf 0}^n) \,\bby{\ell-1}{\sigma'}{}_{\m{T}(\m{V})} \, (q',{\bf u'}) $, and we obtain that:

\begin{align}
\label{chap3.eq.cov-vass-dcfs.1}
\forall m' \in \mathbb{N},\,\, \exists \tau' \in \Delta^*\,\, \text{and} \,\, \exists \mathit{Val}' \in [{\mathit{Loc}(\m{M})}  \ra \mathit{N}]& \,\,\text{s.t.:}\nonumber\\
&  c_{\m{M}}^{\sf init}\, \lby{\tau'}{}_{\m{T}_{[0,2]}(\m{M})}^* \, (q',\bot,\mathit{Val}') \nonumber\\
&  \mathit{Val}'((\gamma'_0,1))=m' \nonumber\\
& \mathit{Val}'((\gamma_i,2))={\bf u'}[i],\forall i \in [1,n] 
\end{align}

 \noindent
 Moreover, we have  $(q',{\bf u'}) \, \by{a}{}_{\m{T}(\m{V})} \, (q,{\bf u}) $. This implies that $\delta(q',a)=(q,{\bf u}-{\bf u}')$, and one of the following cases holds:
 
\medskip
 
 \begin{iteMize}{$\bullet$}
 \item {\bf Case 1:} If ${\bf u'}={\bf u}$,  then  $t= q' \swi_{\m{M}}\, q' \vtl \gamma'_0$,  $t'=\co q',\gamma'_0\cf \, \by{}_{\m{M}}\, \co q, \gamma'_0 \cf \vtr \epsilon$, and $t''=\co q,\gamma_0 \cf \swi_{\m{M}} \co q,\epsilon\cf$.  This implies that for every $\mathit{Val} \in [{\mathit{Loc}(\m{M})} \ra \mathbb{N}]$, $\m{T}_{[0,2]}(\m{M})$ can move from the configuration $(q',\bot,\mathit{Val}+\id{\mathit{Loc}(\m{M})}{\{(\gamma'_0,1)\}})$ to the configuration  $(q,\bot,\mathit{Val}+\id{\mathit{Loc}(\m{M})}{\{(\epsilon,2)\}})$. Now, we can use  Equations \ref{chap3.eq.cov-vass-dcfs.1},  to show that:

 \begin{align}
\forall m \in \mathbb{N},\,\, \exists \tau' \in \Delta^*\,\, \text{and} \,\, \exists \mathit{Val}\in  [{\mathit{Loc}(\m{M})} \ra \mathbb{N}]& \,\,\text{s.t.:}\nonumber\\
&  c_{\m{M}}^{\sf init}\, \lby{\tau'tt't''}{}_{\m{T}_{[0,2]}(\m{M})}^* \, (q,\bot,\mathit{Val}) \nonumber\\
&  \mathit{Val}((\gamma'_0,1))=m \nonumber\\
& \mathit{Val}((\gamma_i,2))={\bf u}[i], \forall i \in [1,n] 
\end{align}

 \item {\bf Case 2:}  If ${\bf u}={\bf u'}[j \sub ({\bf u'}[j]+1)]$ for some $j \in [1,n]$,  then we have that $t= q' \swi_{\m{M}}\, q' \vtl \gamma'_0$,  $t'=\co q',\gamma'_0\cf \, \by{}_{\m{M}}\, \co q, \gamma'_0 \cf \vtr \gamma_j$, and $t''=\co q,\gamma_0 \cf \swi_{\m{M}} \co q,\epsilon\cf$. This implies that for every $\mathit{Val}'' \in [{\mathit{Loc}(\m{M})} \ra \mathbb{N}]$, $\m{T}_{[0,2]}(\m{M})$ can move from the configuration $(q',\bot,\mathit{Val}''+ \id{\mathit{Loc}(\m{M})}{\{(\gamma'_0,1)\}})$ to the configuration  $(q,\bot,\mathit{Val})$ with $\mathit{Val}=\mathit{Val}''+\id{\mathit{Loc}(\m{M})}{\{(\epsilon,2),(\gamma_j,2)\}}$. 
 Now, we can use  Equations \ref{chap3.eq.cov-vass-dcfs.1},  to show that:

 \begin{align}
\forall m \in \mathbb{N},\,\, \exists \tau' \in \Delta^*\,\, \text{and} \,\, \exists \mathit{Val}\in  [{\mathit{Loc}(\m{M})} \ra \mathbb{N}] & \,\,\text{s.t.:}\nonumber\\
&  c_{\m{M}}^{\sf init}\, \lby{\tau'tt't''}{}_{\m{T}_{[0,2]}(\m{M})}^* \, (q,\bot,\mathit{Val}) \nonumber\\
&  \mathit{Val}((\gamma'_0,1))=m \nonumber\\
& \mathit{Val}((\gamma_i,2))={\bf u}[i], \forall i \in [1,n] 
\end{align}

 \item {\bf Case 3:}  If ${\bf u}={\bf u'}[j \sub ({\bf u'}[j]-1)]$  and ${\bf u'}[j]\geq 1$ for some $j \in [1,n]$, then $t= q' \swi_{\m{M}}\, q \vtl \gamma_j$, and  $t'=\co q,\gamma_j \cf \swi_{\m{M}} \co q,\epsilon\cf$.  This implies that for every $\mathit{Val}'' \in  [{\mathit{Conf}_{\sf loc}(\m{M})} \ra \mathbb{N}]$ such that $\mathit{Val}''((\gamma_j,2))\geq 1$, $\m{T}_{[0,2]}(\m{M})$ can move from the configuration $(q',\bot,\mathit{Val}'')$ to the configuration  $(q,\bot,\mathit{Val})$ with $\mathit{Val}=\mathit{Val}''+\id{\mathit{Loc}(\m{M})}{\{(\epsilon,3)\}}-\id{\mathit{Loc}(\m{M})}{\{((\gamma_j,2)\}}$. 
%
%
Now, we can use  Equations \ref{chap3.eq.cov-vass-dcfs.1},  to show that:
 
 \begin{align}
\forall m \in \mathbb{N},\,\, \exists \tau' \in \Delta^*\,\, \text{and} \,\, \exists \mathit{Val}\in  [{\mathit{Loc}(\m{M})} \ra \mathbb{N}] & \,\,\text{s.t.:}\nonumber\\
&  c_{\m{M}}^{\sf init}\, \lby{\tau'tt'}{}_{\m{T}_{[0,2]}(\m{M})}^* \, (q,\bot,\mathit{Val}) \nonumber\\
&  \mathit{Val}((\gamma'_0,1))=m \nonumber\\
& \mathit{Val}((\gamma_i,2))={\bf u'}[i]\,\, \,,\, \forall i \in [1,n] 
\end{align}

\noindent
(This is possible since $\mathit{Val'}((\gamma_j,2))={\bf u'}[j] \geq 1$.)  \end{iteMize}
\qed

 \paragraph{\bf The Only if direction of Lemma \ref{chap3.lem.vass.dcfs.dir1}:}
 In the following,  we show that  if $q \in Q$ is $2$-bounded reachable by   $\m{M}$, then  $q$  is reachable by  $\m{V}$.

\begin{lem}
\label{chap3.lemm.cov.dcfs.vssa-dir2}
  If $c_{\m{M}}^{\sf init} \, \lby{\tau}{}_{\m{T}_{[0,2]}(\m{M})}^* \, c$ for some $\tau \in \Delta^*$ and  $c \in \mathit{Conf}(\m{M})$  such that $\mathit{State}(c) \in Q$, then the following conditions are satisfied:
 
 \begin{enumerate}[\em(1)]
 \item  $\mathit{Active}(c) \in (\{\bot\} \cup (\{(\gamma'_0,1)\}) \cup ((\Gamma \setminus \{\gamma_0,\gamma'_0\}) \times \{2\}))$,
 \item $\mathit{Idle}(c)((\gamma_0,i))=0$ for all $i \in \mathbb{N}$,
 \item $\mathit{Idle}(c)((\gamma_j,i))=0$ for all $j \in [1,n]$ and $i \neq 2$,
 \item $\mathit{Idle}(c)((\gamma'_0,i))=0$ for all $i \neq 1$, and
 \item  there  is  $\sigma \in \Sigma^*$ with $(q_0,{\bf 0}^n)
   \,\lby{\sigma}{}_{\m{T}(\m{V})}^* \, (q,{\bf u})$, where $q=\mathit{State}(c)$ and  ${\bf u}[j]=\mathit{Idle}(c)((\gamma_j,2))$ for all $j \in [1,n]$.
\end{enumerate}  
\end{lem}  
  
\proof
Again, we use induction.
Let us  assume that $c_{\m{M}}^{\sf init} \, \lby{\tau}{}_{\m{T}_{[0,2]}(\m{M})}^* \, c$ for some $\tau \in \Delta^*$ and  $c \in \mathit{Conf}(\m{M})$  such that $\mathit{State}(c) \in Q$. Then, from the definition of $\m{M}$, there are $\tau_1, \tau_2 \in \Delta^*$ and  $m \in \mathbb{N}$ such that $\tau=\tau_1 \tau_2$ and $c_{\m{M}}^{\sf init} \, \lby{\tau_1}{}_{\m{T}_{[0,2]}(\m{M})}^* \, (q_0,\bot,\mathit{Val}_0) \, \lby{\tau_2}{}_{\m{T}_{[0,2]}(\m{M})}^*  c$
with $\mathit{Val}_0((\gamma'_0,1))=m$ and $\mathit{Val}_0((\alpha,j))=0$ for all $(\alpha,j) \in \Gamma \times \mathbb{N}$ such that $(\alpha,j) \neq (\gamma'_0,1)$.

\noindent
Since $(q_0,\bot,\mathit{Val}_0) \, \lby{\tau_2}{}_{\m{T}_{[0,2]}(\m{M})}^*  c$, then there is $\ell \in \mathbb{N}$ such that $(q_0,\bot,\mathit{Val}_0)\bby{\ell}{\tau_2}{}_{\m{T}_{[0,2]}(\m{M})}\,   c$. To prove Lemma  \ref{chap3.lemm.cov.dcfs.vssa-dir2}, we proceed by induction on $\ell$.

\noindent
{\bf Basis.} $\ell=0$. Then, $\tau_2=\epsilon$, $c=(q_0,\bot,\mathit{Val}_0)$. By taking $\sigma=\epsilon$ and ${\bf u}={\bf 0}^n$, we have $(q_0,{\bf 0}^n) \,\lby{\sigma}{}_{\m{T}(\m{V})}^* \, (q,{\bf u})$ with  $q=\mathit{State}(c)$ and ${\bf u}[i]=\mathit{Idle}(c)((\gamma_i,2))=0$ for all $i \in [1,n]$. Moreover, we have  $\mathit{Idle}(c)((\alpha,j))=0$ for all $(\alpha,j) \in \Gamma \times \mathbb{N}$ such that $(\alpha,j) \neq (\gamma'_0,1)$.

\noindent
{\bf Step.} $\ell>0$. Then, there are $\tau' \in \Delta^*$, $t \in \Delta$, and $c' \in \mathit{Conf}(\m{M})$ such that $\tau_2=\tau't$ and 
$(q_0,\bot,\mathit{Val}_0)\bby{\ell-1}{\tau'}{}_{\m{T}_{[0,2]}(\m{M})}  \,c' \,\by{t}{}_{\m{T}_{[0,2]}(\m{M})} \, c$.

\medskip 
From the definition of $\m{M}$, it is not hard  to prove that $\mathit{State}(c') \in Q$.

\noindent
We apply the induction hypothesis to $(q_0,\bot,\mathit{Val}_0)\bby{\ell-1}{\tau'}{}_{\m{T}_{[0,2]}(\m{M})}  \,c' $, and we obtain that the following conditions are satisfied:

 \begin{iteMize}{$\bullet$} 
 \item  $\mathit{Active}(c') \in (\{\bot\} \cup (\{(\gamma'_0,1)\}) \cup ((\Gamma \setminus \{\gamma_0,\gamma'_0\}) \times \{2\}))$,
 \item $\mathit{Idle}(c')((\gamma_0,i))=0$ for all $i \in \mathbb{N}$,
 \item $\mathit{Idle}(c')((\gamma_j,i))=0$ for all $j \in [1,n]$ and $i \neq 2$,
 \item $\mathit{Idle}(c')((\gamma'_0,i))=0$ for all $i \neq 1$, and
 \item  there  is  $\sigma' \in \Sigma^*$ such that: 
 
 \begin{equation}
 \label{eq.appendix.chap3.lemm.cov.bsr.dir20}
  (q_0,{\bf 0}^n) \,\lby{\sigma'}{}_{\m{T}(\m{V})}^* \, (q',{\bf u}')
  \end{equation}

\noindent
where $q'=\mathit{State}(c')$ and  ${\bf u}'[j]=\mathit{Idle}(c')((\gamma_j,2))$ for all $j \in [1,n]$.
\end{iteMize}

\noindent
Moreover, we have $ c' \by{t}{}_{\m{T}_{[0,2]}(\m{M})} \, c$. This implies that one of the following cases holds:

\begin{iteMize}{$\bullet$}
\item {\bf Case 1:} $t= \co q', \gamma'_0 \cf \by{}_{\m{M}} \co q,\gamma'_0 \cf \vtr \epsilon$. Then, $\mathit{State}(c)=q$, $\mathit{State}(c')=q'$, $\mathit{Active}(c)=\mathit{Active}(c')=(\gamma'_0,1)$, and $\mathit{Idle}(c)=\mathit{Idle}(c')$. This implies that the conditions 1-4 of Lemma \ref{chap3.lemm.cov.dcfs.vssa-dir2} are satisfied. 
Moreover, from the definition of $\m{M}$, there is $a \in \Sigma$  such that $\delta(q',a)=(q,{\bf 0}^n)$ since we have $t= \co q', \gamma'_0 \cf \by{}_{\m{M}} \co q,\gamma'_0 \cf \vtr \epsilon$. This implies that $\m{T}(\m{V})$ can reach the configuration $(q,{\bf u}')$ from the configuration $(q',{\bf u}')$. I.e., we have the following computation of $\m{T}({V})$:

\begin{equation}
\label{eq.appendix.chap3.lemm.cov.bsr.dir201}
(q',{\bf u}') \by{a}{}_{\m{T}(\m{V})} \, (q, {\bf u})
\end{equation}

with ${\bf u}={\bf u'}$.

\noindent
Putting together Equation \ref{eq.appendix.chap3.lemm.cov.bsr.dir20} and \ref{eq.appendix.chap3.lemm.cov.bsr.dir201}, we obtain that:

\begin{equation}
(q_0,{\bf 0}^n) \,\lby{\sigma'a}{}_{\m{T}(\m{V})}^* \, (q,{\bf u})
\end{equation}

\noindent
Now, we can use the fact that ${\bf u}={\bf u'}$, $\mathit{Idle}(c)=\mathit{Idle}(c')$, and ${\bf u}'[j]=\mathit{Idle}(c')((\gamma_j,2))$ for all $j \in [1,n]$, to show that ${\bf u}[j]=\mathit{Idle}(c)((\gamma_j,2))$ for all $j \in [1,n]$. 


\item {\bf Case 2:} $t= \co q', \gamma'_0 \cf \by{}_{\m{M}} \co q,\gamma'_0 \cf \vtr \gamma_{k}$ for some $k \in [1,n]$. Then, $\mathit{State}(c)=q$, $\mathit{State}(c')=q'$, $\mathit{Active}(c)=\mathit{Active}(c')=(\gamma'_0,1)$, and $\mathit{Idle}(c)=\mathit{Idle}(c')+\id{\mathit{Loc}(\m{M})}{\{(\gamma_k,2)\}}$. This implies that the conditions 1-4 of Lemma \ref{chap3.lemm.cov.dcfs.vssa-dir2} are satisfied. 
Moreover, from the definition of $\m{M}$, there is $a \in \Sigma$  such that $\delta(q',a)=(q,{\bf 0}^n[k \sub 1])$ since we have $t= \co q', \gamma'_0 \cf \by{}_{\m{M}} \co q,\gamma'_0 \cf \vtr \gamma_k$.
 This implies that $\m{T}(\m{V})$ can reach the configuration $(q,{\bf u})$ from the configuration $(q',{\bf u}')$ with ${\bf u}={\bf u'}[k \sub ({\bf u'}[k]+1)]$. I.e., we have the following computation of $\m{T}({V})$:

\begin{equation}
\label{eq.appendix.chap3.lemm.cov.bsr.dir302}
(q',{\bf u}') \by{a}{}_{\m{T}(\m{V})} \, (q, {\bf u})
\end{equation}

\noindent
Putting together Equation \ref{eq.appendix.chap3.lemm.cov.bsr.dir20} and \ref{eq.appendix.chap3.lemm.cov.bsr.dir302}, we obtain that:

\begin{equation}
(q_0,{\bf 0}^n) \,\lby{\sigma'a}{}_{\m{T}(\m{V})}^* \, (q,{\bf u})
\end{equation}

\noindent
Now, we can use the fact that ${\bf u}={\bf u'}[k \sub ({\bf u'}[k]+1)]$, $\mathit{Idle}(c)=\mathit{Idle}(c')+\id{\mathit{Loc}(\m{M})}{\{(\gamma_k,2)\}}$, and ${\bf u}'[j]=\mathit{Idle}(c')((\gamma_j,2))$ for all $j \in [1,n]$, to show that ${\bf u}[j]=\mathit{Idle}(c)((\gamma_j,2))$ for all $j \in [1,n]$.


\item {\bf Case 3:} $t= \co q', \gamma_k \cf \swi{}_{\m{M}} \co q',\epsilon \cf $ for some $k \in [1,n]$. Then, $\mathit{State}(c)=\mathit{State}(c')=q'$, $\mathit{Active}(c)=\bot$, $\mathit{Active}(c')=(\gamma_k,2)$, and $\mathit{Idle}(c)=\mathit{Idle}(c')+\id{\mathit{Loc}(\m{M})}{\{(\epsilon,3)\}}$. This implies that the conditions 1-4 of Lemma \ref{chap3.lemm.cov.dcfs.vssa-dir2} are satisfied. 
Moreover, by taking $q=q'$, $\sigma=\sigma'$, and ${\bf u}={\bf u'}$, we have:

\begin{equation}
(q_0,{\bf 0}^n) \,\lby{\sigma}{}_{\m{T}(\m{V})}^* \, (q,{\bf u})
\end{equation}

\noindent
Now, we can use the fact  that ${\bf u}={\bf u'}$, $\mathit{Idle}(c)=\mathit{Idle}(c')+\id{\mathit{Loc}(\m{M})}{\{(\epsilon,3)\}}$, and ${\bf u}'[j]=\mathit{Idle}(c')((\gamma_j,2))$ for all $j \in [1,n]$, to show that ${\bf u}[j]=\mathit{Idle}(c)((\gamma_j,2))$ for all $j \in [1,n]$.


\item {\bf Case 4:} $t= q' \swi_{\m{M}} q \vtl \gamma_{k}$ for some $k \in [1,n]$. Then, $\mathit{State}(c)=q$, $\mathit{State}(c')=q'$, $\mathit{Active}(c)=(\gamma_k,2)$, $\mathit{Active}(c')=\bot$, $\mathit{Idle}(c')((\gamma_k,2)) \geq 1$, and $\mathit{Idle}(c)=\mathit{Idle}(c')-\id{\mathit{Loc}(\m{M})}{\{(\gamma_k,2)\}}$. This implies that the conditions 1-4 of Lemma \ref{chap3.lemm.cov.dcfs.vssa-dir2} are satisfied. 
Moreover, from the definition of $\m{M}$, there is $a \in \Sigma$  such that $\delta(q',a)=(q,{\bf 0}^n[k \sub -1])$.
 This implies that $\m{T}(\m{V})$ can reach the configuration $(q,{\bf u})$ from the configuration $(q',{\bf u}')$ with ${\bf u}={\bf u'}[k \sub ({\bf u'}[k]-1)]$ since $\mathit{Idle}(c')((\gamma_k,2))={\bf u'}[k]\geq 1$. I.e., we have the following computation of $\m{T}({V})$:

\begin{equation}
\label{eq.appendix.chap3.lemm.cov.bsr.dir303}
(q',{\bf u}') \by{a}{}_{\m{T}(\m{V})} \, (q, {\bf u})
\end{equation}

\noindent
Putting together Equation \ref{eq.appendix.chap3.lemm.cov.bsr.dir20} and \ref{eq.appendix.chap3.lemm.cov.bsr.dir303}, we obtain that:

\begin{equation}
(q_0,{\bf 0}^n) \,\lby{\sigma'a}{}_{\m{T}(\m{V})}^* \, (q,{\bf u})
\end{equation}

\noindent
Now, we can use that ${\bf u}={\bf u'}[k \sub ({\bf u'}[k]-1)]$, $\mathit{Idle}(c)=\mathit{Idle}(c')-\id{\mathit{Loc}(\m{M})}{\{(\gamma_k,2)\}}$, and ${\bf u}'[j]=\mathit{Idle}(c')((\gamma_j,2))$ for all $j \in [1,n]$, to show that ${\bf u}[j]=\mathit{Idle}(c)((\gamma_j,2))$ for all $j \in [1,n]$. 


\item {\bf Case 5:} $t= \co q', \gamma'_0 \cf \swi_{\m{M}} \co q', \epsilon \cf$. Then, $\mathit{State}(c)=\mathit{State}(c')=q'$, $\mathit{Active}(c')=(\gamma'_0,1)$, $\mathit{Active}(c)=\bot$, and $\mathit{Idle}(c)=\mathit{Idle}(c')+\id{\mathit{Loc}(\m{M})}{\{(\epsilon,2)\}}$. This implies that the conditions 1-4 of Lemma \ref{chap3.lemm.cov.dcfs.vssa-dir2} are satisfied. By taking ${\bf u}={\bf u}'$, $\sigma=\sigma'$, and $q=q'$, we have that 
 $\m{T}(\m{V})$ can reach the configuration $(q,{\bf u})$ from the configuration $(q',{\bf u}')$. I.e., we have the following computation of $\m{T}({V})$:

\begin{equation}
(q_0,{\bf 0}^n) \,\lby{\sigma}{}_{\m{T}(\m{V})}^* \, (q,{\bf u})
\end{equation}

\noindent
Now, we can use that ${\bf u}={\bf u'}$, $\mathit{Idle}(c)=\mathit{Idle}(c')+\id{\mathit{Loc}(\m{M})}{\{(\epsilon,2)\}}$, and ${\bf u}'[j]=\mathit{Idle}(c')((\gamma_j,2))$ for all $j \in [1,n]$, to show that ${\bf u}[j]=\mathit{Idle}(c)((\gamma_j,2))$ for all $j \in [1,n]$.

\item {\bf Case 6:} $t= q' \swi_{\m{M}} q' \vtl \gamma'_{0}$. Then, $\mathit{State}(c)=\mathit{State}(c')=q'$, $\mathit{Active}(c)=(\gamma'_0,1)$, $\mathit{Active}(c')=\bot$, $\mathit{Idle}(c')((\gamma'_0,1)) \geq 1$, and $\mathit{Idle}(c)=\mathit{Idle}(c')-\id{\mathit{Loc}(\m{M})}{\{(\gamma'_0,1)\}}$. This implies that the conditions 1-4 of Lemma \ref{chap3.lemm.cov.dcfs.vssa-dir2} are satisfied. By taking ${\bf u}={\bf u}'$, $\sigma=\sigma'$, and $q=q'$, we have that 
 $\m{T}(\m{V})$ can reach the configuration $(q,{\bf u})$ from the configuration $(q',{\bf u}')$. I.e., we have the following computation of $\m{T}({V})$:

\begin{equation}
(q_0,{\bf 0}^n) \,\lby{\sigma}{}_{\m{T}(\m{V})}^* \, (q,{\bf u})
\end{equation}

\noindent
Now, we can use that ${\bf u}={\bf u'}$, $\mathit{Idle}(c)=\mathit{Idle}(c')-\id{\mathit{Loc}(\m{M})}{\{(\gamma'_0,1)\}}$, and ${\bf u}'[j]=\mathit{Idle}(c')((\gamma_j,2))$ for all $j \in [1,n]$, to show that ${\bf u}[j]=\mathit{Idle}(c)((\gamma_j,2))$ for all $j \in [1,n]$.\qed
\end{iteMize} 

\noindent Hence, Lemma \ref{chap3.lem.vass.dcfs.dir1} is an immediate consequence of Lemma \ref{chap3.lemm.cov.dcfs.vssa-dir1} and Lemma \ref{chap3.lemm.cov.dcfs.vssa-dir2}.  \qed

\section{The proof of Lemma \ref{ssrproblem-pda-relation}}
\label{appendix.ssrproblem-pda-relation}

\noindent
{\bf Lemma \ref{ssrproblem-pda-relation}.}
A state $q \in F$  is $k$-stratified reachable by $\m{M}$ if and only if there is  $\sigma_i \in \Sigma_{i}^*$ for all $i \in [0,k]$ such that:

\begin{iteMize}{$\bullet$}
\item $\sigma_0 \sigma_1 \cdots \sigma_k \in \mathit{Traces}_{\m{T}(\m{P})}(\{(q_0,\bot)\},F \times \{\bot\})$, and
\item $|\sigma_i|_{(\gamma,i,\vtl)} \leq |\sigma_{i-1}|_{(\gamma,i,\vtr)}$ for all $\gamma \in \Gamma$ and $i \in [0,k]$ where $\sigma_{-1}=(\gamma_0,0,\vtr)$.
\end{iteMize}

\medskip

\proof
To prove Lemma  \ref{ssrproblem-pda-relation}, we  need first  to  define a simulation relation $\mu$ between $\m{M}$ and $\m{P}$ that maps any configuration of $\m{M}$ to a configuration of $\m{P}$.

\begin{defi}
Let $\mu$ be a  function from $Q \times \Gamma_{\m{P}} \times \mathit{Loc}(\m{M})$ to $\mathit{Conf}(\m{P})$ such that for every $c \in Q \times \Gamma_{\m{P}} \times \mathit{Loc}(\m{M})$, $\mu(c)=(\mathit{State}(c), \mathit{Active}(c))$.
\end{defi}

\subsubsection*{The Only if direction of Lemma \ref{ssrproblem-pda-relation}:}  In the following, we show that   if  there is a state $q \in F$ such that $q$  is $k$-stratified reachable by $\m{M}$, then there is  $\sigma_i \in \Sigma_{i}^*$ for all $i \in [0,k]$ such that $\sigma_0 \sigma_1 \cdots \sigma_k \in \mathit{Traces}_{\m{T}(\m{P})}(\{(q_0,\bot)\},F \times \{\bot\})$, and  $|\sigma_i|_{(\gamma,i,\vtl)} \leq |\sigma_{i-1}|_{(\gamma,i,\vtr)}$ for all $\gamma \in \Gamma$ and $i \in [0,k]$ where $\sigma_{-1}=(\gamma_0,0,\vtr)$.

  To this aim, we first prove that if there is a run $\,c\,\lby{\tau_i}{}_{\m{T}_{\{i\}}(\m{M})}^* \, c'$    ( where $\m{M}$  executes only threads with switch number   $i \in [0,k]$), then there  is  a run   $\mu(c)\lby{\sigma_i}{}_{\m{T}(\m{P})}^*\, \mu(c') $ of $\m{P}$ such that: $(1)$ $\sigma_i \in \Sigma_i^*$, $(2)$ the number of occurrences of  $(\gamma,i,\vtl)$ in $\sigma_i$ is equal to the number of  activated  threads by $\m{M}$  with local configuration $(\gamma,i)$,  and $(3)$ the number of created/added threads   by $\m{M}$   with local configuration $(\gamma,i+1)$  is equal to the number of occurrence of  $(\gamma,i+1,\vtr)$ in $\sigma_i$.

%

\begin{lem}
\label{lem.presbu.dir1.sec4}
For every $i \in [0,k]$, and $c,c' \in  \big(Q \times \Gamma_{\m{P}} \times \mathit{Loc}(\m{M})\big)$, if there is  $\tau_i \in \Delta^*$  such that  $c\,\lby{\tau_i}{}_{\m{T}_{\{i\}}(\m{M})}^* \, c'$, then  there is  $\sigma_i \in \Sigma_i^*$ such that: 
\begin{enumerate}[\em(1)]
\item $\mu(c)\lby{\sigma_i}{}_{\m{T}(\m{P})}^*\, \mu(c') $.
 \item $\mathit{Idle}(c)((\alpha,i)) \geq |\sigma_i|_{(\alpha,i,\vtl)}$  for all $\alpha \in \Gamma_\epsilon$.

 \item  $\mathit{Idle}(c')((\alpha,i))=\mathit{Idle}(c)((\alpha,i))- |\sigma_i|_{(\alpha,i,\vtl)}$ and $\mathit{Idle}(c')((\alpha,i+1))=\mathit{Idle}(c)((\alpha,i+1))+ |\sigma_i|_{(\alpha,i+1,\vtr)}$ for all $\alpha \in \Gamma_{\epsilon}$.
 
 \item  $\mathit{Idle}(c')((\alpha,j))=\mathit{Idle}(c)((\alpha,j))$   for all $(\alpha,j) \in \Gamma_{\epsilon} \times [0,k+1]$ such that $j \notin \{i,i+1\}$.
 \end{enumerate}

\end{lem}

\proof
Assume that $c\,\bby{\ell}{\tau_i}{}_{\m{T}_{\{i\}}(\m{M})} \, c'$ for some $\ell \in \mathbb{N}$. We proceed by induction on $\ell$.

\noindent
{\bf Basis. } $\ell=0$. Then, $\tau_i=\epsilon$ and $c=c'$. By taking  $\sigma_i=\epsilon$, we have all the  conditions of Lemma \ref{lem.presbu.dir1.sec4} are fulfilled.

\noindent
{\bf Step.} $\ell>0$. Then there are  $\tau'_i \in \Delta^*$, $t \in \Delta$, $c'' \in \big(Q \times \Gamma_{\m{P}} \times \mathit{Loc}(\m{M})\big)$ such that:
\begin{equation}
c\,\bby{\ell-1}{\tau'_i}{}_{\m{T}_{\{i\}}(\m{M})} \, c'' \,\by{t}{}_{\m{T}_{\{i\}}(\m{M})} \, c'
\end{equation}
We apply the induction hypothesis to the run
$c\,\bby{\ell-1}{\tau'_i}{}_{\m{T}_{\{i\}}(\m{M})} \, c'' $, and we
obtain that there is  $\sigma'_i \in \Sigma_i^* $  such that:

\begin{iteMize}{$\bullet$}
\item $\mu(c)\lby{\sigma'_i}{}_{\m{T}(\m{P})}^*\, \mu(c'') $.
 \item $\mathit{Idle}(c)((\alpha,i)) \geq |\sigma'_i|_{(\alpha,i,\vtl)}$  for all $\alpha \in \Gamma_\epsilon$.

 \item  $\mathit{Idle}(c'')((\alpha,i))=\mathit{Idle}(c)((\alpha,i))- |\sigma'_i|_{(\alpha,i,\vtl)}$ and $\mathit{Idle}(c'')((\alpha,i+1))=\mathit{Idle}(c)((\alpha,i+1))+ |\sigma'_i|_{(\alpha,i+1,\vtr)}$ for all $\alpha \in \Gamma_{\epsilon}$.
 
 \item  $\mathit{Idle}(c'')((\alpha,j))=\mathit{Idle}(c)((\alpha,j))$   for all $(\alpha,j) \in \Gamma_{\epsilon} \times [0,k+1]$ such that $j \notin \{i,i+1\}$.
\end{iteMize}

 Since we have $ c'' \,\by{t}{}_{\m{T}_{\{i\}}(\m{M})} \, c'$, one of the followings cases holds:

\begin{iteMize}{$\bullet$}
\item {\bf Case 1:} $t=\co q'', \gamma \cf \by{}_{\m{M}} \co q', u \cf \vtr \epsilon $. Then $\mathit{State}(c'')=q''$, $\mathit{State}(c')=q'$, $\mathit{Active}(c'')=(\gamma,i)$, $\mathit{Active}(c')=(u,i)$, and $\mathit{Idle}(c')=\mathit{Idle}(c'')$.
 Moreover, from the definition of $\m{P}$, we have  $\co q'',(\gamma,i) \cf \by{(\epsilon,i,-)}_{\m{P}} \co q',(u,i)\cf$. This implies that $\m{T}(\m{P})$ has the following run:
\begin{equation}
\label{appendix.eq.H.presbu.dir1.01}
(q'',(\gamma,i)) \by{(\epsilon,i,-)}{}_{\m{T}(\m{P})} (q',(u,i))
\end{equation}
We can use the definition of $\mu$ to show that $\mu(c'')=(q'',(\gamma,i))$ and $\mu(c')=(q',(u,i))$.  Then, let  $\sigma_i=\sigma'_i (\epsilon,i,-)$.
Putting  together the equation $\mu(c)\lby{\sigma'_i}{}_{\m{T}(\m{P})}^*\, \mu(c'') $  and Equation \ref{appendix.eq.H.presbu.dir1.01}, we obtain that:
\begin{equation}
\mu(c)\lby{\sigma_i}{}_{\m{T}(\m{P})}^*\, \mu(c') 
\end{equation}
 Then, we can use  the fact that   $\mathit{Idle}(c')=\mathit{Idle}(c'')$  and  $\sigma_i=\sigma'_i (\epsilon,i,-)$ to show that all  the conditions   of Lemma \ref{lem.presbu.dir1.sec4} are fulfilled.

\item {\bf Case 2:} $t=\co q'', \gamma \cf \by{}_{\m{M}} \co q', u \cf \vtr \alpha $ with $\alpha \in \Gamma$. Then $\mathit{State}(c'')=q''$, $\mathit{State}(c')=q'$, $\mathit{Active}(c'')=(\gamma,i)$, $\mathit{Active}{(c')}=(u,i)$, and $\mathit{Idle}(c')=\mathit{Idle}(c'')+\id{\mathit{Loc}(\m{M})}{\{(\alpha,i+1)\}}$.
 Moreover, from the definition of $\m{P}$, we have  $\co q'',(\gamma,i) \cf \by{(\alpha,i+1,\vtr)}_{\m{P}} \co q',(u,i)\cf$. This implies that $\m{T}(\m{P})$ has the following run:
\begin{equation}
\label{appendix.eq.H.presbu.dir1.0206}
(q'',(\gamma,i)) \by{(\alpha,i+1,\vtr)}{}_{\m{T}(\m{P})} (q',(u,i))
\end{equation}
We can use  the definition of the function $\mu$  to show that $\mu(c'')=(q'',(\gamma,i))$ and $\mu(c')=(q',(u,i))$.  Then,   let  $\sigma_i=\sigma'_i (\alpha,i+1,\vtr)$.

Putting  together the equation $\mu(c)\lby{\sigma'_i}{}_{\m{T}(\m{P})}^*\, \mu(c'') $  and Equation \ref{appendix.eq.H.presbu.dir1.0206}, we obtain that:
\begin{equation}
\mu(c)\lby{\sigma_i}{}_{\m{T}(\m{P})}^*\, \mu(c') 
\end{equation}
Then, we can use  the fact that   $\mathit{Idle}(c')=\mathit{Idle}(c'')+\id{\mathit{Loc}(\m{M})}{\{(\alpha,i+1)\}}$  and $\sigma_i=\sigma'_i (\alpha,i+1,\vtr)$ to show that all  the conditions   of Lemma \ref{lem.presbu.dir1.sec4} are fulfilled.

\item {\bf Case 3:} $t=\co q'', \gamma \cf \swi_{\m{M}} \co q', u \cf$. Then $\mathit{State}(c'')=q''$, $\mathit{State}(c')=q'$, $\mathit{Active}(c'')=(\gamma,i)$, $\mathit{Active}({c'})=\bot$, and $\mathit{Idle}(c')=\mathit{Idle}(c'')+\id{\mathit{Loc}(\m{M})}{\{(u,i+1)\}}$.
 Moreover, from the definition of $\m{P}$, we have  $\co q'',(\gamma,i) \cf \by{(u,i+1,\vtr)}_{\m{P}} \co q',\bot\cf$. This implies that $\m{T}(\m{P})$ has the following run:
\begin{equation}
\label{appendix.eq.H.presbu.dir1.0205}
(q'',(\gamma,i)) \by{(u,i+1,\vtr)}{}_{\m{T}(\m{P})} (q',\bot)
\end{equation}
Using the definition of the function $\mu$, it is easy to observe that $\mu(c'')=(q'',(\gamma,i))$ and $\mu(c')=(q',\bot)$.  Then,   let  $\sigma_i=\sigma'_i (u,i+1,\vtr)$. 
Putting  together the equation $\mu(c)\lby{\sigma'_i}{}_{\m{T}(\m{P})}^*\, \mu(c'') $  and Equation \ref{appendix.eq.H.presbu.dir1.0205}, we obtain that:
\begin{equation}
\mu(c)\lby{\sigma_i}{}_{\m{T}(\m{P})}^*\, \mu(c') 
\end{equation}
 Then, we can use the fact  that   $\mathit{Idle}(c')=\mathit{Idle}(c'')+\id{\mathit{Loc}(\m{M})}{\{(u,i+1)\}}$  and $\sigma_i=\sigma'_i (u,i+1,\vtr)$ to show that all  the conditions   of Lemma \ref{lem.presbu.dir1.sec4} are fulfilled.

\item {\bf Case 4:} $t= q''\swi_{\m{M}}  q' \vtl \gamma$. Then $\mathit{State}(c'')=q''$, $\mathit{State}(c')=q'$, $\mathit{Active}(c'')=\bot$, $\mathit{Active}({c'})=(\gamma,i)$, $\mathit{Idle}(c'')((\gamma,i)) \geq 1$, and $\mathit{Idle}(c')=\mathit{Idle}(c'')-\id{\mathit{Loc}(\m{M})}{\{(\gamma,i)\}}$.
 Moreover, from the definition of $\m{P}$, we have  $\co q'',\bot \cf \by{(\gamma,i,\vtl)}_{\m{P}} \co q',(\gamma,i)\cf$. This implies that $\m{T}(\m{P})$ has the following run:
\begin{equation}
\label{appendix.eq.H.presbu.dir1.0204}
(q'',\bot) \by{(\gamma,i,\vtl)}{}_{\m{T}(\m{P})} (q',(\gamma,i))
\end{equation}
Using the definition of the function $\mu$, it is easy to observe that $\mu(c'')=(q'',\bot)$ and $\mu(c')=(q',(\gamma,i))$.  Then,   let  $\sigma_i=\sigma'_i (\gamma,i,\vtl)$. 
Putting  together the equation $\mu(c)\lby{\sigma_i}{}_{\m{T}(\m{P})}^*\, \mu(c'') $  and Equation \ref{appendix.eq.H.presbu.dir1.0204}, we obtain that:
\begin{equation}
\mu(c)\lby{\sigma_i}{}_{\m{T}(\m{P})}^*\, \mu(c') 
\end{equation}
 Then, we can use the fact  that   $\mathit{Idle}(c')=\mathit{Idle}(c'')-\id{\mathit{Loc}(\m{M})}{\{(\gamma,i)\}}$  and $\sigma_i=\sigma'_i (\gamma,i,\vtl)$ to show that all  the conditions   of Lemma \ref{lem.presbu.dir1.sec4} are fulfilled.\qed
\end{iteMize}

\noindent Now, we are ready to prove the only if direction of Lemma  \ref{ssrproblem-pda-relation}.

\begin{lem}
\label{if-direction001}
If $q \in F$ is $k$-stratified reachable by $\m{M}$, then there is  $\sigma_i \in \Sigma_{i}^*$ for all $i \in [0,k]$ such that $\sigma_0 \sigma_1 \cdots \sigma_k \in \mathit{Traces}_{\m{T}(\m{P})}(\{(q_0,\bot)\},F \times \{\bot\})$, and  $|\sigma_i|_{(\gamma,i,\vtl)} \leq |\sigma_{i-1}|_{(\gamma,i,\vtr)}$ for all $\gamma \in \Gamma$ and $i \in [0,k]$ where $\sigma_{-1}=(\gamma_0,0,\vtr)$.  
\end{lem}

\proof
Let us assume  that there is a state $q \in F$ such that  $q$  is $k$-stratified reachability  by $\m{M}$.  Then, there are $\tau_0,\tau_1,\ldots, \tau_k \in \Delta^*$, and  $c_1, \ldots, c_{k+1} \in \mathit{Conf}(\m{M})$ such that $\mathit{State}(c_{k+1})=q$, $\mathit{Active}(c_{k+1})=\bot$, and we have: 
\begin{equation}
c^{\sf init}_{\m{M}}  \;\lby{\tau_0}{}_{\m{T}_{\{0\}}(\m{M})}^* \, c_1 \;\lby{\tau_1}{}_{\m{T}_{\{1\}}(\m{M})}^*\,  \cdots \;\lby{\tau_{k-1}}{}_{\m{T}_{\{k-1\}}(\m{M})}^* \, c_k \;\lby{\tau_k}{}_{\m{T}_{\{k\}}(\m{M})}^* \, c_{k+1}
\end{equation}
Notice that all the configurations  $c_{\m{M}}^{\sf init}, c_1,c_2, \ldots, c_{k+1}$ are in $\big(Q \times \Gamma_{\m{P}} \times \mathit{Loc}(\m{M})\big)$ by definition. Then, we can use Lemma \ref{lem.presbu.dir1.sec4},  to show that there are $\sigma_i  \in \Sigma_i^*$ for all $i \in [0,k]$ such that:
\begin{equation}
\mu(c^{\sf init}_{\m{M}})  \;\lby{\sigma_0}{}_{\m{T}(\m{P})}^* \, \mu(c_{1}) \;\lby{\sigma_1}{}_{\m{T}(\m{P})}^*\,  \cdots \;\lby{\sigma_{k-1}}{}_{\m{T}(\m{P})}^* \, \mu(c_{k}) \;\lby{\sigma_k}{}_{\m{T}(\m{P})}^* \, \mu(c_{k+1})
\end{equation}
Then,  we obtain   $\sigma_0 \sigma_1 \cdots \sigma_k \in  \mathit{Traces}_{\m{T}(\m{P})}(\{(q_0,\bot)\},F \times \{\bot\})$ since $\mu(c^{\sf init}_{\m{M}}) =(q_0,\bot)$,  $\mathit{State}(c_{k+1})=q \in F$, and $\mathit{Active}(c_{k+1})=\bot$ (i.e., $\mu(c_{k+1}) \in F \times \{\bot\}$).  Moreover, we can use  the fact that  $\mathit{Idle}(c_{\m{M}}^{\sf init})=\id{\mathit{Loc}(\m{M})}{\{(\gamma_0,0)\}}$ and  the second condition   of Lemma \ref{lem.presbu.dir1.sec4}, to prove that for every $\alpha \in \Gamma_{\epsilon}$, we have that $|\sigma_0|_{(\alpha,i,\vtl)} \leq |\sigma_{-1}|_{(\alpha,i,\vtr)}$ with $\sigma_{-1}=(\gamma_0,0,\vtr)$.

Conversely, we can use the conditions (3) and (4)  of Lemma \ref{lem.presbu.dir1.sec4}, to prove that for every $j \in [1,k+1]$ and every $\alpha \in \Gamma_{\epsilon}$, we have
$\mathit{Idle}(c_{j})((\alpha,j))=|\sigma_{j-1}|_{(\alpha,j,\vtr)} \,\,\text{and}\,\, \mathit{Idle}(c_{j})((\alpha,j+1))=0$. So,   for every $j \in [1,k]$ and $\alpha \in \Gamma_{\epsilon}$, we can use  the fact that $ \mathit{Idle}(c_{j})((\alpha,j))=|\sigma_{j-1}|_{(\alpha,j,\vtr)}$ and the second condition   of Lemma \ref{lem.presbu.dir1.sec4} (i.e.,  $\mathit{Idle}(c_j)((\alpha,j)) \geq |\sigma_j|_{(\alpha,j,\vtl)}$), to prove that  $|\sigma_{j-1}|_{(\alpha,j,\vtr)} \geq |\sigma_j|_{(\alpha,j,\vtl)}$.  
\qed

\subsubsection*{The  if direction of Lemma \ref{ssrproblem-pda-relation}:}  In the following, we prove that    if there is $\sigma_i \in \Sigma_{i}^*$ for all $i \in [0,k]$ such that $\sigma_0 \sigma_1 \cdots \sigma_k \in \mathit{Traces}_{\m{T}(\m{P})}(\{(q_0,\bot)\},F \times \{\bot\})$, and  $|\sigma_i|_{(\gamma,i,\vtl)} \leq |\sigma_{i-1}|_{(\gamma,i,\vtr)}$ for all $\gamma \in \Gamma$ and $i \in [0,k]$ where $\sigma_{-1}=(\gamma_0,0,\vtr)$, then   there is a state $q \in F$ such $q$ is   $k$-stratified reachable by $\m{M}$.

To this aim, we first  show that for every configuration $c \in \big(Q \times \Gamma_{\m{P}} \times \mathit{Loc}(\m{M})\big)$ and $\sigma_i \in \Sigma_i^*$, if there is a run $ \mu(c) \, \lby{\sigma_i}{}_{\m{T}(\m{P})}^*\, \varsigma'$ for some $\varsigma' \in \mathit{Conf}(\m{P})$ and  the number of occurrences of $(\gamma,i,\vtl)$ in $\sigma_i$ is less than the number of pending thread in $c$ with local configuration $(\gamma,i)$, then there are $c' \in \mathit{Conf}(\m{M})$ and a run     $c\,\lby{\tau_i}{}_{\m{T}_{\{i\}}(\m{M})}^* \, c'$ such that: $(1)$ $\mu(c')=\varsigma'$, $(2)$ the number of occurrences of  $(\gamma,i,\vtl)$ in $\sigma_i$ is equal to the number of  activated  threads by $\m{M}$  with local configuration $(\gamma,i)$, and $(3)$ the number of created/added threads   by $\m{M}$   with local configuration $(\gamma,i+1)$  is equal to the number of occurrence of  $(\gamma,i+1,\vtr)$ in $\sigma_i$.

\begin{lem}
\label{lem.presbu.dir2.sec4}
For every $i \in [0,k]$, $\varsigma, \varsigma' \in \mathit{Conf}(\m{P})$, $c \in \big(Q \times \Gamma_{\m{P}} \times \mathit{Loc}(\m{M})\big)$,  and $\sigma_i \in \Sigma_i^*$,  if $\varsigma \, \lby{\sigma_i}{}_{\m{T}(\m{P})}^*\, \varsigma'$, $\mu(c)=\varsigma$,  and  $\mathit{Idle}(c)((\alpha,i)) \geq |\sigma_i|_{(\alpha,i,\vtl)}$ for all $\alpha \in \Gamma_{\epsilon}$, then  there are $\tau_i \in \Delta^*$ and  $c' \in \big(Q \times \Gamma_{\m{P}} \times \mathit{Loc}(\m{M})\big)$ such that:

\begin{enumerate}[\em(1)]

\item $\mu(c')=\varsigma'$.
 \item   $c\,\lby{\tau_i}{}_{\m{T}_{\{i\}}(\m{M})}^* \, c'$. 
 
 \item  $\mathit{Idle}(c')((\alpha,i))=\mathit{Idle}(c)((\alpha,i))- |\sigma_i|_{(\alpha,i,\vtl)}$ and $\mathit{Idle}(c')((\alpha,i+1))=\mathit{Idle}(c)((\alpha,i+1))+ |\sigma_i|_{(\alpha,i+1,\vtr)} $ for all $\alpha \in \Gamma_{\epsilon}$.
 
 \item  $\mathit{Idle}(c')((\alpha,j))=\mathit{Idle}(c)((\alpha,j))$   for all $(\alpha,j) \in \Gamma_{\epsilon} \times [0,k+1]$ such that $j \notin \{i,i+1\}$.
 \end{enumerate}

\end{lem}

\proof
Assume that 
 $\varsigma \, \bby{\ell}{\sigma_i}{}_{\m{T}(\m{P})}\, \varsigma'$ for some $\ell \in \mathbb{N}$,   $\mu(c)=\varsigma$,  and  $\mathit{Idle}(c)((\alpha,i)) \geq |\sigma_i|_{(\alpha,i,\vtl)}$ for all $\alpha \in \Gamma_{\epsilon}$.   The proof is done  by induction on $\ell$.

\medskip
\noindent
{\bf Basis.}  $\ell=0$. Then, $\varsigma=\varsigma'$, $\sigma_i=\epsilon$. By taking $c'=c$ and $\tau_i=\epsilon$, all the conditions  of Lemma \ref{lem.presbu.dir2.sec4} are fulfilled.

\medskip
\noindent
{\bf Step.} $\ell>0$. Then, there are $\sigma'_i \in \Sigma_i^*$, $a \in \Sigma_i $, and $\varsigma'' \in \mathit{Conf}(\m{P})$ such that $\sigma_i=\sigma'_i a$ and $\varsigma \bby{\ell}{\sigma'_i}{}_{\m{T}(\m{P})}\, \varsigma'' \, \by{a}{}_{\m{T}(\m{P})}\, \varsigma'$.

We apply the induction hypothesis to $\varsigma \bby{\ell}{\sigma'_i}{}_{\m{T}(\m{P})}\, \varsigma''$ since $\mathit{Idle}(c)((\alpha,i)) \geq |\sigma_i|_{(\alpha,i,\vtl)} \geq |\sigma'_i|_{(\alpha,i,\vtl)}$ for all $\alpha \in \Gamma_{\epsilon}$, and we obtain that there are $\tau'_i \in \Delta^*$ and  $c'' \in \big(Q \times \Gamma_{\m{P}} \times \mathit{Loc}(\m{M})\big)$ such that:

\begin{iteMize}{$\bullet$}
\item $\mu(c'')=\varsigma''$.
 \item   $c\,\lby{\tau'_i}{}_{\m{T}_{\{i\}}(\m{M})}^* \, c''$. 
 
 \item  $\mathit{Idle}(c'')((\alpha,i))=\mathit{Idle}(c)((\alpha,i))- |\sigma'_i|_{(\alpha,i,\vtl)}$ and $\mathit{Idle}(c'')((\alpha,i+1))=\mathit{Idle}(c)((\alpha,i+1))+ |\sigma'_i|_{(\alpha,i+1,\vtr)} $ for all $\alpha \in \Gamma_{\epsilon}$.
 
 \item  $\mathit{Idle}(c'')((\alpha,j))=\mathit{Idle}(c)((\alpha,j))$   for all $(\alpha,j) \in \Gamma_{\epsilon} \times [0,k+1]$ such that $j \notin \{i,i+1\}$.
 \end{iteMize}

Since we have $\varsigma'' \, \by{a}{}_{\m{T}(\m{P})}\, \varsigma'$, one of the following cases holds:

\begin{iteMize}{$\bullet$}
\item {\bf Case 1:} $a=(\epsilon,i,-)$. Then, there are  $q,q ' \in Q$, $\gamma \in \Gamma$, and $u \in \Gamma_{\epsilon}$ such that $\co q, (\gamma,i) \cf \by{(\epsilon,i,-)}_{\m{P}} \co q', (u,i) \cf$, $\varsigma''=(q,(\gamma,i))$, and $\varsigma'=(q',(u,i))$. Since $\mu(c'')=\varsigma''$, we have $\mathit{State}(c'')=q$ and $\mathit{Active}(c'')=(\gamma,i)$. Moreover, we have $t=\co q,\gamma \cf \by{}_{\m{M}} \co q',u\cf \vtr \epsilon$.   By taking $c'= (q',(u,i),\mathit{Idle(c'')})$, we have $c'' \by{t}_{\m{T}_i(\m{M})} \, c'$.

Now, we can use  the definition of the function $\mu$ to show that  $\mu(c')=(q',(u,i))=\varsigma'$.   Let  $\tau_i=\tau'_i t$. We can put together the equation $c\lby{\tau'_i}{}_{\m{T}_i(\m{M})}^*\, c'' $  and  the equation $c'' \by{t}_{\m{T}_i(\m{M})} \, c'$ to obtain the following run of $\m{T}(\m{P})$:
\begin{equation}
c\, \lby{\tau_i}{}_{\m{T}(\m{P})}^*\, c' 
\end{equation}
 Then, we can use  the fact that   $\mathit{Idle}(c')=\mathit{Idle}(c'')$  and $\sigma_i=\sigma'_i(\epsilon,i,-)$ to show that  the conditions 4-5  of Lemma \ref{lem.presbu.dir2.sec4} are fulfilled.

\item {\bf Case 2:} $a=(\alpha,i+1,\vtr)$ and $\varsigma' \notin (Q \times  (\{\bot\})$. Then, there are  $q,q ' \in Q$, $\gamma \in \Gamma$, and $u \in \Gamma_{\epsilon}$ such that $\co q, (\gamma,i) \cf \by{a}_{\m{P}} \co q', (u,i) \cf$, $\varsigma''=(q,(\gamma,i))$, and $\varsigma'=(q',(u,i))$. Since $\mu(c'')=\varsigma''$, we have  $\mathit{State}(c'')=q$ and  $\mathit{Active}(c'')=(\gamma,i)$. Moreover,  we have $t=\co q,\gamma \cf \by{}_{\m{M}} \co q',u\cf \vtr \alpha$.   By taking $c'= (q',(u,i),\mathit{Idle(c'')}+\id{\mathit{Loc}(\m{M})}{\{(\alpha,i+1)\}})$, we have $c'' \by{t}_{\m{T}_i(\m{M})} \, c'$.

Then, we can use  the definition of the function $\mu$ to show that  $\mu(c')=(q',(u,i))=\varsigma'$.   Let  $\tau_i=\tau'_i t$.  We can put  together the equation $c\lby{\tau'_i}{}_{\m{T}_i(\m{M})}^*\, c'' $  and  the equation $c'' \by{t}_{\m{T}_i(\m{M})} \, c'$ to obtain the following run of $\m{T}(\m{P})$:
\begin{equation}
c\,\lby{\tau_i}{}_{\m{T}(\m{P})}^*\, c' 
\end{equation}
 Then, we can use the fact  that   $\mathit{Idle}(c')=\mathit{Idle}(c'')+\id{\mathit{Loc}(\m{M})}{\{(\alpha,i+1)\}}$  and $\sigma_i=\sigma'_i (\alpha,i+1,\vtr)$ to show that  the conditions 4-5  of Lemma \ref{lem.presbu.dir2.sec4} are fulfilled.

\item {\bf Case 3:} $a=(u,i+1,\vtr)$ and $\varsigma' \in (Q \times  (\{\bot\})$. Then, there are  $q,q ' \in Q$ and  $\gamma \in \Gamma$ such that $\co q, (\gamma,i) \cf \by{a}_{\m{P}} \co q', \bot \cf$, $\varsigma''=(q,(\gamma,i))$, and $\varsigma'=(q',\bot)$. Since $\mu(c'')=\varsigma''$, we have $\mathit{State}(c'')=q$ and $\mathit{Active}(c'')=(\gamma,i)$. Moreover,  we have $t=\co q,\gamma \cf \swi_{\m{M}} \co q',u\cf$.   By taking $c'= (q',\bot,\mathit{Idle(c'')}+\id{\mathit{Loc}(\m{M})}{\{(u,i+1)\}})$, we have $c'' \by{t}_{\m{T}_i(\m{M})} \, c'$.

Now, we can use  the definition of the function $\mu$ to show that  $\mu(c')=(q',\bot)=\varsigma'$.  
Let  $\tau_i=\tau'_i t$.  We can put  together the equation $c\lby{\tau'_i}{}_{\m{T}_i(\m{M})}^*\, c'' $  and the equation $c'' \by{t}_{\m{T}_i(\m{M})} \, c'$ to obtain the following run of $\m{T}(\m{P})$:
\begin{equation}
c\,\lby{\tau_i}{}_{\m{T}(\m{P})}^*\, c' 
\end{equation}
 Then, we can use the fact that   $\mathit{Idle}(c')=\mathit{Idle}(c'')+\id{\mathit{Loc}(\m{M})}{\{(u,i+1)\}}$  and $\sigma_i=\sigma'_i (u,i+1,\vtr)$ to show that  the conditions 4-5  of Lemma \ref{lem.presbu.dir2.sec4} are fulfilled.
 \medskip

\item {\bf Case 4:} $a=(\gamma,i,\vtl)$. Then, there are  $q,q ' \in Q$  such that $\co q, \bot \cf \by{a}_{\m{P}} \co q', (\gamma,i) \cf$, $\varsigma''=(q,\bot)$, and $\varsigma'=(q',(\gamma,i))$. Since $\mu(c'')=\varsigma''$,  $\mathit{State}(c'')=q$ and $\mathit{Active}(c'')=\bot$. In addiction, we have  $\mathit{Idle}(c'')((\gamma,i)) \geq 1$ since $\mathit{idle}(c'')((\gamma,i))=\mathit{Idle}(c)-|\sigma'_i|_{(\gamma,i,\vtr)}$, $\mathit{Idle}(c) \geq |\sigma_i|_{(\gamma,i,\vtl)}$, and $|\sigma_i|_{(\gamma,i,\vtl)}=|\sigma'_i|_{(\gamma,i,\vtl)}+1$.
Moreover,  we have $t=q  \swi_{\m{M}}  q' \vtl \gamma$.    Then, by taking $c'= (q',(\gamma,i),\mathit{Idle(c'')}-\id{\mathit{Loc}(\m{M})}{\{(\gamma,i)\}})$, we have $c'' \by{t}_{\m{T}_i(\m{M})} \, c'$.

Now, we can use  the definition of the function $\mu$ to show  that  $\mu(c')=(q',\bot)=\varsigma'$.  Let  $\tau_i=\tau'_i t$. Then, we can put  together the equation $c\lby{\tau'_i}{}_{\m{T}_i(\m{M})}^*\, c'' $  and the equation $c'' \by{t}_{\m{T}_i(\m{M})} \, c'$ to obtain the following run of $\m{T}(\m{P})$:
\begin{equation}
c\,\lby{\tau_i}{}_{\m{T}(\m{P})}^*\, c' 
\end{equation}
 Then, we can use the fact  that   $\mathit{Idle}(c')=\mathit{Idle}(c'')-\id{\mathit{Loc}(\m{M})}{\{(\gamma,i)\}}$  and $\sigma_i=\sigma'_i (\gamma,i,\vtl)$ to show that  the conditions 4-5  of Lemma \ref{lem.presbu.dir2.sec4} are fulfilled.\qed
\end{iteMize} 

\noindent Now, we are ready to prove the if direction of Lemma \ref{lem.presbu.dir2.sec4}:

\begin{lem}
\label{if-direction000}
There is $\sigma_i \in \Sigma_{i}^*$ for all $i \in [0,k]$ such that
$\sigma_0 \sigma_1 \cdots \sigma_k \in
\mathit{Traces}_{\m{T}(\m{P})}(\{(q_0,\bot)\},$ $F \times \{\bot\})$, and
$|\sigma_i|_{(\gamma,i,\vtl)} \leq |\sigma_{i-1}|_{(\gamma,i,\vtr)}$
for all $\gamma \in \Gamma$ and $i \in [0,k]$ with
$\sigma_{-1}=(\gamma_0,0,\vtr)$, then there is a state $q \in F$ such
that $q$ is $k$-stratified reachable by $\m{M}$.
\end{lem}

\proof
Let us assume now that there are   $\sigma_i \in \Sigma_{i}^*$ for all $i \in [0,k]$ such that $\sigma_0 \sigma_1 \cdots \sigma_k \in \mathit{Traces}_{\m{T}(\m{P})}(\{(q_0,\bot)\},F \times \{\bot\})$, and  $|\sigma_i|_{(\gamma,i,\vtl)} \leq |\sigma_{i-1}|_{(\gamma,i,\vtr)}$ for all $\gamma \in \Gamma$ and $i \in [0,k]$ with  $\sigma_{-1}=(\gamma_0,0,\vtr)$. Then, there are $\varsigma_0,\varsigma_1, \ldots, \varsigma_{k+1}$ $ \in \mathit{Conf}(\m{P})$ such that: $(i)$ $\varsigma_0=(q_0,\bot)$ and  $\varsigma_{k+1} \in F \times \{\bot\}$, and  $(ii)$ we have the following run of $\m{T}(\m{P})$: 
 $$\varsigma_0\, \lby{\sigma_0}{}_{\m{T}(\m{P})}^*\, \varsigma_1\, \lby{\sigma_1}{}_{\m{T}(\m{P})}^*\,\varsigma_2 \,  \cdots \,  \varsigma_{k-1} \lby{\sigma_{k-1}}{}_{\m{T}(\m{P})}^*\,\varsigma_k \lby{\sigma_k}{}_{\m{T}(\m{P})}^*\, \varsigma_{k+1}$$
Then, we can apply  Lemma \ref{lem.presbu.dir2.sec4}  to  $\mu(c_{\m{M}}^{\sf init})=\varsigma_0$, $\varsigma_0\, \lby{\sigma_0}{}_{\m{T}(\m{P})}^*\, \varsigma_1$, and $\mathit{Idle}(c_{\m{M}}^{\sf init}) ((\alpha,0)) \geq |\sigma_0|_{(\alpha,0,\vtl)}$ for all $\alpha \in \Gamma_{\epsilon}$, to prove   that    there are $\tau_0\in \Delta^*$ and $c_1 \in \big(Q \times \Gamma_{\m{P}} \times \mathit{Loc}(\m{M})\big)$ such that:

\begin{iteMize}{$\bullet$}
\item $\mu(c_1)=\varsigma_1$.
 \item   $c_{\m{M}}^{\sf init}\,\lby{\tau_0}{}_{\m{T}_{\{0\}}(\m{M})}^* \, c_1$. 
 
 \item  $\mathit{Idle}(c_1)((\alpha,0))=\mathit{Idle}(c_{\m{M}}^{\sf init})((\alpha,0))- |\sigma_0|_{(\alpha,0,\vtl)}$ and $\mathit{Idle}(c_1)((\alpha,1))= |\sigma_0|_{(\alpha,1,\vtr)} $ for all $\alpha \in \Gamma_{\epsilon}$.
 
 \item  $\mathit{Idle}(c_1)((\alpha,j))=\mathit{Idle}(c_{\m{M}}^{\sf init})((\alpha,j))=0$   for all $(\alpha,j) \in \Gamma_{\epsilon} \times [0,k+1]$ such that $j \notin \{0,1\}$.
 \end{iteMize}

Now, we can apply Lemma \ref{lem.presbu.dir2.sec4}  to  $\mu(c_1)=\varsigma_1$, $\varsigma_1\, \lby{\sigma_1}{}_{\m{T}(\m{P})}^*\, \varsigma_2$, and $\mathit{Idle}(c_1) ((\alpha,1))=|\sigma_0|_{(\alpha,1,\vtr)}  \geq |\sigma_1|_{(\alpha,1,\vtl)}$ for all $\alpha \in \Gamma_{\epsilon}$, to show that    there are $\tau_1\in \Delta^*$ and $c_2 \in \big(Q \times \Gamma_{\m{P}} \times \mathit{Loc}(\m{M})\big)$ such that:

\begin{iteMize}{$\bullet$}
\item $\mu(c_2)=\varsigma_2$.
 \item   $c_1\,\lby{\tau_1}{}_{\m{T}_{\{1\}}(\m{M})}^* \, c_2$. 
 
 \item  $\mathit{Idle}(c_2)((\alpha,1))=\mathit{Idle}(c_1)((\alpha,1))- |\sigma_1|_{(\alpha,1,\vtl)}$ and $\mathit{Idle}(c_2)((\alpha,2))= |\sigma_1|_{(\alpha,2,\vtr)} $ for all $\alpha \in \Gamma_{\epsilon}$.
 
 \item  $\mathit{Idle}(c_2)((\alpha,j))=\mathit{Idle}(c_1)((\alpha,j))$   for all $(\alpha,j) \in \Gamma_{\epsilon} \times [0,k+1]$ such that $j \notin \{1,2\}$.
 \end{iteMize}

So, we can apply step by step Lemma  \ref{lem.presbu.dir2.sec4}  to prove that there are $\tau_0,\ldots,\tau_k \in \Delta^*$ and $c_0,c_1,\ldots,c_{k+1} \in \big(Q \times \Gamma_{\m{P}} \times \mathit{Loc}(\m{M})\big)$ such that: $(1)$ $c_0=c_{\m{M}}^{\sf init}$, $(2)$  $\mu(c_{i})=\varsigma_i$ for all  $i \in [0,k+1]$, and $(3)$ $c_{0}  \;\lby{\tau_0}{}_{\m{T}_{\{0\}}(\m{M})}^* \, c_1 \;\lby{\tau_1}{}_{\m{T}_{\{1\}}(\m{M})}^*\,  \cdots \;\lby{\tau_{k-1}}{}_{\m{T}_{\{k-1\}}(\m{M})}^* \, c_k \;\lby{\tau_k}{}_{\m{T}_{\{k\}}(\m{M})}^* \, c_{k+1}$.
Moreover, we have $\mathit{State}(c_{k+1}) \in F$ and  $\mathit{Active}(c_{k+1})=\bot$ since $\varsigma_{k+1} \in F \times \{\bot\}$ and $\mu(c_{k+1})=\varsigma_{k+1}$. This implies that $\mathit{State}(c_{k+1}) \in F$  is $k$-stratified reachable by $\m{M}$. 
\qed

Lemma  \ref{ssrproblem-pda-relation} is an immediate consequence of Lemma \ref{if-direction001} and Lemma \ref{if-direction000}.
\qed


\section{The proof of Lemma \ref{dcfs->dcps-1}}
\label{sec.proof.lemma}

   The proof  of Lemma \ref{dcfs->dcps-1} is structured as follows: First, we establish the relation between a computation  of a thread of  $\m{M}_{\sf fs}$ and a run of  $\m{S}_{(q,\gamma)}$. Then, we give the relation between  a computation  of a thread of $\m{M}$ and a run of  $\m{P}_{(q,\gamma)}$. Due to the link   between the set of  runs of   $\m{P}_{(q,\gamma)}$ and the set of runs of  $\m{A}_{(q,\gamma)}$, these  two relations permit us to construct for every thread computation of $\m{M}$ an ``equivalent'' thread computation of $\m{M}_{\sf fs}$ and vice-versa.
Then, we consider a DCPS $\m{M}_{\cup}$ which is the union of $\m{M}$ and $\m{M}_{\sf fs}$ in the sense that for each thread $T$ with initial configuration $\gamma \in \Gamma$, $\m{A}_{\cup}$ chooses in nondeterministic way to execute the thread $T$ following the transition relation of $\m{M}_{\sf fs}$ or the transition relation of $\m{M}$.

Afterwards, we define the rank of a run of $\m{M}_{\cup}$ from the initial configuration $c_{\m{M}_{\cup}}^{\sf init}$ by the pair $(m,n) \in \mathbb{N} \times \mathbb{N}$ where $m$ is the number of threads involved in the run following the transition relation of $\m{M}$ and $n$ is the number of threads involved in the run following the transition relation of $\m{M}_{\sf fs}$. Observe that  runs of rank $(m,n)$ where $n=0$ (resp. $m=0$) are precisely the runs of $\m{M}$ (resp. $\m{M}_{\sf fs}$). Then, we prove that for any computation of $\m{M}_{\sf \cup}$ (from the initial configuration  $c_{\m{M}_{\cup}}^{\sf init}$) of rank $(m+1,n)$ (resp. $(m,n+1)$), there is a run of $\m{A}_{\cup}$ of rank $(m,n+1)$ (resp. $(m+1,n)$). This run is obtained from the original one by replacing   a thread that follows the transition relation of $\m{M}$ (resp. $\m{M}_{\sf fs}$) by a thread that follows the transition relation of $\m{M}_{\sf fs}$ (resp. $\m{M}$). This is possible  since any  thread of $\m{M}_{\sf ls}$ can be simulated by a thread of $\m{M}_{\sf fs}$ and vice-versa. As an immediate  consequence of the following result  is that, for every $m \in \m{M}$, a state $q$ is $k$-bounded (resp. $k$-stratified) reachable by a run of $\m{M}_{\cup}$ of rank $(m,0)$ (i.e., a run of $\m{M}_{\sf ls}$) if and only if it is $k$-bounded  (resp. $k$-stratified) reachable by a run of $\m{M}_{\cup}$ of rank $(0,m)$ (i.e., a run of $\m{M}_{\sf fs}$). This is precisely  what Lemma \ref{dcfs->dcps-1} says.

%
%
%
%

\subsection{The language of finite state automata $\m{A}_{(q,\gamma)}$}
In the following, we establish the following property about the finite state automata $\m{A}_{(q,\gamma)}$:

\begin{lem}
\label{lem.rel.fsa.sec6}
 Let $i \leq k$. If there are elements $\sigma_0, \ldots,\sigma_i \in
 \Gamma^*$,   $\gamma_0,\ldots,\gamma_i \in \Gamma$, $\alpha \in
 \Gamma_{\epsilon}$, $g_0,p_0,g_1,\ldots,p_{i},$ $ g_{i+1} \in Q$,
 $s_0 \in I_{(g_0,\gamma)}$, and $s \in S_{(g_0,\gamma)}$ such that
\[\sigma_0 (p_0,\gamma_1,g_1) \sigma_1 \cdots \sigma_{i-1}
 (p_{i-1},\gamma_i,g_{i}) \sigma_i (p_i,\alpha,g_{i+1})\in
 \mathit{Traces}_{\m{T}(\m{A}_{(g_0,\gamma)})}(\{s_0\},\{s\}),
\]
  then 
\[\sigma_0 (p_0,\gamma_1,g_1) \sigma_1 \cdots \sigma_{i-1}
(p_{i-1},\gamma_i,g_{i}) \sigma_i (p_i,\epsilon,g_{i+1}) \in
L(\m{A}_{(g_0,\gamma)}).
\]
\end{lem}

\proof 
Since  all the  states in the automaton $\m{A}_{(g_0,\gamma)}$ are co-reachable from the final states, in particular the state $s$,  there is $\nu \in \Sigma^*$ such that:

$$ \sigma_0 (p_0,\gamma_1,g_1) \sigma_1 \cdots \sigma_{i-1} (p_{i-1},\gamma_i,g_{i}) \sigma_i (p_i,\alpha,g_{i+1}) \nu \in L(\m{A}_{(g_0,\gamma)})$$

\noindent
This implies that there are $\sigma'_0, \ldots,\sigma'_i \in \Gamma^*$ and $\nu' \in \Sigma^*$ such that: $(1)$ $\sigma_l \preceq \sigma'_l$ for all $l \in [0,i]$, $(2)$ $\nu \preceq \nu'$,  and $(3)$ we have: 
$$ \sigma'_0 (p_0,\gamma_1,g_1) \sigma'_1 \cdots \sigma'_{i-1} (p_{i-1},\gamma_i,g_{i}) \sigma'_i (p_i,\alpha,g_{i+1}) \nu' \in L(\m{P}_{(g_0,\gamma)})$$

\noindent
Now, we can use the definition of $\m{P}_{(g_0,\gamma)}$ to show that we have:

$$ \sigma'_0 (p_0,\gamma_1,g_1) \sigma'_1 \cdots \sigma'_{i-1} (p_{i-1},\gamma_i,g_{i}) \sigma'_i (p_i,\epsilon,g_{i+1}) \in L(\m{P}_{(g_0,\gamma)})$$

\noindent
In addition, we   can show  that $ \sigma'_0 (p_0,\gamma_1,g_1) \sigma'_1 \cdots \sigma'_{i-1} (p_{i-1},\gamma_i,g_{i}) \sigma'_i (p_i,\epsilon,g_{i+1}) \in L'_{((g_0,\gamma),i+1)}$ since $i \leq k$.  This implies that 
  $\sigma_0 (p_0,\gamma_1,g_1) \sigma_1 \cdots \sigma_{i-1} (p_{i-1},\gamma_i,g_{i}) \sigma_i (p_i,\epsilon,g_{i+1}) \in L(\m{A}_{(g_0,\gamma)})$ since  $\sigma_l \preceq \sigma'_l$ for all $l \in [0,i]$.
\qed

\subsection{The relation between the DCFS $\m{M}_{\sf fs}$ and the FSA $\m{A}_{(p,\gamma)}$} In the following, we establish the link between  the set of runs  of a thread of $\m{M}_{\sf fs}$ without a context switches and the language generated  by  the finite state  automaton $\m{A}_{(q,\gamma)}$.

\begin{lem}
\label{lem.dcfs.fsa.lin2}
Let $j \in \mathbb{N}$,  $s, s' \in S_{\sf fs}^{\sf sm} $, and $\mathit{Val},\mathit{Val'} \in [\mathit{Loc}(\m{M}_{\sf fs}) \ra \mathbb{N}]$.  There is $\tau \in \Delta_{\sf fs}^*$ such that  $(\sharp,({s},j),\mathit{Val})\lby{\tau}{}^*_{\m{T}_{\{j\}}(\m{M}_{\sf fs})}$ $ (\sharp,(s',j),\mathit{Val}+\mathit{Val}'))$ if and only if  there are $q \in Q$, $\gamma \in \Gamma$, and  $\sigma \in \Gamma^*$ such that $s \, \lby{\sigma}{}^*_{\m{T}(\m{A}_{(q,\gamma)})} \,s'$, $\mathit{Val}'((\gamma',j+1))=|\sigma|_{\gamma'}$ for all $\gamma' \in \Gamma$, and $\mathit{Val}'((w,l))=0$ for all $w \in \Gamma_{\sf fs}^*$ and $l \in \mathbb{N}$ such that $(w,l) \notin \Gamma \times \{j+1\}$. 
\end{lem}

\medskip

\proof 
{\em The Only if direction:} Assume that there is $\tau \in \Delta_{\sf fs}^*$ such that  $(\sharp,({s},j),\mathit{Val})\bby{n}{\tau}{}_{\m{T}_{\{j\}}(\m{M}_{\sf fs})}$ $ (\sharp,(s',j),\mathit{Val}+\mathit{Val}'))$ for some $n \in \mathbb{N}$. We proceed by induction on $n$.

\noindent
{\bf Basis.} $n=0$. Then, $\tau=\epsilon$,  $s=s'$, and $\mathit{Val}'=\id{\mathit{Loc}(\m{M}_{\sf fs})}{\emptyset}$. Since $s \in S_{\sf fs}^{\sf sm}$, there is $q \in Q$ and $\gamma \in \Gamma$ such that $s \in S_{(q,\gamma)}$. By taking $\sigma=\epsilon$, we have $s \,\lby{\sigma}{}^*_{\m{T}(\m{A}_{(q,\gamma)})}\, s'$, $\mathit{Val}'((\gamma',j+1))=|\sigma|_{\gamma'}=0$ for all $\gamma' \in \Gamma$, and $\mathit{Val}'((w,l))=0$ for all $w \in \Gamma_{\sf fs}^*$ and $l \in \mathbb{N}$ such that $(w,l) \notin \Gamma \times \{j+1\}$.

\noindent
{\bf Step.} $n>0$. Then, from the definition of $\m{M}_{\sf fs}$, there is $\tau' \in \Delta_{\sf fs}^*$, $t \in \Delta_{\sf fs}$, $s'' \in S_{\sf fs}^{\sf sm}$, and $\mathit{Val''} \in [\mathit{Loc}(\m{M}_{\sf fs})\ra \mathbb{N}]$ such that $\tau=\tau't$, and:
\begin{equation}
 (\sharp,({s},j),\mathit{Val})\bby{n-1}{\tau'}{}_{\m{T}_{\{j\}}(\m{M}_{\sf fs})}(\sharp,(s'',j),\mathit{Val}+\mathit{Val}'')) \by{t}{}_{\m{T}_{\{j\}}(\m{M}_{\sf fs})}(\sharp,(s',j),\mathit{Val}+\mathit{Val}'))
\end{equation}

\noindent
We apply the induction hypothesis to  $(\sharp,({s},j),\mathit{Val})\bby{n-1}{\tau'}{}_{\m{T}_{\{j\}}(\m{M}_{\sf fs})}(\sharp,(s'',j),\mathit{Val}+\mathit{Val}''))$, and we obtain that there are $q \in Q$, $\gamma \in \Gamma$, and $\sigma' \in \Gamma^*$ such that  $s \,\lby{\sigma'}{}^*_{\m{T}(\m{A}_{(q,\gamma)})}\, s''$, $\mathit{Val}''((\gamma',j+1))=|\sigma'|_{\gamma'}$ for all $\gamma' \in \Gamma$, and $\mathit{Val}''((w,l))=0$ for all $w \in \Gamma_{\sf fs}^*$ and $l \in \mathbb{N}$ such that $(w,l) \notin \Gamma \times \{j+1\}$.

\noindent
In addition, from the definition of $\m{M}_{\sf fs}$, $t$ is necessarily of the form $\co \sharp,s' \cf \by{}_{\m{M}} \co \sharp,s'\cf \vtr \alpha$ with $\alpha \in \Gamma$. This implies that $s' \in S_{(q,\gamma)}$ and $s''\, \by{\alpha}_{\m{A}_{(q,\gamma)}} \,s'$. Moreover, we have $\mathit{Val}'=\mathit{Val}''+\id{\mathit{Loc}(\m{M}_{\sf fs})}{\{(\alpha,j+1)\}}$. Then, by taking $\sigma=\sigma' \alpha$, we can show (using the induction hypothesis) that $s \,\lby{\sigma}{}^*_{\m{T}(\m{A}_{(q,\gamma)})}\, s'$,  $\mathit{Val}'((\gamma',j+1))=|\sigma|_{\gamma'}$ for all $\gamma' \in \Gamma$, and $\mathit{Val}'((w,l))=0$ for all $w \in \Gamma_{\sf fs}^*$ and $l \in \mathbb{N}$ such that $(w,l) \notin \Gamma \times \{j+1\}$.

\bigskip

\noindent
{\em The If direction:}  Assume that there are $q \in Q$, $\gamma \in \Gamma$, and  $\sigma \in \Gamma^*$ such that $s \, \bby{n}{\sigma}{}_{\m{T}(\m{A}_{(q,\gamma)})} \,s'$ for some $n \in \mathbb{N}$, $\mathit{Val}'((\gamma',j+1))=|\sigma|_{\gamma'}$ for all $\gamma' \in \Gamma$, and $\mathit{Val}'((w,l))=0$ for all $w \in \Gamma_{\sf fs}^*$ and $l \in \mathbb{N}$ such that $(w,l) \notin \Gamma \times \{j+1\}$.  We proceed by induction on $n$.

\medskip
\noindent
{\bf Basis.} $n=0$. Then, $s=s'$ and $\sigma=\epsilon$. By taking $\tau=\epsilon$, we get  $(\sharp,({s},j),\mathit{Val})\lby{\tau}{}^*_{\m{T}_{\{j\}}(\m{M}_{\sf fs})}$ $ (\sharp,(s',j),\mathit{Val}+\mathit{Val}'))$ since $\mathit{Val}'=\id{\mathit{Loc}(\m{M}_{\sf fs})}{\emptyset}$.

\medskip
\noindent
{\bf Step.} $n>0$. Then, there are $s''$, $\sigma' \in \Gamma^*$, and $\alpha \in \Gamma$ such that $\sigma=\sigma' \alpha$, and:
\begin{equation}
s \, \bby{n-1}{\sigma'}{}_{\m{T}(\m{A}_{(q,\gamma)})} \,s'' \, \by{\alpha}{}_{\m{T}(\m{A}_{(q,\gamma)})} \, s'
\end{equation}

\medskip
\noindent
Let $\mathit{Val}'' \in [\mathit{Loc}(\m{M}) \ra \mathbb{N}]$ such that , $\mathit{Val}''((\gamma',j+1))=|\sigma'|_{\gamma'}$ for all $\gamma' \in \Gamma$, and $\mathit{Val}''((w,l))=0$ for all $w \in \Gamma_{\sf fs}^*$ and $l \in \mathbb{N}$ such that $(w,l) \notin \Gamma \times \{j+1\}$.  Then, 
we apply the induction hypothesis to $s \, \bby{n-1}{\sigma'}{}_{\m{T}(\m{A}_{(q,\gamma)})} \,s'' $ and $\mathit{Val}''$, and we obtain that there is $\tau \in \Delta^*_{\sf fs}$ such that $(\sharp,({s},j),\mathit{Val})\lby{\tau'}{}^*_{\m{T}_{\{j\}}(\m{M}_{\sf fs})}$ $ (\sharp,(s'',j),\mathit{Val}+\mathit{Val}''))$.

\medskip
\noindent
Since $s'' \, \by{\alpha}{}_{\m{T}(\m{A}_{(q,\gamma)})} \, s'
$, we have $t=\co \sharp, s'' \cf \by{}_{\m{M}_{\sf fs}} \, \co \sharp, s'\cf \vtr \alpha$. Then, using the induction hypothesis, we can show that   $(\sharp,({s},j),\mathit{Val})\lby{\tau}{}^*_{\m{T}_{\{j\}}(\m{M}_{\sf fs})}$ $ (\sharp,(s',j),\mathit{Val}+\mathit{Val}'))$ with $\tau=\tau' t$ since we have $\mathit{Val}'=\mathit{Val}''+\id{\mathit{Loc}(\m{M}_{\sf fs})}{\{(\alpha,j+1)\}}$.
\qed

Next, we use Lemma \ref{lem.dcfs.fsa.lin2} to establish the relation between  the set of  languages accepted by the finite state automata $\m{A}_{(q,\gamma)}$  and the set of  runs  of  $\m{M}_{\sf fs}$ between two configurations with no active thread and without context switches.

\begin{lem}  
\label{lem.dcfs.fsa.lin3}
Let $j \in \mathbb{N}$,  $p_1,p_2 \in Q$, $\lambda_1 \in (S_{\sf fs}^{\sf sw} \cup \Gamma) $, $\lambda_2  \in S_{\sf fs}^{\sf sw}$, and $\mathit{Val} \in [\mathit{Loc}(\m{M}_{\sf fs}) \ra \mathbb{N}]$. There is $\tau \in \Delta_{\sf fs}^*$ such that  $|\tau|\geq 1$  and $(p_1,\bot,\id{\mathit{Loc}(\m{M}_{\sf fs})}{\{(\lambda_1,j)\}})\lby{\tau}{}^*_{\m{T}_{\{j\}}(\m{M}_{\sf fs})}\, (p_2,\bot,\mathit{Val}+\id{\mathit{Loc}(\m{M}_{\sf fs})}{\{(\lambda_2,j+1)\}})$ iff   there are $q \in Q$, $\gamma, \gamma_1 \in \Gamma$, $\alpha \in \Gamma_{\epsilon}$, $p'_1,p'_2 \in Q$, $s,s' \in S_{(q,\gamma)}$, and $\sigma \in \Gamma^*$ such that:

\begin{iteMize}{$\bullet$}
\item $p_1 \swi_{\m{M}_{\sf fs}} p'_1 \vtl \lambda_1$, $\lambda_1=(p'_1,(s,\gamma_1))$ if $\lambda_1 \in S_{\sf fs}^{\sf sw}$, and $p'_1=q$ and $s \in I_{(q,\gamma)}$ if $\lambda_1 \in \Gamma$.
\item $s \lby{\sigma (p_2,\alpha,p'_2)}{}^*_{\m{T}(\m{A}_{(q,\gamma)})} s'$ and $\lambda_2=(p'_2,(s',\alpha))$.
\item  $\mathit{Val}((\gamma',j+1))=|\sigma|_{\gamma'}$ for all $\gamma' \in \Gamma$.
\item  $\mathit{Val}((w,l))=0$ for all $w \in \Gamma_{\sf fs}^*$ and $l \in \mathbb{N}$ such that $(w,l) \notin \Gamma \times \{j+1\}$.

\end{iteMize}

\end{lem}

\proof {\em The Only if direction:} Assume 
the existence of some
$\tau \in \Delta_{\sf fs}^*$ such that  $|\tau|\geq 1$  and $(p_1,\bot,\id{\mathit{Loc}(\m{M}_{\sf fs})}{\{(\lambda_1,j)\}})$ $\lby{\tau}{}^*_{\m{T}_{\{j\}}(\m{M}_{\sf fs})}\, (p_2,\bot,\mathit{Val}+\id{\mathit{Loc}(\m{M}_{\sf fs})}{\{(\lambda_2,j+1)\}})$. Then, from the definition of $\m{T}_{\{j\}}(\m{M}_{\sf fs})$, there are $q \in Q$, $\gamma, \gamma_1 \in \Gamma$, $p'_1 \in Q$, $\tau' \in \Delta_{\sf fs}^*$, $s,s'' \in S_{(q,\gamma)}$, and $\sigma \in \Gamma^*$ such that:

\begin{iteMize}{$\bullet$}
\item $(p_1,\bot,\id{\mathit{Loc}(\m{M}_{\sf fs})}{\{(\lambda_1,j)\}})\by{t}{}_{\m{T}_{\{j\}}(\m{M}_{\sf fs})}\, (p'_1,(\lambda_1,j),\id{\mathit{Loc}(\m{M}_{\sf fs})}{\emptyset})$ such that $t= p_1 \swi_{\m{M}_{\sf fs}} p'_1 \vtl \lambda_1$.

\item $(p'_1,(\lambda_1,j),\id{\mathit{Loc}(\m{M}_{\sf fs})}{\emptyset})\by{t'}{}_{\m{T}_{\{j\}}(\m{M}_{\sf fs})}\, (\sharp,(s,j),\id{\mathit{Loc}(\m{M}_{\sf fs})}{\emptyset})$ with $t'= \co p'_1, \lambda_1 \cf \swi_{\m{M}_{\sf fs}} \co\sharp,s\cf \vtr \epsilon$,  $\lambda_1=(p'_1,(s,\gamma_1))$ if $\lambda_1 \in S_{\sf fs}^{\sf sw}$, and $p'_1=q$ and $s \in I_{(q,\gamma)}$ if $\lambda_1 \in \Gamma$.
\item $(\sharp,(s,j),\id{\mathit{Loc}(\m{M}_{\sf fs})}{\emptyset}) \lby{\tau'}{}_{\m{T}_{\{j\}}(\m{M})}^*\, (\sharp,(s'',j),\mathit{Val})$. Then, we apply Lemma \ref{lem.dcfs.fsa.lin2}, and we obtain that there is $\sigma \in \Gamma^*$ such that  $s \,\lby{\sigma}{}^*_{\m{T}(\m{A}_{(q,\gamma)})} s''$, $\mathit{Val}((\gamma',j+1))=|\sigma|_{\gamma'}$ for all $\gamma' \in \Gamma$ and
  $\mathit{Val}((w,l))=0$ for all $w \in \Gamma_{\sf fs}^*$ and $l \in \mathbb{N}$ such that $(w,l) \notin \Gamma \times \{j+1\}$.
  
 \item  $(\sharp,(s'',j),\mathit{Val}) \by{t''}{}_{\m{T}_{\{j\}}(\m{M})}\, (p_2,\bot,\mathit{Val}+\id{\mathit{Loc}(\m{M}_{\sf fs})}{\{(\lambda_2,j+1)\}})$ with $t''= \co \sharp, s'' \cf \swi_{\m{M}_{\sf fs}} \co p_2, \lambda_2 \cf $. From the definition of $\m{M}_{\sf fs}$, this implies that there is $p'_2 \in Q$, $s' \in S_{(q,\gamma)}$, and  $\alpha \in \Gamma_{\epsilon}$ such that $s'' \by{(p_2,\alpha,p'_2)} s'$ and $\lambda_2=(p'_2,(s',\alpha))$.
\end{iteMize}\medskip

\noindent This terminates the proof of the Only if direction.\medskip

\noindent
{\em The If direction:} Assume   that there are $q \in Q$, $\gamma, \gamma_1 \in \Gamma$, $\alpha \in \Gamma_{\epsilon}$, $p'_1,p'_2 \in Q$, $s,s' \in S_{(q,\gamma)}$, and $\sigma \in \Gamma^*$ such that:

\begin{iteMize}{$\bullet$}
\item $t=p_1 \swi_{\m{M}} p'_1 \vtl \lambda_1$, $\lambda_1=(p'_1,(s,\gamma_1))$ if $\lambda_1 \in S_{\sf fs}^{\sf sw}$, and $p'_1=q$ and $s \in I_{(q,\gamma)}$ if $\lambda_1 \in \Gamma$.
\item $s \lby{\sigma (p_2,\alpha,p'_2)}{}^*_{\m{T}(\m{A}_{(q,\gamma)})} s'$ and $\lambda_2=(p'_2,(s',\alpha))$.
\item  $\mathit{Val}((\gamma',j+1))=|\sigma|_{\gamma'}$ for all $\gamma' \in \Gamma$.
\item  $\mathit{Val}((w,l))=0$ for all $w \in \Gamma_{\sf fs}^*$ and $l \in \mathbb{N}$ such that $(w,l) \notin \Gamma \times \{j+1\}$.

\end{iteMize}

Then, from the definition of $\m{T}({\m{M}_{\sf fs}})$, we have the following run:
\begin{equation}
(p_1,\bot,\id{\mathit{Loc}(\m{M}_{\sf fs})}{\{(\lambda_1,j)\}})\by{t}{}_{\m{T}_{\{j\}}(\m{M}_{\sf fs})}\, (p'_1,(\lambda_1,j),\id{\mathit{Loc}(\m{M}_{\sf fs})}{\emptyset})\end{equation} 
Let $t'= \co p'_1, \lambda_1 \cf \swi_{\m{M}_{\sf fs}} \co\sharp,s\cf \vtr \epsilon$. Then, we have the following run of $\m{T}_{\{j\}}(\m{M}_{\sf fs})$:

\begin{equation}
(p'_1,(\lambda_1,j),\id{\mathit{Loc}(\m{M}_{\sf fs})}{\emptyset})\by{t'}{}_{\m{T}_{\{j\}}(\m{M}_{\sf fs})}\, (\sharp,(s,j),\id{\mathit{Loc}(\m{M}_{\sf fs})}{\emptyset})
\end{equation}

\medskip
\noindent
Let $s'' \in S_{(q,\gamma)}$ such that $s \lby{\sigma }{}^*_{\m{T}(\m{A}_{(q,\gamma)})} s''$ and $s'' \by{ (p_2,\alpha,p'_2)}{}_{\m{T}(\m{A}_{(q,\gamma)})} s'$. Then, we can apply Lemma \ref{lem.dcfs.fsa.lin2}, to prove that there is $\tau' \in \Delta^*_{\sf fs}$ such that:
\begin{equation}
 (\sharp,(s,j),\id{\mathit{Loc}(\m{M}_{\sf fs})}{\emptyset}) \lby{\tau'}{}_{\m{T}_{\{j\}}(\m{M})}^*\, (\sharp,(s'',j),\mathit{Val})
\end{equation}
Since $s'' \by{ (p_2,\alpha,p'_2)}{}_{\m{T}(\m{A}_{(q,\gamma)})} s'$, we have $\co \sharp, s'' \cf \swi_{\m{M}_{\sf fs}} \co p_2, \lambda_2\cf$. This implies that  $\m{T}_{\{j\}}(\m{M}_{\sf fs})$ has  the following run:
\begin{equation}
(\sharp,(s'',j),\mathit{Val}) \by{t''}{}_{\m{T}_{\{j\}}(\m{M})}\, (p_2,\bot,\mathit{Val}+\id{\mathit{Loc}(\m{M}_{\sf fs})}{\{(\lambda_2,j+1)\}})
\end{equation}

\medskip
\noindent
This terminates the proof of the If direction.
\qed

\subsection{The relation between the DCPS $\m{M}$ and the PDA $\m{P}_{(p,\gamma)}$} In the following, we establish the link between  the set of runs  of a thread of $\m{M}$ and the language generated  by  the pushdown automaton $\m{P}_{(q,\gamma)}$.

\begin{lem}
\label{lem.perfect-pda-lin1}
Let $j \in \mathbb{N}$, $q \in Q$, $\gamma \in \Gamma$,  $p_1,p_2 \in Q$, $w_1, w_2 \in \Gamma^*$, and $\mathit{Val},\mathit{Val'} \in [\mathit{Loc}(\m{M}) \ra \mathbb{N}]$. There is $\tau \in \Delta^*$ such that  $(p_1,({w_1},j),\mathit{Val})\lby{\tau}{}^*_{\m{T}_{\{j\}}(\m{M})}$ $ (p_2,(w_2,j),\mathit{Val}+\mathit{Val}'))$ for some  iff there is $\sigma \in \Gamma^*$ such that $(p_1,w_1) \lby{\sigma}{}^*_{\m{P}(q,\gamma)} (p_2,w_2)$, $\mathit{Val}'((\gamma',j+1))=|\sigma|_{\gamma'}$ for all $\gamma' \in \Gamma$, and $\mathit{Val}'((w,l))=0$ for all $w \in \Gamma_{\epsilon}$ and $l \in \mathbb{N}$ such that $(w,l) \notin \Gamma \times \{j+1\}$. 
\end{lem}

\medskip

\proof {\em The Only if direction:} Assume that there is $\tau \in
\Delta^*$ and some $n \in \mathbb{N}$ such that
$(p_1,({w_1},j),\mathit{Val})\bby{n}{\tau}_{\m{T}_{\{j\}}(\m{M})}
(p_2,(w_2,j),\mathit{Val}+\mathit{Val}'))$. We proceed by induction on
$n$.

\medskip
\noindent
{\bf Basis.} $n=0$. This implies that $p_1=p_2$, $w_1=w_2$, $\tau=\epsilon$, and $\mathit{Val'}=\id{\mathit{Loc}(\m{M})}{\emptyset}$. By taking $\sigma=\epsilon$, we have  $(p_1,w_1) \lby{\sigma}_{\m{P}(q,\gamma)} (p_2,w_2)$, $\mathit{Val}'((\gamma',j+1))=|\sigma|_{\gamma'}=0$ for all $\gamma' \in \Gamma$, and $\mathit{Val}'((w,l))=0$ for all $w \in \Gamma_{\epsilon}$ and $l \in \mathbb{N}$ such that $(w,l) \notin \Gamma \times \{j+1\}$.

\medskip
\noindent
{\bf Step.} $n>0$. From the definition of $\m{M}$, this implies that there are $p \in Q$, $w'_1 \in \Gamma^*$, $\mathit{Val}'' \in [\mathit{Loc}(\m{M}) \ra \mathbb{N}]$, $\tau' \in \Delta^*$, and $t \in \Delta$ such that $\tau=\tau' t$ and:
\begin{equation}
(p_1,({w_1},j),\mathit{Val})\bby{n-1}{\tau'}{}_{\m{T}_{\{j\}}(\m{M})} (p,({w'_1},j),\mathit{Val}+\mathit{Val}'')\by{t}{}_{\m{T}_{\{j\}}(\m{M})} (p_2,(w_2,j),\mathit{Val}+\mathit{Val}')
\end{equation}
We apply the induction hypothesis to $(p_1,({w_1},j),\mathit{Val})\lby{\tau'}{}^*_{\m{T}_{\{j\}}(\m{M})} (p,({w'_1},j),\mathit{Val}+\mathit{Val}'')$, and we obtain that there is $\sigma' \in \Gamma^*$ such that
$(p_1,w_1) \lby{\sigma'}{}^*_{\m{T}(\m{P}_{(q,\gamma)})} (p,w'_1)$, $\mathit{Val}''((\gamma',j+1))=|\sigma'|_{\gamma'}$ for all $\gamma' \in \Gamma$, and $\mathit{Val}''((w,l))=0$ for all $w \in \Gamma_{\epsilon}$ and $l \in \mathbb{N}$ such that $(w,l) \notin \Gamma \times \{j+1\}$. 

In addition, from the definition of $\m{M}$, the transition $t$ is necessarily of the form $\co p, \gamma_1 \cf \by{}_{\m{M}}$ $\co p_2,u\cf \vtr \alpha$ with $\gamma_1 \in \Gamma$ and $\alpha \in \Gamma_{\epsilon}$ such that $w'_1=\gamma_1 v$ and $w_2=u v$ for some $v \in \Gamma^*$.  This implies that $\co p,\gamma_1 \cf \by{\alpha}_{\m{P}_{(q,\gamma)}} \co p_2,u\cf$, and so, $(p,w'_1) \by{\alpha}_{\m{T}(\m{P}_{(q,\gamma)})} (p_2,w_1)$. Moreover, we have $\mathit{Val}'=\mathit{Val}''+\id{\mathit{Loc}(\m{M})}{\{(\alpha,j+1)\}}$. Then, by taking $\sigma=\sigma' \alpha$, we can easily  show (using the induction hypothesis) that $(p_1,w_1) \lby{\sigma}_{\m{P}(q,\gamma)} (p_2,w_2)$, $\mathit{Val}'((\gamma',j+1))=|\sigma|_{\gamma'}$ for all $\gamma' \in \Gamma$, and $\mathit{Val}'((w,l))=0$ for all $w \in \Gamma_{\epsilon}$ and $l \in \mathbb{N}$ such that $(w,l) \notin \Gamma \times \{j+1\}$. 

\medskip

\noindent
{\em The If direction:} Assume that there is $\sigma \in \Gamma^*$ such that $(p_1,w_1) \bby{n}{\sigma}_{\m{P}(q,\gamma)} (p_2,w_2)$, $\mathit{Val}'((\gamma',j+1))=|\sigma|_{\gamma'}$ for all $\gamma' \in \Gamma$, and $\mathit{Val}'((w,l))=0$ for all $w \in \Gamma_{\epsilon}$ and $l \in \mathbb{N}$ such that $(w,l) \notin \Gamma \times \{j+1\}$. We proceed by induction on $n$.

\medskip
\noindent
{\bf Basis.} $n=0$. This implies  that $p_1=p_2$, $w_1=w_2$, $\sigma=\epsilon$, and $\mathit{Val'}=\id{\mathit{Loc}(\m{M})}{\emptyset}$. By taking $\tau=\epsilon$, we have $(p_1,({w_1},j),\mathit{Val})\lby{\tau}{}^*_{\m{T}_{\{j\}}(\m{M})}$ $ (p_2,(w_2,j),\mathit{Val}+\mathit{Val}'))$.

\medskip
\noindent
{\bf Step.} $n>0$. Then, from the definition of $\m{P}_{(q,\gamma)}$, there are $p \in Q$, $w'_1 \in \Gamma^*$, $\sigma' \in \Gamma^*$, and  $\alpha \in \Gamma_{\epsilon}$ such that $\sigma=\sigma' \alpha$, and:
\begin{equation}
(p_1,w_1)\bby{n-1}{\sigma'}{}_{\m{T}(\m{P}_{(q,\gamma)})} (p,w'_1) \by{\alpha}{}_{\m{T}(\m{P}_{(q,\gamma)})} (p_2,w_2)
\end{equation}
Let $\mathit{Val}'' \in [\mathit{Loc}(\m{M}) \ra \mathbb{N}]$ such that  $\mathit{Val}''((\gamma',j+1))=|\sigma'|_{\gamma'}$ for all $\gamma' \in \Gamma$, and $\mathit{Val}''((w,l))=0$ for all $w \in \Gamma_{\epsilon}$ and $l \in \mathbb{N}$ such that $(w,l) \notin \Gamma \times \{j+1\}$.

Then, we apply the induction hypothesis to $(p_1,w_1)\bby{n-1}{\sigma'}{}_{\m{T}(\m{P}_{(q,\gamma)})} (p,w'_1)$ and $\mathit{Val''}$, and we obtain that there is $\tau' \in \Delta^*$ such that $(p_1,({w_1},j),\mathit{Val})\lby{\tau'}{}^*_{\m{T}_{\{j\}}(\m{M})}$ $ (p,(w'_1,j),\mathit{Val}+\mathit{Val}''))$

Since $(p,w'_1) \by{\alpha}{}_{\m{T}(\m{P}_{(q,\gamma)})} (p_2,w_2)$,
there are elements $\gamma_1 \in \Gamma$ and $u \in \Gamma^*$ such that $t=\co p, \gamma_1 \cf \by{}_{\m{M}} \co p_2, u\cf$ $ \vtr \alpha$, $w'_1=\gamma_1 v$, and $w_2=u v$ for some $v \in \Gamma^*$. Then, using the induction hypothesis, we can easily  show that $(p_1,({w_1},j),\mathit{Val})\lby{\tau}{}^*_{\m{T}_{\{j\}}(\m{M})}$ $ (p_2,(w_2,j),\mathit{Val}+\mathit{Val}'))$ where $\tau=\tau't$.
\qed

Next, we use Lemma \ref{lem.perfect-pda-lin1} to establish the relation between  the set of  languages accepted by the pushdown   automata $\m{P}_{(q,\gamma)}$  and the set of  runs  of  $\m{M}$ between two configurations with no active thread and without context switches.

\begin{lem}
\label{lem.perfect.pda.lin2}
Let $j \in \mathbb{N}$, $q \in Q$, $\gamma \in \Gamma$,  $p_1,p_2 \in Q$, $w_1 \in \Gamma^*$, and $\mathit{Val} \in [\mathit{Loc}(\m{M}) \ra \mathbb{N}]$. There is $\tau \in \Delta^*$ such that  $|\tau|\geq 1$  and $(p_1,\bot,\id{\mathit{Loc}(\m{M})}{\{(w_1,j)\}})\lby{\tau}{}^*_{\m{T}_{\{j\}}(\m{M})}\, (p_2,\bot,\mathit{Val})$ iff   there are $p'_1 \in Q$, $\gamma_1 \in \Gamma$,  $w_2 \in \Gamma^*$, $\sigma \in \Gamma^*$, $\mathit{Val'} \in [\mathit{Loc}(\m{M}) \ra \mathbb{N}]$ such that:

\begin{iteMize}{$\bullet$}
\item $p_1 \swi_{\m{M}} p'_1 \vtl \gamma_1$ and $w_1=\gamma_1 v_1$ for some $v_1 \in \Gamma^*$.
\item $\mathit{Val}=\mathit{Val}'+\id{\mathit{Loc}(\m{M})}{\{(w_2,j+1)\}}$.
\item $(p'_1,w_1) \lby{\sigma (p_2,\alpha,p'_2)}{}^*_{\m{T}(\m{P}_{(q,\gamma)})} (p'_2,w_2)$ for all $p'_2 \in Q$  and $\alpha \in \Gamma_{\epsilon}$   such that $w_2=\alpha v_2$ for some $v_2 \in \Gamma^*$.
\item  $\mathit{Val}'((\gamma',j+1))=|\sigma|_{\gamma'}$ for all $\gamma' \in \Gamma$.
\item  $\mathit{Val}'((w,l))=0$ for all $w \in \Gamma_{\epsilon}$ and $l \in \mathbb{N}$ such that $(w,l) \notin \Gamma \times \{j+1\}$.

\end{iteMize}
\end{lem}

\medskip

\proof
{\em The Only if direction:} Assume that $(p_1,\bot,\id{\mathit{Loc}(\m{M})}{\{(w_1,j)\}})\lby{\tau}{}^*_{\m{T}_{\{j\}}(\m{M})}\, (p_2,\bot,\mathit{Val}))$  for some $\tau \in \Delta^*$ such that $|\tau|\geq 1$. Then, from the definition of $\m{T}_{\{j\}}(\m{M})$, there are $p'_1,p' \in Q$, $\gamma_1,\gamma_2 \in \Gamma$, $w'_2, u \in \Gamma^*$, $\tau' \in \Delta^*$, $\mathit{Val}' \in [\mathit{Loc}(\m{M}) \ra \mathbb{N}]$ such that:

\begin{iteMize}{$\bullet$}
\item $(p_1,\bot,\id{\mathit{Loc}(\m{M})}{\{(w_1,j)\}})\by{t}{}_{\m{T}_{\{j\}}(\m{M})}\, (p'_1,(w_1,j),\id{\mathit{Loc}(\m{M})}{\emptyset})$ such that $t= p_1 \swi_{\m{M}} p'_1 \vtl \gamma_1$. This implies that $w_1=\gamma_1 v_1$ for some $v_1 \in \Gamma^*$.
\item  $ (p'_1,(w_1,j),\id{\mathit{Loc}(\m{M})}{\emptyset})\lby{\tau'}{}^*_{\m{T}_{\{j\}}(\m{M})}\, (p',(w'_2,j),\mathit{Val'})$. Then, we can apply Lemma \ref{lem.perfect-pda-lin1}, to show that there is $\sigma \in \Gamma^*$ such that $(p'_1,w_1) \lby{\sigma}{}^*_{\m{P}(q,\gamma)} (p',w'_2)$, $\mathit{Val}'((\gamma',j+1))=|\sigma|_{\gamma'}$ for all $\gamma' \in \Gamma$, and $\mathit{Val}'((w,l))=0$ for all $w \in \Gamma_{\epsilon}$ and $l \in \mathbb{N}$ such that $(w,l) \notin \Gamma \times \{j+1\}$. 
\item $(p',(w'_2,j),\mathit{Val}') \by{t'}{}_{\m{T}_{\{j\}}(\m{M})}\, (p_2,\bot,\id{\mathit{Loc}(\m{M})}{\{w_2,j+1\}}+\mathit{Val}')$ such that: $(1)$ $\mathit{Val}=\id{\mathit{Loc}(\m{M})}{\{w_2,j+1\}}+\mathit{Val}'$, $(2)$ $t'= \co p',\gamma_2 \cf \swi_{\m{M}} \co  p_2, u\cf$, and $(3)$ $w'_2=\gamma_2 v$ and $w_2=uv$ for some $v \in \Gamma^*$. Using the  definition of $\m{P}_{(q,\gamma)}$, we have $(p',w'_2) \lby{(p_2,\alpha,p'_2)}{}_{\m{P}(q,\gamma)}^* (p'_2,w_2)$  for all $p'_2 \in Q$  and $\alpha \in \Gamma_{\epsilon}$   such that $w_2=\alpha v_2$ for some $v_2 \in \Gamma^*$.
\end{iteMize}\medskip

\noindent This terminates the proof of the only if direction of Lemma
\ref{lem.perfect.pda.lin2}

\medskip
\noindent
{\em The If direction:} Assume that  there are $p'_1 \in Q$, $\gamma_1 \in \Gamma$,  $w_2 \in \Gamma^*$, $\sigma \in \Gamma^*$, $\mathit{Val'} \in [\mathit{Loc}(\m{M}) \ra \mathbb{N}]$ such that:

\begin{iteMize}{$\bullet$}
\item $t=p_1 \swi_{\m{M}} p'_1 \vtl \gamma_1$ and $w_1=\gamma_1 v_1$ for some $v_1 \in \Gamma^*$.
\item $\mathit{Val}=\mathit{Val}'+\id{\mathit{Loc}(\m{M})}{\{(w_2,j+1)\}}$.
\item $(p'_1,w_1) \lby{\sigma (p_2,\alpha,p'_2)}{}^*_{\m{P}(q,\gamma)} (p'_2,w_2)$ for all $p'_2 \in Q$  and $\alpha \in \Gamma_{\epsilon}$   such that $w_2=\alpha v_2$ for some $v_2 \in \Gamma^*$.
\item  $\mathit{Val}'((\gamma',j+1))=|\sigma|_{\gamma'}$ for all $\gamma' \in \Gamma$.
\item  $\mathit{Val}'((w,l))=0$ for all $w \in \Gamma_{\epsilon}$ and $l \in \mathbb{N}$ such that $(w,l) \notin \Gamma \times \{j+1\}$

\end{iteMize}

\medskip
\noindent
Since $t=_1 \swi_{\m{M}} p'_1 \vtl \gamma_1$ and $w_1=\gamma_1 v_1$ for some $v_1 \in \Gamma^*$, we have the following run of $\m{T}_{\{j\}}(\m{M})$:
\begin{equation}
\label{append.sec7.eq.007}
(p_1,\bot,\id{\mathit{Loc}(\m{M})}{\{(w_1,j)\}})\by{t}{}_{\m{T}_{\{j\}}(\m{M})}\, (p'_1,(w_1,j),\id{\mathit{Loc}(\m{M})}{\emptyset})
\end{equation}

\medskip
\noindent
Let $p' \in Q$ and $w'_2 \in \Gamma^*$ such that $(p'_1,w_1) \lby{\sigma}{}_{\m{P}_{(q,\gamma)}}^* (p',w'_2)$ and $(p',w'_2) \by{(p_2,\epsilon,p'_2)}{}_{\m{P}_{(q,\gamma)}} (p'_2,w_2)$ for some $p'_2 \in Q$. 
Now, we can apply  Lemma  \ref{lem.perfect-pda-lin1} to $(p'_1,w_1) \lby{\sigma}{}_{\m{P}_{(q,\gamma)}}^* (p',w'_1)$ show that there is $\tau' \in \Delta^*$ such that: 
\begin{equation}
\label{append.sec7.eq.008}
(p'_1,(w_1,j),\id{\mathit{Loc}(\m{M})}{\emptyset})\lby{\tau'}{}^*_{\m{T}_{\{j\}}(\m{M})}\, (p',(w'_2,j),\mathit{Val'})
\end{equation}

\medskip
\noindent
Since $(p',w'_2) \by{(p_2,\epsilon,p'_2)}{}_{\m{P}_{(q,\gamma)}} (p'_2,w_2)$, we can use the definition of $\m{P}_{(q,\gamma)}$, to show that there are $\gamma_2 \in \Gamma$ and $u \in \Gamma^*$ such that $t'=\co p',\gamma_2 \cf \swi_{\m{M}}  \co p_2 ,u \cf$ such that $w'_2=\gamma_2 v$ and $w_2 =u v$ for some $v \in \Gamma^*$. So, $\m{T}_{j}(\m{M})$ has  the following run:
\begin{equation}
\label{append.sec7.eq.009}
(p',(w'_2,j),\mathit{Val}') \by{t'}{}_{\m{T}_{\{j\}}(\m{M})}\, (p_2,\bot,\id{\mathit{Loc}(\m{M})}{\{w_2,j+1\}}+\mathit{Val}')
\end{equation}

\medskip
\noindent
Now, we can put together Equation \ref{append.sec7.eq.007}, Equation \ref{append.sec7.eq.008}, and Equation \ref{append.sec7.eq.009}, and we obtain that    $(p_1,\bot,\id{\mathit{Loc}(\m{M})}{\{(w_1,j)\}})\lby{\tau}{}^*_{\m{T}_{\{j\}}(\m{M})}\, (p_2,\bot,\mathit{Val})$ with $\tau=t\tau' t'$.
\qed

%
\subsection{From the DCPS $\m{M}$ to the  DCPS $\m{M}_{\sf pf}$} In order to be able to distinguish between pending threads of $\m{M}$ that has been  activated   at least one time from the other ones, we need to define a DCPS $\m{M}_{\sf pf}$ (which is just a copy of $\m{M}$)   that uses,  in addition to the stack alphabet $\Gamma$,  a new stack alphabet $\Gamma'$, which  is a copy    $\Gamma$, to process threads. 
 Let $\Gamma'$ be a stack alphabet such that  $\Gamma' \cap \Gamma_{\sf fs}=\emptyset$ and there is a bijective   function ${\sf f}$ from $\Gamma$ to $\Gamma'$. This function ${\sf f}$ is extended to words over $\Gamma$ in the natural way: ${\sf f}(\epsilon)=\epsilon$ and ${\sf f}(u \cdot v)={\sf f}(u) \cdot {\sf f}(v)$ for all $u,v \in \Gamma^*$.  Moreover, we define the function   ${\sf h}$ from  $\Gamma_{\sf pf}$ to  $\Gamma$ such that  ${\sf h}(\gamma)={\sf h}({\sf f}(\gamma))=\gamma$ for all $\gamma \in \Gamma$. The function ${\sf h}$ is extended in the usual way to words.

In the following, we define the DCPS $\m{M}_{\sf pf}$ obtained from $\m{M}$ by using $\Gamma'$ to process threads instead of  $\Gamma$. Let   $\m{M}_{\sf pf}=(Q,\Gamma_{\sf pf}, \Delta_{\sf pf},q_0,\gamma_0,F )$  be a DCPS  where $\Gamma_{\sf pf}=\Gamma \cup \Gamma'$ and $\Delta'_{\sf pf}$ is the smallest transition relation satisfying the following conditions:

\begin{iteMize}{$\bullet$}
\item {\em Initialize:} For every $q \in Q$ and $\gamma \in \Gamma$, we have $\co q,\gamma \cf \by{}_{\m{M}_{\sf pf}} \co q, {\sf f}(\gamma) \cf \vtr \epsilon$.
\item  {\em Spawn:} For every $\co q,\gamma \cf \by{}_{\m{M}} \co q',u\cf \vtr \alpha$,  we have $\co q,{\sf f}(\gamma) \cf \by{}_{\m{M}_{\sf pf}} \co q',{\sf f}(u)\cf \vtr \alpha$.

\item  {\em Interrupt:} For every $\co q,\gamma \cf \swi_{\m{M}} \co q',u\cf $, we have  $\co q,{\sf f}(\gamma) \cf \swi_{\m{M}_{\sf pf}} \co q',{\sf f}(u)\cf $.
\item {\em Dispatch:} For every $ q \swi_{\m{M}} q' \vtl \gamma$,  we have $ q \swi_{\m{M}_{\sf pf}} q' \vtl \gamma$ and $ q \swi_{\m{M}_{\sf pf}} q' \vtl {\sf f}(\gamma)$.
\end{iteMize}

Then, the relation between a thread execution of $\m{M}$ and  a thread execution of $\m{M}_{\sf pf}$ is given by the following lemma:

\begin{lem}
\label{lemm.pf-stack-differ}
Let $j \in \mathbb{N}$, $p_1,p_2 \in Q$, $w_1 \in ((\Gamma')^* \cup
\Gamma)$, $w_2 \in (\Gamma')^*$,  $\mathit{Val} \in
      [\mathit{Loc}(\m{M}) \ra \mathbb{N}]$, and $\mathit{Val}' \in
      [\mathit{Loc}(\m{M}_{\sf pf}) \ra \mathbb{N}]$. There is $\tau'
      \in \Delta_{\sf pf}^*$ such that 
\[(p_1,\bot, \id{\mathit{Loc}(\m{M}_{\sf pf})}{\{(w_1,j)\}})\,
\lby{\tau'}{}_{\m{T}_{j}(\m{M}_{\sf pf})}\, (p_2,\bot,\mathit{Val}'+
\id{\mathit{Loc}(\m{M}_{\sf pf})}{\{(w_2,j+1)\}})
\]
  if and only if there is  $\tau \in \Delta^*$ such that 
\[(p_1,\bot, \id{\mathit{Loc}(\m{M}_{\sf pf})}{\{({\sf
    h}(w_1),j)\}})\, \lby{\tau}{}_{\m{T}_{j}(\m{M})}\,
(p_2,\bot,\mathit{Val}+ \id{\mathit{Loc}(\m{M}_{\sf pf})}{\{({\sf
    h}(w_2),j+1)\}})
\]
 and $\mathit{Val}'((\gamma',j+1))=\mathit{Val}((\gamma',j+1))$ for all $\gamma' \in \Gamma$, and $\mathit{Val}'((w',l))=\mathit{Val}((w,l))=0$ for all $w \in \Gamma^*$, $w' \in \Gamma_{\sf pf}^*$, and $l \in \mathbb{N}$ such that $(w,l), (w',l) \notin \Gamma \times \{j+1\}$.
\end{lem}

Moreover, we can  show that the BSR$[k]$ problem for $\m{M}$ is reducible to  its corresponding problem for $\m{M}_{\sf pf}$.

\begin{lem}
For every $k \in \mathbb{N}$, a state $q \in Q$ is $k$-bounded reachable  by  $\m{M}$ if and only if  $q$ is  $k$-bounded reachable  by  $\m{M}_{\sf pf}$.
\end{lem}

\subsection{From the DCPS $\m{M}_{\sf pf}$ and the DCFS $\m{M}_{\sf fs}$ to the DCPS $\m{M}_{\cup}$}

We define the DCPS  $\m{M}_{\cup}=(Q_{\cup},\Gamma_{\cup},\Delta_{\cup},q_0,\gamma_0,F)$ as the union of $\m{M}_{\sf pf}$ and $\m{M}_{\sf fs}$ where: $(1)$ $Q_{\cup}=Q_{\sf fs}$, $(2)$ $\Gamma_{\cup}=\Gamma_{\sf pf} \cup \Gamma_{\sf fs} $ is a finite set of stack symbols, and $(3)$ $\Delta_{\cup}=\Delta_{\sf pf} \cup \Delta_{\sf fs}$ is the transition relation.

Now, we are ready to define the rank of a run of $\m{T}({\m{M}_{\cup}})$. Intuitively, the number of threads that are simulated according to $\Delta_{\sf fs}$ (resp. $\Delta_{\sf ls}$) is given by the number of pending threads with stack  configuration in $ {S}_{\sf fs}^{\sf sw}  \times \mathbb{N}$ (resp. $(\Gamma')^* \times \mathbb{N}$). Formally, we have:

\begin{defi}(The rank of a run of $\m{M}_{\cup}$) Let $\rho=c_{\m{M}_{\cup}}^{\sf init} \lby{\tau}{}^*_{\m{T}_{[0,k]}(\m{M}_{\cup})} \, c$ be a run of $\m{T}_{[0,k]}(\m{M}_{\cup})$ such that  $\mathit{Active}(c)=\bot$. The rank of $\rho $, denoted by $\mathit{rank}(\rho)$, is defined by the pair $(m,n)$ with $m=\sum_{(w,j) \in (\Gamma')^* \times \mathbb{N}} \mathit{Idle}(c) ((w,j))$ and $m=\sum_{(\lambda,j) \in S_{\sf fs}^{\sf sw} \times \mathbb{N}} \mathit{Idle}(c) ((\lambda,j))$. 

\end{defi}

\subsection{From a run of rank $(m+1,n)$ of $\m{M}_{\cup}$ to a run of rank $(m,n+1)$ of $\m{M}_{\cup}$}
In the following, we establish that given a run of $\m{M}_{\cup}$ such that there is one thread executed following the set of transitions $\Delta_{\sf pf}$, we can compute a run of $\m{M}_{\cup}$ where the execution of this thread is replaced by an execution of a thread following the set of transitions $\Delta_{\sf fs}$. To this aim, we need first to prove  Lemma \ref{lem.rplc.dcps.dcfs.dir1} which states that for any run of a thread of $\m{M}_{\sf pf}$, we can construct an equivalent run of a thread of $\m{M}_{\sf fs}$.

\begin{lem}
\label{lem.rplc.dcps.dcfs.dir1}
Let  $\gamma \in \Gamma$ and $i,j \in \mathbb{N} $ such that 
$i+j \leq k$. If there are $p_0,p'_0,\ldots,p_i,p'_i \in Q$, $w_1,\ldots,w_{i+1} \in (\Gamma')^*$, $\tau_0,\ldots,\tau_i \in \Delta_{\cup}^*$, and $\mathit{Val}'_0,\ldots,\mathit{Val}'_i \in [\mathit{Loc}(\m{M}_{\cup}) \ra \mathbb{N}]$ such that:

\begin{iteMize}{$\bullet$}
\item $(p_0,\bot,\id{\mathit{Loc}(\m{M}_{\cup})}{\{(\gamma,j)\}}) \lby{\tau_0}{}_{\m{T}_{\{j\}}(\m{M}_{\cup})}^* (p'_0,\bot,\mathit{Val}'_0+\id{\mathit{Loc}(\m{M}_{\cup})}{\{(w_1,j+1)\}})$.

\item For every $l \in [1,i]$, $(p_l,\bot,\id{\mathit{Loc}(\m{M}_{\cup})}{\{(w_l,j+l)\}}) \lby{\tau_l}{}_{\m{T}_{\{j+l\}}(\m{M}_{\cup})}^* (p'_l,\bot,\mathit{Val}'_l+\id{\mathit{Loc}(\m{M}_{\cup})}{\{(w_{l+1},j+l+1)\}})$.
\end{iteMize}

Then, there are $\lambda_1,\ldots, \lambda_{i+1} \in S_{\sf fs}^{\sf sw}$  and $\tau'_0,\ldots,\tau'_i \in \Delta_{\cup} ^*$ such that:
\begin{iteMize}{$\bullet$}
\item $(p_0,\bot,\id{\mathit{Loc}(\m{M}_{\cup})}{\{(\gamma,j)\}}) \lby{\tau'_0}{}_{\m{T}_{\{j\}}(\m{M}_{\cup})}^* (p'_0,\bot,\mathit{Val}'_0+\id{\mathit{Loc}(\m{M}_{\cup})}{\{(\lambda_1,j+1)\}})$.

\item For every $l \in [1,i]$, $(p_l,\bot,\id{\mathit{Loc}(\m{M}_{\cup})}{\{(\lambda_l,j+l)\}}) \lby{\tau'_l}{}_{\m{T}_{\{j+l\}}(\m{M}_{\cup})}^* (p'_l,\bot,\mathit{Val}'_l+\id{\mathit{Loc}(\m{M}_{\cup})}{\{(\lambda_{l+1},j+l+1)\}})$.
\end{iteMize}
\end{lem}

\proof
Let us assume there are $p_0,p'_0,\ldots,p_i,p'_i \in Q$, $w_1,\ldots,w_{i+1} \in (\Gamma')^*$, $\tau_0,\ldots,\tau_i \in \Delta_{\cup}^*$, and $\mathit{Val}'_0,\ldots,\mathit{Val}'_i \in [\mathit{Loc}(\m{M}_{\cup}) \ra \mathbb{N}]$ such that:

\begin{iteMize}{$\bullet$}
\item $(p_0,\bot,\id{\mathit{Loc}(\m{M}_{\cup})}{\{(\gamma,j)\}}) \lby{\tau_0}{}_{\m{T}_{\{j\}}(\m{M}_{\cup})}^* (p'_0,\bot,\mathit{Val}'_0+\id{\mathit{Loc}(\m{M}_{\cup})}{\{(w_1,j+1)\}})$.

\item For every $l \in [1,i]$, $(p_l,\bot,\id{\mathit{Loc}(\m{M}_{\cup})}{\{(w_l,j+l)\}}) \lby{\tau_l}{}_{\m{T}_{\{j+l\}}(\m{M}_{\cup})}^* (p'_l,\bot,\mathit{Val}'_l+\id{\mathit{Loc}(\m{M}_{\cup})}{\{(w_{l+1},j+l+1)\}})$.
\end{iteMize}

From now, we confuse the system $\m{M}_{\cup}$  and $\m{M}_{\sf pf}$ (resp. $\m{M}_{\cup}$  and $\m{M}_{\sf fs}$) when $\m{M}_{\cup}$ behaves according to the set of transitions $\Delta_{\sf pf}$ (resp. $\Delta_{\sf fs}$). 

Then, we apply    Lemma \ref{lem.perfect.pda.lin2}  and Lemma \ref{lemm.pf-stack-differ}  to  show that there are $\sigma_0, \ldots,\sigma_i \in \Gamma^*$, $\gamma_0,\ldots,\gamma_i \in \Gamma$, and $g_0,\ldots,g_{i+1} \in Q$ such that:

\begin{iteMize}{$\bullet$}
\item $\gamma_0=\gamma$.
\item For every $l \in [0,i]$, $p_l \swi_{\m{M}_{\cup}} g_l \vtl \gamma_l$.
\item $\sigma_0 (p'_0,\gamma_1,g_1) \sigma_1 \cdots \sigma_{i-1} (p'_{i-1},\gamma_i,g_{i}) \sigma_i (p'_i,\epsilon,g_{i+1})$ in $L'_{((g_0,\gamma),i+1)}$.
\item For every $l \in [0,i]$,  $\mathit{Val}'_l((\gamma',j+l+1))=|\sigma_l|_{\gamma'}$ for all $\gamma' \in \Gamma$.
\item   For every $l \in [0,i]$, $\mathit{Val}'_l((w,l))=0$ for all $w \in \Gamma_{\cup}^*$ and $l \in \mathbb{N}$ such that $(w,l) \notin \Gamma \times \{j+l+1\}$.
\end{iteMize}\smallskip

\noindent Since $L'_{((g_0,\gamma),i+1)} \subseteq L(\m{A}_{(g_0,\gamma)})$, there are  $s_0,\ldots,s_{i+1} \in S_{(g_0,\gamma)}$ such that: $(1)$ $s_0 \in I_{(g_0,\gamma)}$, $(2)$    $s_l\lby{\sigma_l(p'_l,\gamma_{l+1},g_{l+1})}{}_{\m{T}(\m{A}_{(g_0,\gamma)})}^* s_{l+1}$ for all $l \in [0,i[$, and $(3)$  $s_i\lby{\sigma_l(p'_{i},\epsilon,g_{i+1})}{}_{\m{T}(\m{A}_{(g_0,\gamma)})}^* s_{i+1}$.

Let $\lambda_l=(g_l,(s_{l},\gamma_l))$ for all $l \in [1,i]$ and $\lambda_{i+1}=(g_{i+1},(s_{i+1}, \epsilon))$.  Since $p_l \swi_{\m{M}_{\cup}} g_l \vtl \gamma_l$ for all  $l \in [1,i]$, we  can use the definition of $\m{M}_{\sf fs}$ to show  that   $p_l \swi_{\m{M}_{\cup}} g_l \vtl \lambda_l$ for all $l \in [1,i]$.

Now, we can apply Lemma \ref{lem.dcfs.fsa.lin2} to prove that there are $\tau'_0,\ldots,\tau'_i \in \Delta_{\cup} ^*$ such that:
\begin{iteMize}{$\bullet$}
\item $(p_0,\bot,\id{\mathit{Loc}(\m{M}_{\cup})}{\{(\gamma,j)\}}) \lby{\tau'_0}{}_{\m{T}_{\{j\}}(\m{M}_{\cup})}^* (p'_0,\bot,\mathit{Val}'_0+\id{\mathit{Loc}(\m{M}_{\cup})}{\{(\lambda_1,j+1)\}})$.

\item For every $l \in [1,i]$, $(p_l,\bot,\id{\mathit{Loc}(\m{M}_{\cup})}{\{(\lambda_l,j+l)\}}) \lby{\tau'_l}{}_{\m{T}_{\{j+l\}}(\m{M}_{\cup})}^* (p'_l,\bot,\mathit{Val}'_l+\id{\mathit{Loc}(\m{M}_{\cup})}{\{(\lambda_{l+1},j+l+1)\}})$.
\end{iteMize}
\qed

Next, we  show that  if  some state $q$ is $k$-bounded reachable by  a   run of $\m{M}_{\cup}$ of rank $(m+1,n)$, then $q$ is $k$-bounded reachable by  a   run of $\m{M}_{\cup}$ of rank $(m,n+1)$.

\begin{lem}
\label{lemm.sec6.1.dcps-dcfs-dir1}
Let $(m,n) \in \mathbb{N} \times \mathbb{N}$,   $c_{\m{M}_{\cup}}^{\sf init} \lby{\tau}{}^*_{\m{T}_{[0,k]}(\m{M}_{\cup})} \, c$ be a run   of rank $(m+1,n)$ such that  $\mathit{Active}(c)=\bot$. Then, there is  a  run $c_{\m{M}_{\cup}}^{\sf init} \lby{\tau'}{}^*_{\m{T}_{[0,k]}(\m{M}_{\cup})} \, c'$  of rank $(m,n+1)$ such that $\mathit{Active}(c')=\bot$, and $\mathit{State}(c')=\mathit{State}(c)$.\end{lem}

\medskip

\proof

Let us assume that $c_{\m{M}_{\cup}}^{\sf init}
\lby{\tau}{}^*_{\m{T}_{[0,k]}(\m{M}_{\cup})} \, c$ is a run of rank
$(m+1,n)$ with $\mathit{Active}(c)=\bot$. Then, 
by the definition of DCPSs there are
$i,j \in \mathbb{N}$,
$\gamma \in \Gamma$, $p_0,p'_0,\ldots,p'_i,p_{i+1} \in Q$,
$w_1,\ldots,w_{i+1} \in (\Gamma')^*$,
$\kappa_0,\tau_0,\kappa_1,\tau_1,\ldots,\tau_i,\kappa_{i+1} \in
\Delta_{\cup}^*$, and
$\mathit{Val}_0,\mathit{Val}'_0,\ldots,\mathit{Val}'_i,\mathit{Val}_{i+1}
\in [\mathit{Loc}(\m{M}_{\cup}) \ra \mathbb{N}]$ such that the
following conditions are satisfied:

\begin{iteMize}{$\bullet$}
\item $i+j \leq k$.
\item $\tau=\kappa_0\tau_0\kappa_1\tau_1\cdots \tau_i\kappa_{i+1}$.

\item $\mathit{State}(c)=p_{i+1}$ and  $\mathit{Idle}=Val_{l}+\id{\mathit{Loc}(\m{M}_{\cup})}{\{(w_{l+1},j+l+1)\}}$.
\item $c_{\m{M}_{\cup}}^{\sf init} \lby{\kappa_0}{}^*_{\m{T}_{[0,k]}(\m{M}_{\cup})} (p_0,\bot,\mathit{Val}_0 +\id{\mathit{Loc}(\m{M}_{\cup})}{\{(\gamma,j)\}})$.

\item $(p_0,\bot,\id{\mathit{Loc}(\m{M}_{\cup})}{\{(\gamma,j)\}}) \lby{\tau_0}{}_{\m{T}_{\{j\}}(\m{M}_{\cup})}^* (p'_0,\bot,\mathit{Val}'_0+\id{\mathit{Loc}(\m{M}_{\cup})}{\{(w_1,j+1)\}})$.

\item For every $l \in [1,i+1]$, $(p'_{l-1},\bot,\mathit{Val}'_{l-1}+\mathit{Val}_{l-1}) \lby{\kappa_l}{}_{\m{T}_{[0,k]}(\m{M}_{\cup})}^* (p_l,\bot,\mathit{Val}_{l})$.

\item For every $l \in [1,i]$, $(p_l,\bot,\id{\mathit{Loc}(\m{M}_{\cup})}{\{(w_l,j+l)\}}) \lby{\tau_l}{}_{\m{T}_{\{j+l\}}(\m{M}_{\cup})}^* (p'_l,\bot,\mathit{Val}'_l+\id{\mathit{Loc}(\m{M}_{\cup})}{\{(w_{l+1},j+l+1)\}})$.

\end{iteMize}\medskip

Now, we can apply Lemma \ref{lem.rplc.dcps.dcfs.dir1}  to show that
there are $\lambda_1,\ldots, \lambda_{i+1} \in S_{\sf fs}^{\sf sw}$
as well as $\tau'_0,\ldots,\tau'_i \in \Delta_{\cup} ^*$ such that:
\begin{iteMize}{$\bullet$}
\item $(p_0,\bot,\id{\mathit{Loc}(\m{M}_{\cup})}{\{(\gamma,j)\}}) \lby{\tau'_0}{}_{\m{T}_{\{j\}}(\m{M}_{\cup})}^* (p'_0,\bot,\mathit{Val}'_0+\id{\mathit{Loc}(\m{M}_{\cup})}{\{(\lambda_1,j+1)\}})$.

\item For every $l \in [1,i]$, $(p_l,\bot,\id{\mathit{Loc}(\m{M}_{\cup})}{\{(\lambda_l,j+l)\}}) \lby{\tau'_l}{}_{\m{T}_{\{j+l\}}(\m{M}_{\cup})}^* (p'_l,\bot,\mathit{Val}'_l+\id{\mathit{Loc}(\m{M}_{\cup})}{\{(\lambda_{l+1},j+l+1)\}})$.

\end{iteMize}

\medskip
\noindent
Then, we can use  the definition of DCPSs to show that there is  a  run $c_{\m{M}_{\cup}}^{\sf init} \lby{\tau'}{}^*_{\m{T}_{[0,k]}(\m{M}_{\cup})} \, c'$  of rank $(m,n+1)$ such that  $\mathit{Active}(c')=\bot$ and $\mathit{State}(c')=\mathit{State}(c)$. 
\qed

\subsection{From a run of rank $(m,n+1)$ of $\m{M}_{\cup}$ to a run of rank $(m+1,n)$ of $\m{M}_{\cup}$}
In the following, we establish that given a run of $\m{M}_{\cup}$ such that there is one thread executed following the set of transitions $\Delta_{\sf fs}$, we can compute a run of $\m{M}_{\cup}$ where the execution of this thread is replaced by an execution of a thread following the set of transitions $\Delta_{\sf pf}$. To this aim, we need first to prove Lemma \ref{lem.rplc.dcps.dcfs.dir2} which states  that for any run of a thread of $\m{M}_{\sf fs}$, we can construct an equivalent run of a thread of $\m{M}_{\sf pf}$.


\begin{lem}
\label{lem.rplc.dcps.dcfs.dir2}
Let  $\gamma \in \Gamma$ and $i,j \in \mathbb{N} $ such that 
$i+j \leq k$. If there are $p_0,p'_0,\ldots,p_i,p'_i \in Q$, $\lambda_1,\ldots, \lambda_{i+1} \in S_{\sf fs}^{\sf sw}$,  $\tau'_0,\ldots,\tau'_i \in \Delta_{\cup} ^*$, and $\mathit{Val}'_0,\ldots,\mathit{Val}'_i \in [\mathit{Loc}(\m{M}_{\cup}) \ra \mathbb{N}]$ such that:

\begin{iteMize}{$\bullet$}

\item $(p_0,\bot,\id{\mathit{Loc}(\m{M}_{\cup})}{\{(\gamma,j)\}}) \lby{\tau'_0}{}_{\m{T}_{\{j\}}(\m{M}_{\cup})}^* (p'_0,\bot,\mathit{Val}'_0+\id{\mathit{Loc}(\m{M}_{\cup})}{\{(\lambda_1,j+1)\}})$.

\item For every $l \in [1,i]$, $(p_l,\bot,\id{\mathit{Loc}(\m{M}_{\cup})}{\{(\lambda_l,j+l)\}}) \lby{\tau'_l}{}_{\m{T}_{\{j+l\}}(\m{M}_{\cup})}^* (p'_l,\bot,\mathit{Val}'_l+\id{\mathit{Loc}(\m{M}_{\cup})}{\{(\lambda_{l+1},j+l+1)\}})$.
\end{iteMize}

Then, there are elements $w_1,\ldots,w_{i+1} \in (\Gamma')^*$,  $\tau_0,\ldots,\tau_i \in \Delta_{\cup}^*$, and  $\mathit{Val}''_0,\ldots,\mathit{Val}''_i \in [\mathit{Loc}(\m{M}_{\cup}) \ra \mathbb{N}]$ such that:

\begin{iteMize}{$\bullet$}

\item $(p_0,\bot,\id{\mathit{Loc}(\m{M}_{\cup})}{\{(\gamma,j)\}}) \lby{\tau_0}{}_{\m{T}_{\{j\}}(\m{M}_{\cup})}^* (p'_0,\bot,\mathit{Val}''_0+\id{\mathit{Loc}(\m{M}_{\cup})}{\{(w_1,j+1)\}})$.

\item For every $l \in [1,i]$, $(p_l,\bot,\id{\mathit{Loc}(\m{M}_{\cup})}{\{(w_l,j+l)\}}) \lby{\tau_l}{}_{\m{T}_{\{j+l\}}(\m{M}_{\cup})}^* (p'_l,\bot,\mathit{Val}''_l+\id{\mathit{Loc}(\m{M}_{\cup})}{\{(w_{l+1},j+l+1)\}})$.
\item For every $l \in [0,i]$, $\mathit{Val}'_l \leq \mathit{Val}''_l$.

\end{iteMize}
\end{lem}

\proof
Let us assume that there are $p_0,p'_0,\ldots,p_i,p'_i \in Q$, $\lambda_1,\ldots, \lambda_{i+1} \in S_{\sf fs}^{\sf sw}$,  $\tau'_0,\ldots,\tau'_i \in \Delta_{\cup} ^*$, and $\mathit{Val}'_0,\ldots,\mathit{Val}'_i \in [\mathit{Loc}(\m{M}_{\cup}) \ra \mathbb{N}]$ such that:

\begin{iteMize}{$\bullet$}

\item $(p_0,\bot,\id{\mathit{Loc}(\m{M}_{\cup})}{\{(\gamma,j)\}}) \lby{\tau'_0}{}_{\m{T}_{\{j\}}(\m{M}_{\cup})}^* (p'_0,\bot,\mathit{Val}'_0+\id{\mathit{Loc}(\m{M}_{\cup})}{\{(\lambda_1,j+1)\}})$.

\item For every $l \in [1,i]$, $(p_l,\bot,\id{\mathit{Loc}(\m{M}_{\cup})}{\{(\lambda_l,j+l)\}}) \lby{\tau'_l}{}_{\m{T}_{\{j+l\}}(\m{M}_{\cup})}^* (p'_l,\bot,\mathit{Val}'_l+\id{\mathit{Loc}(\m{M}_{\cup})}{\{(\lambda_{l+1},j+l+1)\}})$.
\end{iteMize}

Then, we apply    Lemma \ref{lem.dcfs.fsa.lin2}   to  show that there are $\sigma_0, \ldots,\sigma_i \in \Gamma^*$,   $\gamma_0,\ldots,\gamma_i \in \Gamma$, $\alpha \in \Gamma_{\epsilon}$, $g_0,\ldots,g_{i+1} \in Q$, $s_0 \in I_{(g_0,\gamma)}$, and $s \in S_{(g_0,\gamma)}$ such that:

\begin{iteMize}{$\bullet$}
\item $\gamma_0=\gamma$.
\item For every $l \in [0,i]$, $p_l \swi_{\m{M}_{\cup}} g_l \vtl \gamma_l$.
\item $\sigma_0 (p'_0,\gamma_1,g_1) \sigma_1 \cdots \sigma_{i-1} (p'_{i-1},\gamma_i,g_{i}) \sigma_i (p'_i,\alpha,g_{i+1})$ in $\mathit{Traces}_{\m{T}(\m{A}_{(g_0,\gamma)})}(\{s_0\},\{s\})$.
\item For every $l \in [0,i]$,  $\mathit{Val}'_l((\gamma',j+l+1))=|\sigma_l|_{\gamma'}$ for all $\gamma' \in \Gamma$.
\item   For every $l \in [0,i]$, $\mathit{Val}'_l((w,l))=0$ for all $w \in \Gamma_{\cup}^*$ and $l \in \mathbb{N}$ such that $(w,l) \notin \Gamma \times \{j+l+1\}$.
\end{iteMize}

\noindent On the other hand, we can use the Lemma
\ref{lem.rel.fsa.sec6} to show that 
\[\sigma_0 (p'_0,\gamma_1,g_1) \sigma_1 \cdots \sigma_{i-1}
(p'_{i-1},\gamma_i,g_{i})\sigma_i (p'_i,\epsilon,g_{i+1})\in
L(\m{A}_{(g_0,\gamma)}).
\]
 Now, we can use  the definition of $L(\m{A}_{(g_0,\gamma)})$ to  show that  there are  $\sigma'_0,\ldots,\sigma'_i \in \Gamma^*$ such that  $\sigma'_0 (p'_0,\gamma_1,g_1) \sigma'_1 \cdots \sigma'_{i-1} (p'_{i-1},\gamma_i,g_{i}) \sigma'_i (p'_i,\epsilon,g_{i+1}) \in L'_{((g_0,\gamma),i+1)}$ and $\sigma_l \preceq \sigma'_l$ for all $l \in [0,i]$.

 Then, we can apply  \ref{lem.perfect.pda.lin2}  and Lemma \ref{lemm.pf-stack-differ} to prove that there are $w_1,\ldots,w_{i+1} \in (\Gamma')^*$,  $\tau_0,\ldots,\tau_i \in \Delta_{\cup}^*$,  and $\mathit{Val}''_0,\ldots,\mathit{Val}''_i \in [\mathit{Loc}(\m{M}_{\cup}) \ra \mathbb{N}]$ such that:
\begin{iteMize}{$\bullet$}

\item $(p_0,\bot,\id{\mathit{Loc}(\m{M}_{\cup})}{\{(\gamma,j)\}}) \lby{\tau_0}{}_{\m{T}_{\{j\}}(\m{M}_{\cup})}^* (p'_0,\bot,\mathit{Val}''_0+\id{\mathit{Loc}(\m{M}_{\cup})}{\{(w_1,j+1)\}})$.

\item For every $l \in [1,i]$, $(p_l,\bot,\id{\mathit{Loc}(\m{M}_{\cup})}{\{(w_l,j+l)\}}) \lby{\tau_l}{}_{\m{T}_{\{j+l\}}(\m{M}_{\cup})}^* (p'_l,\bot,\mathit{Val}''_l+\id{\mathit{Loc}(\m{M}_{\cup})}{\{(w_{l+1},j+l+1)\}})$.
\item For every $l \in [0,i]$, $\mathit{Val}'_l \leq \mathit{Val}''_l$.
\item For every $l \in [0,i]$,  $\mathit{Val}''_l((\gamma',j+l+1))=|\sigma'_l|_{\gamma'}$ for all $\gamma' \in \Gamma$.
\item   For every $l \in [0,i]$, $\mathit{Val}''_l((w,l))=0$ for all $w \in \Gamma_{\cup}^*$ and $l \in \mathbb{N}$ such that $(w,l) \notin \Gamma \times \{j+l+1\}$.\qed
\end{iteMize}\medskip

\noindent Next, we  show that  if  some state $q$ is $k$-bounded reachable by  a   run of $\m{M}_{\cup}$ of rank $(m,n+1)$, then $q$ is $k$-bounded reachable by  a   run of $\m{M}_{\cup}$ of rank $(m+1,n)$.

\begin{lem}
\label{lemm.sec6.1.dcps-dcfs-dir3}
Let $(m,n) \in \mathbb{N} \times \mathbb{N}$,   $c_{\m{M}_{\cup}}^{\sf init} \lby{\tau}{}^*_{\m{T}_{[0,k]}(\m{M}_{\cup})} \, c$ be a run   of rank $(m,n+1)$ such that  $\mathit{Active}(c)=\bot$. Then, there is  a  run $c_{\m{M}_{\cup}}^{\sf init} \lby{\tau'}{}^*_{\m{T}_{[0,k]}(\m{M}_{\cup})} \, c'$  of rank $(m+1,n)$ such that $\mathit{Active}(c')=\bot$, and $\mathit{State}(c')=\mathit{State}(c)$.\end{lem}

\proof
Let us assume that $c_{\m{M}_{\cup}}^{\sf init} \lby{\tau}{}^*_{\m{T}_{[0,k]}(\m{M}_{\cup})} \, c$ is  a run of rank $(m,n+1)$ such that  $\mathit{Active}(c)=\bot$. Then,   we can use the definition of DCPSs and Lemma \ref{lem.rplc.dcps.dcfs.dir2} to show that there is  a  run $c_{\m{M}_{\cup}}^{\sf init} \lby{\tau'}{}^*_{\m{T}_{[0,k]}(\m{M}_{\cup})} \, c'$  of rank $(m+1,n)$ such that  $\mathit{Active}(c')=\bot$, and $\mathit{State}(c')=\mathit{State}(c)$.
\qed

\end{document}